\documentclass{jfm}

\usepackage{graphicx}
\usepackage{newtxtext}
\usepackage{newtxmath}
\usepackage{natbib}
\usepackage{hyperref}
\usepackage{import}
\usepackage{xcolor}
\usepackage[acronym]{glossaries}
\usepackage{subfig}
\usepackage{pdflscape}
\usepackage{floatrow}
\hypersetup{
    colorlinks = true,
    urlcolor   = blue,
    citecolor  = black,
    linkcolor  = black
}

\newcommand{\RomanNumeralCaps}[1]
\linenumbers

\newacronym{sbli}{SBLI}{shock wave/boundary layer interaction}
\newacronym{rans}{RANS}{Reynolds-averaged Navier--Stokes simulation}
\newacronym{piv}{PIV}{particle image velocimetry}
\newacronym{les}{LES}{large eddy simulation}
\newacronym{kh}{K--H}{Kelvin--Helmholtz}
\newacronym{iles}{ILES}{implicit large eddy simulation}
\newacronym{lda}{LDA}{laser Doppler anemometry}
\newacronym{dns}{DNS}{direct numerical simulation}
\newacronym{hpc}{HPC}{high performance computing}
\newacronym{vg}{VG}{vortex generator}
\newacronym{mvg}{MVG}{microvortex generator}
\newacronym{ibm}{IBM}{immersed boundary method}
\newacronym{streams}{\textsc{STREAmS~2.0}}{Supersonic TuRbulEnt Accelerated navier stokes Solver 2.0}
\newacronym{rms}{rms}{root mean square}
\newacronym{psd}{PSD}{power spectral density}
\newacronym{weno}{WENO}{weighted essentially non-oscillatory}
\newacronym{pdf}{pdf}{probability density function}
\newacronym{gpu}{GPU}{graphics processing unit}
\newacronym{gfm}{GFM}{Gallery of Fluid Motion}
\newacronym{dfd}{DFD}{Division of Fluid Dynamics}
\newacronym{rmi}{RMI}{Richtmyer--Meshkov instability}
\newacronym{dmd}{DMD}{dynamic mode decomposition}

\definecolor{lightgr}{rgb}{0.85,0.85,0.85}
\definecolor{darkgr}{rgb}{0.7,0.7,0.7}
\definecolor{grey1}{rgb}{0.2 0.2 0.2}
\definecolor{grey2}{rgb}{0.45 0.45 0.45}
\definecolor{grey3}{rgb}{0.6 0.6 0.6}
\definecolor{myred}{rgb}{0.0 0.0 0.0} 
\definecolor{myorange}{rgb}{1.000000000000000 0.501960784313725 0.0}
\definecolor{lightbl}{rgb}{0.0 0.666666666666667 1.000000000000000}
\definecolor{lightgr}{rgb}{0.333333333333333 0.666666666666667 0.0}

\title{High-fidelity simulations of microramp-controlled \\
shock wave/boundary layer interaction}

\author{G. Della Posta\aff{1}
  \corresp{\email{giacomo.dellaposta@uniroma1.it}},
  E. Martelli\aff{2},
  F. Salvadore\aff{3}
 \and M. Bernardini\aff{1}}

\affiliation{\aff{1}Department of Mechanical and Aerospace Engineering, Sapienza University of Rome, via Eudossiana 18, 00184, Rome (RM), Italy
\aff{2} Department of Mechanical and Aerospace Engineering, Politecnico di Torino, Corso Duca degli Abruzzi 24, 10129, Torino (TO), Italy
\aff{3} HPC Department, CINECA, via dei Tizii 6/B, 00185, Rome (RM), Italy}

\begin{document}
\maketitle

\begin{abstract}
Microvortex generators (MVGs) are a promising solution to control shock wave/turbulent boundary layer interactions (SBLIs), especially in supersonic inlets. In this study, we examine the effects of a microramp VG on an SBLI generated by an oblique shock wave and a turbulent boundary layer using direct numerical simulations (DNSs). Two cases, with and without the presence of a microramp, are compared in terms of their mean and unsteady flow features at free-stream Mach number equal to 2 and friction Reynolds number at the inviscid shock impingement equal to 600. The long integration period allows us to assess how microramps affect the typical low-frequency unsteadiness observed in SBLIs. The analysis shows that the three-dimensional microramp wake alters dramatically the interaction region, inducing a significant spanwise modulation and topology change of the separation. For example, tornado-like structures redistribute the flow in both the spanwise and wall-normal directions inside the recirculation region. The increase in momentum close to the wall by the ramp vortices effectively delays the onset of the separation, and thus the separation length, but at the same time leads to a significant increase in the intensity of the wall-pressure fluctuations. We then characterise the mutual interaction between the arch-like vortices around the ramp wake and the SBLI. The specific spanwise vorticity shows that these vortices follow the edge of the separation and their intensity, apart from mean compressibility effects, is not affected by the shocks. The shocks, instead, are deformed in shape by the periodic impingement of the vortices, although the spectral analysis did not reveal any significant trace of their shedding frequency in the separation region. These Kelvin-Helmholtz vortices, however, may be relevant in the closure of the separation bubble. Fourier analysis also shows a constant increase, in both value and magnitude, in the low-frequency peak all along the span, suggesting that the motion of the separation shock remains coherent while being disturbed by the arch-like vortices and oscillating at a higher frequency in absolute terms. 
\end{abstract}

\begin{keywords}
Supersonic flow, turbulent boundary layers, shock waves, turbulence simulation, flow control, vortex interactions.
\end{keywords}
%
\section{Introduction}
\label{sec:introduction}

It is widely recognised that the interaction between shock waves and boundary layers poses significant challenges for aerospace systems. 
Increased thermo-mechanical loads, shock-induced separation, amplified  
pressure losses, and intermittent, low-frequency flow unsteadiness that 
may interact with structures represent only some of the possible hazardous effects 
\gls{sbli} may cause. 
For this reason, significant research efforts
during the last decade have focused on possible control solutions aiming 
at cancelling, or at least mitigating, some of the detrimental consequences of \gls{sbli}.

Among the proposed control remedies, \glspl{mvg} -- and in particular microramps --
are promising passive devices, smaller than the boundary layer thickness,
that generate a system of trailing vortices energising the boundary layer \citep{mccormick1993shock, anderson2006optimal}. 
Such vortices bring high-momentum fluid closer to the wall,
which makes the velocity profile fuller and hence
more resistant to the separation induced by the following shock wave impingement. 
Researchers have demonstrated that \glspl{mvg} have the potential to reduce 
shock-induced separation, even if some aspects of the flow generated 
are still unclear due to the complexity of the new interaction taking place 
between the shock and the incoming flow \citep{lu2012microvortex, panaras2015micro, titchener2015review}.

Several studies have first clarified the organisation of the wake behind 
a microramp immersed into a turbulent boundary layer. 
Both experimental \citep{sun2012three} and numerical \citep{lee2009supersonic} results 
regarding the mean flow field revealed a low-momentum region in the wake, 
associated with the so-called primary vortex pair. 
Indeed, the two vortices developing at the sides of the \gls{mvg} converge at the trailing edge 
and then proceed approximately in parallel in the streamwise direction. 
Besides adding momentum to the region close to the wall, the primary vortices
are associated with the presence of secondary vortices at the bottom and top corners at the side walls of the ramp and mutually induce a lift-up at the symmetry plane that gradually pushes the wake far from the wall after the ramp \citep{babinsky2009microramp, lu2010experimental}. 
The vortex pair decays slowly and continues to energise the boundary layer even far downstream of the main interaction, when \gls{sbli} is present \citep{ghosh2008rans}. 
The differences in the mean wake development have been quantified for several geometrical and flow parameters,  
like the relative height of the ramp \citep{giepman2015effects, babinsky2009microramp, tambe2021relation}, 
the Reynolds number \citep{dellaposta2023direct_reynolds, salvadore2023direct}, and 
the Mach number \citep{giepman2016mach, dellaposta2023direct_mach}. 

The instantaneous flow structure is, instead, more controversial. 
Researchers agree about the periodic shedding from the ramp 
of almost-toroidal vortical structures quickly developing around 
the low-momentum region as a consequence of \gls{kh} instability, 
but they debate on their precise nature. 
On one side, \gls{kh} vortices are seen as closed vortex rings with 
connected filaments in the bottom part \citep{li2010supersonic}. 
On the other side, it is assumed that the legs of the top arch are not connected and, instead, 
become parallel to the wall in their bottom part, thus forming hairpin vortices \citep{blinde2009effects}.
Through the analysis of \gls{dns} results in terms of mean vorticity field and characteristic-based \gls{dmd}, 
\cite{dellaposta2023direct_mach} characterised the properties of the top arch-like vortical structures for a range of Mach numbers, 
but -- as in \cite{bo2012experimental} -- did not detect the bottom vortex cores that should be associated with 
the closure of the vortex rings at the symmetry plane. 
Results thus suggest that wall turbulence dominates over the \gls{kh} instability of the bottom shear layer 
and prevents the formation of closed vortex rings even after a large distance from the ramp, 
where the wake is further from the wall because of the lift-up from the primary vortex pair.  
Moreover, the analysis of the near wake in \cite{dellaposta2023direct_reynolds}
highlighted a strong connection between the vortical structures around the wake and those 
inside of it close to the ramp trailing edge, as a consequence of the internal convolution 
of the primary vortices at the sides of the ramp.  
Various conceptual models have been proposed 
describing the evolution of the instantaneous vortical organisation 
\citep{blinde2009effects, bo2012experimental, sun2014decay, dellaposta2023direct_reynolds}
but a clear and definite understanding is still missing. 

The introduction in the field of a shock wave impinging on the perturbed boundary layer 
complicates further the scenario. 
The first effect of \glspl{mvg} on \gls{sbli} is the overall reduction of the extent of the separation region, 
observed in both experiments \citep{blinde2009effects, babinsky2009microramp, giepman2014flow} and 
simulations \citep{grebert2018simulations, sun2019direct}. 
The shock foot is generally displaced downstream, and the separation length is reduced compared to the uncontrolled case. 
In some experimental works, reversed flow even disappears at some spanwise sections \citep{giepman2014flow}, 
although reduced resolution close to the wall affects the results. 
In general, however, a notable difference 
is that the separation is modulated by \glspl{mvg} in the spanwise direction, 
with alternating regions of reduced and increased skin friction. 
Indeed, the microramp wake induces a strong three-dimensionality of the flow impacting the shock wake that 
alters completely the topology of the interaction region compared to the traditional 2D \gls{sbli}. 
Besides presenting the spanwise modulation of the separation length, 
oil-flow visualisations in \cite{babinsky2009microramp} and skin-friction lines based on \gls{les} 
results in \cite{grebert2023microramp} agree with reporting the formation of tornado-like structures lifting  
the flow from the wall inside the recirculation bubble and transporting it downstream. In particular, 
the numerical results of \cite{grebert2023microramp} showed that different regions 
in the span upstream of the interaction contribute differently to the feeding of the separation 
and that the separation bubble is mostly fed by the flow coming from the sides of the ramp wake. 
However, understanding if and how these tornado-like structures have an active role in enhancing or 
attenuating the separation is not straightforward. 

Another relevant effect is the periodic disruption of the shock surface 
due to the arch-like vortices shed by the microramp, 
first observed in the \gls{iles} results of \cite{li2010supersonic}.
When impacting the separation region, the primary vortex pair from the ramp and the arch-like vortices around it 
are mildly affected by the shocks and remain on top of the separation region, 
thus initially rising and later descending towards the wall after the reflected shock. 
\cite{yan2013study} showed that baroclinic sources of vorticity are negligible 
and that the interaction between the shocks and the arch-like vortices 
does not influence the structure of the latter much, also confirmed by \cite{dong2018spectrum}. 
The authors thus suggest that the vortices around the microramp wake keep their shapes and vorticity magnitude and travel undisturbed as the shock is absent. 
Hence, arch-like vortices seem to be unaffected by shocks, although the shocks are affected by the
arch-like vortices, as they generate periodic ``bumps'' in the separation shock surface.

In addition, \glspl{mvg} have an impact on the typical low-frequency unsteadiness associated with \gls{sbli} as well. 
As a matter of fact, \glspl{mvg} influence all the mechanisms that have been hypothesised to determine and affect this phenomenon. 
On the one hand, \glspl{mvg} change dramatically the flow upstream of the interaction, altering the effective thickness of the 
incoming boundary layer, increasing the anisotropy of the incoming turbulence, and altering the 
structure of the typical low-speed streaks colliding with the shocks \citep{lee2010microramps, bo2012experimental, salvadore2023direct}, 
and thus potentially influencing the upstream mechanism proposed by \cite{ganapathisubramani2009low}. 
On the other hand, if downstream mechanisms are considered:
the wake of the microramps completely alters the shape of the separation bubble and thus may affect potential feedback mechanisms like those 
proposed by \cite{pirozzoli2006direct} and \cite{adler2018dynamic}; the additional transfer of momentum towards the wall from the primary vortex pair and the change in the topology of the 
separation region may affect the mass balance and the properties of ﬂuid entrainment in the mixing layer generated downstream of the
separation shock, which is central in the model of \cite{piponniau2009simple}; 
the change in the shape of the bubble and the vortex pair surviving even after the separation may interfere 
with the G\"{o}rtler vortices developing within the separation region and at reattachment observed in \cite{priebe2016low} and \cite{pasquariello2017unsteady}. \\
Based on the variance of the shock foot position, \cite{giepman2014flow} observed that the spanwise modulation 
of the separation bubble has a beneficial effect on the reflected shock unsteadiness. 
This observation is also confirmed by the spectral analysis of the wall pressure from the \gls{les} results 
in \cite{grebert2018simulations, grebert2023microramp}, which shows a slight increase in the dominant peak frequency 
of the shock unsteadiness when averaging wall-pressure spectra in the spanwise direction. However, 
as in the case of the separation, considerable differences are present at different spanwise sections, 
although there does not seem to be a trace of the arch-like shedding frequency identified by \cite{bo2012experimental} even at the symmetry plane.
The authors also confirmed the results of \cite{adler2018dynamic} that, 
for the uncontrolled \gls{sbli}, the recirculation bubble is modulated by a mode at 
a Strouhal number based on the separation length $St_{L_{sep}} = 0.1$, 
an order of magnitude larger than that of the shock unsteadiness. 
For the controlled case, instead, the recirculation bubble is synchronised with the 
shock motion at a Strouhal number of approximately $St_{L_{sep}} = 0.05$. 
However, no physical explanation was provided regarding the mechanism behind this synchronisation. 

The above literature review indicates that relevant research questions about 
the control of \gls{sbli} through \glspl{mvg} remain open. 
The interpretation and description of the mean and instantaneous wake from the microramp is still debated. 
The effects of the 3D and unsteady changes in the separation topology are unclear. 
Regarding the low-frequency unsteadiness, there is no clear physical explanation of the observed increase in the distinctive low frequency in the controlled case, and further analysis is necessary to characterise and understand the effects of the periodic disruption of the reflected shock surface associated with the passage of the arch-like vortices. 
In addition, despite being known that the shock unsteadiness is strongly 
intermittent \citep{dolling2001fifty}, 
little is known regarding if and how microramps affect this intermittency. 
\cite{bernardini2023unsteadiness} showed through wavelet analysis that 
the broadband shock motion can be interpreted as the result of a collection of 
sparse events in time, each characterised by its temporal scale. 
How these events are affected by the unsteady and 3D changes in \gls{sbli} 
due to \glspl{mvg} is yet to be discussed. 

From a methodological point of view, \gls{dns} data are relatively scarce and limited to 
compression-ramp \glspl{sbli} in hypersonic flow conditions only \citep{sun2019direct, sun2020wake}, 
despite the complex three-dimensional and unsteady nature of the flow may benefit from a description without 
simplifying and modelling assumptions. 
Indeed, several studies \citep{ghosh2008rans} proved that \gls{rans} methodology with traditional 
eddy-viscosity turbulence models is completely inadequate even to predict only the mean field, 
but also that \gls{iles} methods fail to accurately reproduce experimental data \citep{lee2010microramps}, 
which in turn suffer from known limitations in resolution and data accessibility.

Given this scenario, this work examines the \gls{dns} database of a \gls{sbli} 
generated by an oblique shock wave impinging on a turbulent boundary layer in the 
presence of an infinite array of microramp \glspl{mvg}. 
The turbulent boundary layer has a free-stream Mach number $M_\infty = U_\infty/a_\infty = 2.28$ and 
a friction Reynolds number at the shock impingement location $Re_\tau = \rho_w \delta u_\tau/\mu_w \approx 600$, 
where $U_\infty$ is the free-stream velocity, $a_\infty$ is the free-stream speed of sound, $\rho_w$ is the density at the wall, $\delta$ is the boundary layer thickness, 
$u_\tau = \sqrt{\tau_w/\rho_w}$ is the friction velocity, $\tau_w$ is the wall shear stress, 
and $\mu_w$ is the dynamic viscosity at the wall. 
The shock generator has an angle $\alpha = 9.5^\circ$, while the microramp 
geometry and position are based on the optimal setup indicated by \citet{anderson2006optimal}.

The analysis compares the results of two simulations, the uncontrolled \gls{sbli}, 
indicated in the following as USBLI, and the controlled one, indicated in the following as CSBLI. 
The qualitative and quantitative effects of the microramps on \gls{sbli}
are assessed in terms of both the mean and the instantaneous flow field. 
Moreover, the long integration period covered by the simulations 
makes it possible to consider the effects of microramps on the \gls{sbli} low-frequency 
unsteadiness, which is studied using both traditional Fourier spectra
and wavelet analysis in the time/frequency domain.  

The paper is organised as follows: Section \ref{sec:method} presents the 
methodology and numerical setup of the simulations; Section \ref{sec:dataset} describes the 
database generated and the validation carried out; Section \ref{sec:results} 
presents the analysis of the results; finally, Section \ref{sec:conclusions} 
reports some final comments.
\section{Methodology and numerical setup}
\label{sec:method}
\glspl{dns} have been carried out using
\textsc{STREAmS~2.0}\footnote{\url{https://github.com/STREAmS-CFD/STREAmS-2}}  \citep{bernardini2021streams,bernardini2023streams, sathyanarayana2023highspeed}.
\textsc{STREAmS} is an open-source, finite-difference flow solver
designed to solve the compressible Navier--Stokes 
equations for a perfect, heat-conducting gas in canonical 
wall-bounded turbulent high-speed flows. 
The solver is oriented to modern CPU-GPU \gls{hpc} platforms and has been 
extensively validated in recent works \citep{dellaposta2023high, yu2023direct}.

The convective terms are discretized through a hybrid energy-conservative/shock-capturing approach in a locally conservative form. 
In smooth regions, stability is maintained using a sixth-order, central, energy-preserving flux formulation, avoiding the need for additional numerical diffusivity. 
Near shock waves, a fifth-order \gls{weno} reconstruction~\citep{jiang1996efficient} is applied to calculate numerical fluxes at cell faces, using a Lax--Friedrichs flux vector splitting. A shock sensor assesses the local smoothness of the solution, determining the presence of discontinuities where central and \gls{weno} schemes switch. Viscous terms are approximated using sixth-order, central finite-difference formulas. Time advancement is achieved through a third-order, low-storage Runge-Kutta scheme~\citep{spalart1991spectral}.

\begin{figure*}[t]
     \centering
     \includegraphics[width=\textwidth]{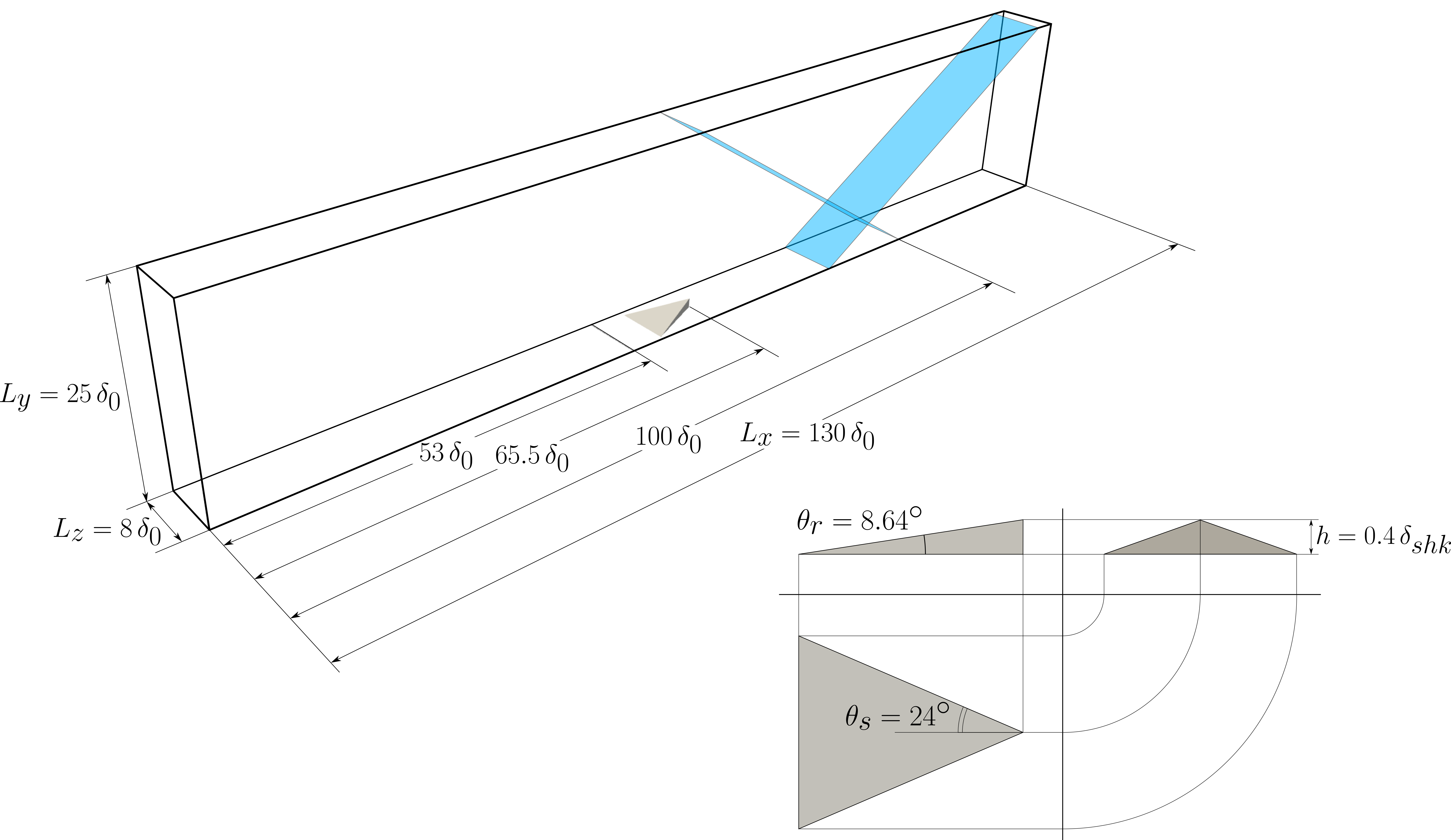} 
     \caption{Domain with sizes and orthogonal projections of the microramp.}
     \label{fig:domain_ramp}
\end{figure*}

The size of the computational domain adopted is 
$L_x/\delta_0 \times L_y/\delta_0 \times L_z/\delta_0 = 130 \times 25 \times 8$, 
with $\delta_0$ being the thickness of the boundary layer at the inflow. 
The geometrical setup of the microramp is based on the optimal configuration 
defined by \cite{anderson2006optimal} (see figure~\ref{fig:domain_ramp} and table~\ref{tab:sbli_geom}), 
with a ratio between the ramp height and the boundary layer thickness at the inviscid shock impingement location 
equal to $h/\delta_{shk}=0.40$ ($h/\delta_0 = 1.073$) and a distance from the ramp trailing edge 
and the inviscid shock impingement location equal to $14.16\,\delta_{shk}$.
The microramp is centred in the spanwise direction ($z_{TE} = L_z/2$) and 
is simulated using a ghost-point-forcing \gls{ibm}~\citep{piquet2016comparative} 
already validated in previous works~\citep{dellaposta2023high}.
Lateral periodic boundary conditions allow representing an infinite array of 
microramps with the Anderson's optimal lateral spacing $s/h = L_z/h = 7.46$, 
despite considering a single microramp only in the computational domain.

At the outflow and the top boundaries, non-reflecting conditions 
are imposed by performing a characteristic decomposition 
in the direction normal to the boundaries~\citep{poinsot1992boundary}. 
Inviscid shock relations are set at the top boundary to impose the presence of the shock 
generator, such that the inviscid shock impingement $x_{imp}$ at the wall is at $100\,\delta_0$ from the inlet. 
A characteristic wave decomposition is also employed at the bottom no-slip wall, where the 
wall temperature is set to the recovery value of the incoming boundary layer to impose weak adiabatic wall conditions.
A recycling-rescaling procedure~\citep{lund1998generation} is used to provide 
the inflow with suitable turbulent fluctuations. 
The recycling station is placed at 53 $\delta_0$ from the inflow to guarantee
a sufficient decorrelation between the inlet and the recycling station.

The mesh is uniform in the wall-parallel directions, corresponding to a viscous-scaled spacing of $\Delta x^+ \approx 4.9$ and $\Delta z^+ \approx 4.8$, in the streamwise and spanwise directions, respectively. 
A stretching is used, instead, in the wall-normal direction, corresponding to a wall-spacing in the range $\Delta y^+\approx0.61$--$0.70$. Unless specified, the coordinates are expressed in terms of the separation length of the uncontrolled case $L_{sep}^u$, according to which {$x^* = (x - x_{imp})/L_{sep}^u$}, {$y^* = y/L_{sep}^u$}, {$(z - z_{TE})/L_{sep}^u$}. 
A total time of 3108 $L_{sep}^u/U_\infty$ for the USBLI case 
and of 1595 $L_{sep}^u/U_\infty$ for the CSBLI case has been recorded, 
with $L_{sep}^u$ being the separation length of the uncontrolled case. 
Considering that the non-dimensional characteristic low frequency typically stands 
in the range $St = f\, L_{sep}^u/U_\infty \in [0.02, 0.06]$, the period considered 
corresponds to 60--190 and 30--100 cycles for the USBLI and CSBLI cases respectively. 
The nondimensional sampling frequency recorded is equal to $L_{sep}^u/(U_\infty \Delta t) = 2.0139$. 
\section{Numerical data set and validation}
\label{sec:dataset}

The main parameters of the numerical database analysed in this work are reported 
in table~\ref{tab:sbli_flow}. 
The data set includes a baseline simulation with the undisturbed \gls{sbli}, 
which is used as a reference to assess the effects of the microramps, and the 
same \gls{sbli} controlled by the microramp.
The friction Reynolds number of the incoming boundary layer based on the
properties at the shock impingement location is approximately 600, 
corresponding to momentum-based Reynolds numbers 
$Re_{\delta_2} = \rho_\infty U_\infty \theta / \mu_w$ 
equal to 1935 and 
$Re_{\theta} = \rho_\infty U_\infty \theta / \mu_\infty$ 
equal to 3205. 
The free-stream Mach number $M_\infty$ is equal to 2.28, 
while the shock generator has an angle of $9.5^\circ$.

\begin{table}[t]
 \begin{center}
\def~{\hphantom{0}}
  \begin{tabular}{ccccccc}
    \hline\noalign{\smallskip}
 	$M_\infty$ & $\alpha$ & $Re_{\tau}$ & $Re_{\theta}$ & $Re_{\delta_2}$ & $N_x \times N_y \times N_z$ & $\frac{T\,U_\infty}{L_{sep}^u}$\\
    \noalign{\smallskip}\hline\noalign{\smallskip}
 	2.28 &  9.5$^\circ$ & \begin{tabular}{@{}c@{}} 505 (TE) \\ 599 (shk) \end{tabular} & \begin{tabular}{@{}c@{}} 2584 (TE) \\ 3205 (shk) \end{tabular} & \begin{tabular}{@{}c@{}} 1560 (TE) \\ 1935 (shk) \end{tabular} & 6144 $\times$ 512 $\times$ 384 & \begin{tabular}{@{}c@{}}3108 (USBLI) \\ 1595 (CSBLI) \end{tabular} \\
    \noalign{\smallskip}\hline
  \end{tabular}
  \caption{Main flow parameters of the numerical database. $\Delta t$ is the sampling time step used to record unsteady data, 
  $T$ is the total time considered for statistics.\\
  TE: property at the microramp trailing edge, shk: property at the inviscid shock impingement, USBLI: uncontrolled \gls{sbli}, CSBLI: microramp-controlled \gls{sbli}. \label{tab:sbli_flow}}
 \end{center}
\end{table}
\begin{table}[t]
 \begin{center}
\def~{\hphantom{0}}
  \begin{tabular}{ccccc}
    \hline\noalign{\smallskip}
 	$h/\delta_{shk}$ & $\theta_s$ & $\theta_r$ & $d/\delta_{shk}$ & $s/h$\\
    \noalign{\smallskip}\hline\noalign{\smallskip}
    0.40 & 24$^\circ$ & 8.64$^\circ$ & 14.16 & 7.46\\
    \noalign{\smallskip}\hline
  \end{tabular}
  \caption{Main geometrical parameters of the microramp based on the optimal configuration of \cite{anderson2006optimal} (see figure \ref{fig:domain_ramp}). \label{tab:sbli_geom}}
 \end{center}
\end{table}

%
\begin{figure}[h]
     \centering
     \subfloat[\label{fig:ramp_wake_uwake}]{
     \includegraphics[width=0.49\textwidth]{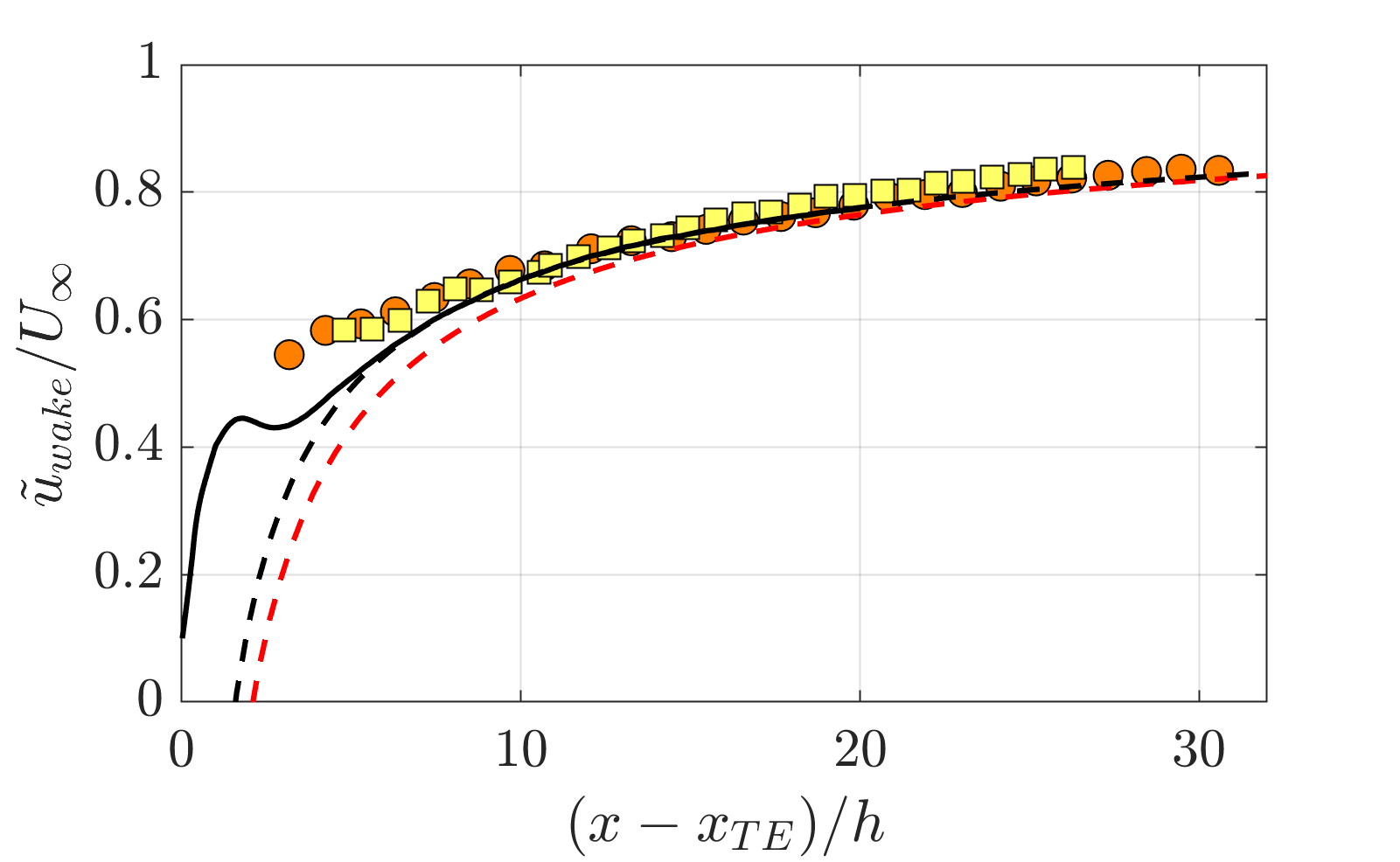}} 
     \subfloat[\label{fig:ramp_wake_vmax}]{
     \includegraphics[width=0.49\textwidth]{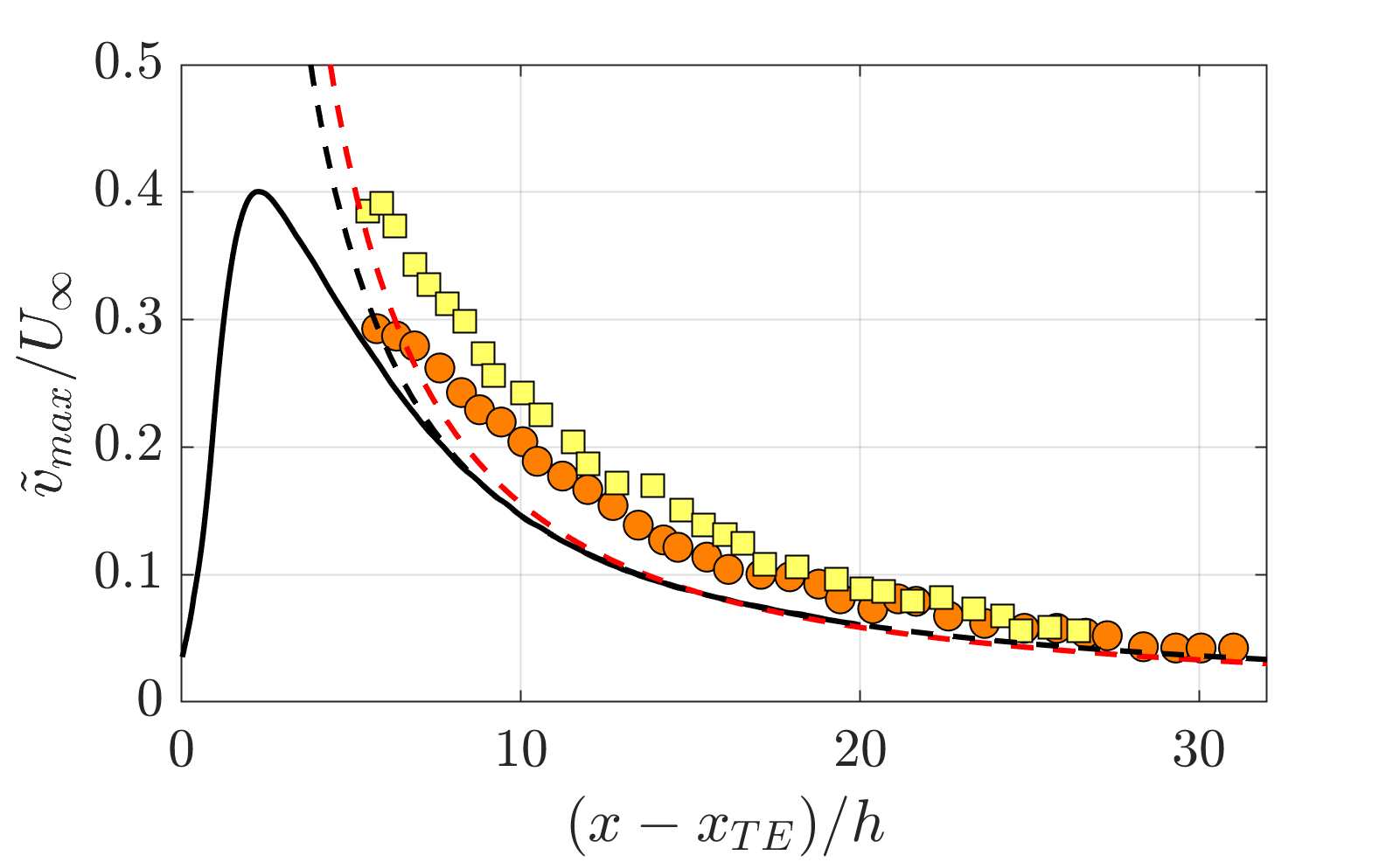}}
     \caption{Streamwise evolution of (a) wake velocity and (b) upwash velocity at the symmetry plane. Experimental data from \citet{giepman2016mach} (orange circles $h/\delta = 0.35$, yellow squares $h/\delta = 0.46$), present data (solid black line), data fitting of LES results from \citet{grebert2023microramp} (dashed red line), data fitting of present data (dashed black line). 
     }
     \label{fig:ramp_wake}
\end{figure}
%

To validate the development of the wake behind the microramp, figure~\ref{fig:ramp_wake} reports two wake properties along the streamwise direction typically considered for microramps and which have been demonstrated to be rather independent of the flow conditions in terms of Mach and Reynolds numbers. 
In particular, figure~\ref{fig:ramp_wake_uwake} reports the 
streamwise evolution of the Favre-averaged wake velocity $\tilde{u}_{wake}/U_\infty$, 
which is defined as the velocity in correspondence with the minimum difference between the controlled and the uncontrolled velocity profiles at the symmetry plane. 
The wake velocity represents a measure of the intensity of the 
low-momentum region generated by the microramp and progressively increases proceeding 
downstream as the wake decays by the action of molecular and turbulent mixing. 
Present results are compared with the experimental measurements of 
\citet{giepman2015effects} for two values of the relative height of the microramp similar 
to the case under study and with the empirical fitting proposed by 
\citet{grebert2023microramp} and based on \gls{les} data. 
The data agree well with the results from the literature, and the coefficients resulting from a power-law fitting differ only slightly from the ones proposed by \citet{grebert2023microramp}, which may be related to the different flow conditions considered in the two cases.
Figure~\ref{fig:ramp_wake_vmax} shows instead a similar comparison for the peak Favre-averaged lift-up velocity $\tilde{v}_{max}/U_\infty$, which is the maximum 
value of the vertical velocity component along the wall-normal coordinate at the symmetry plane.
The peak lift-up velocity represents a measure of the intensity of the lift-up mutually induced by the primary vortex pair at the symmetry plane, which gradually pushes the wake farther from the wall. As for the wake velocity, the agreement with the literature results is satisfactory, with initial discrepancies mainly related to low-Reynolds number effects \citep{dellaposta2023direct_reynolds}. 

\section{Results}
\label{sec:results}

\begin{figure}[t]
     \centering
     \includegraphics[width=\textwidth]{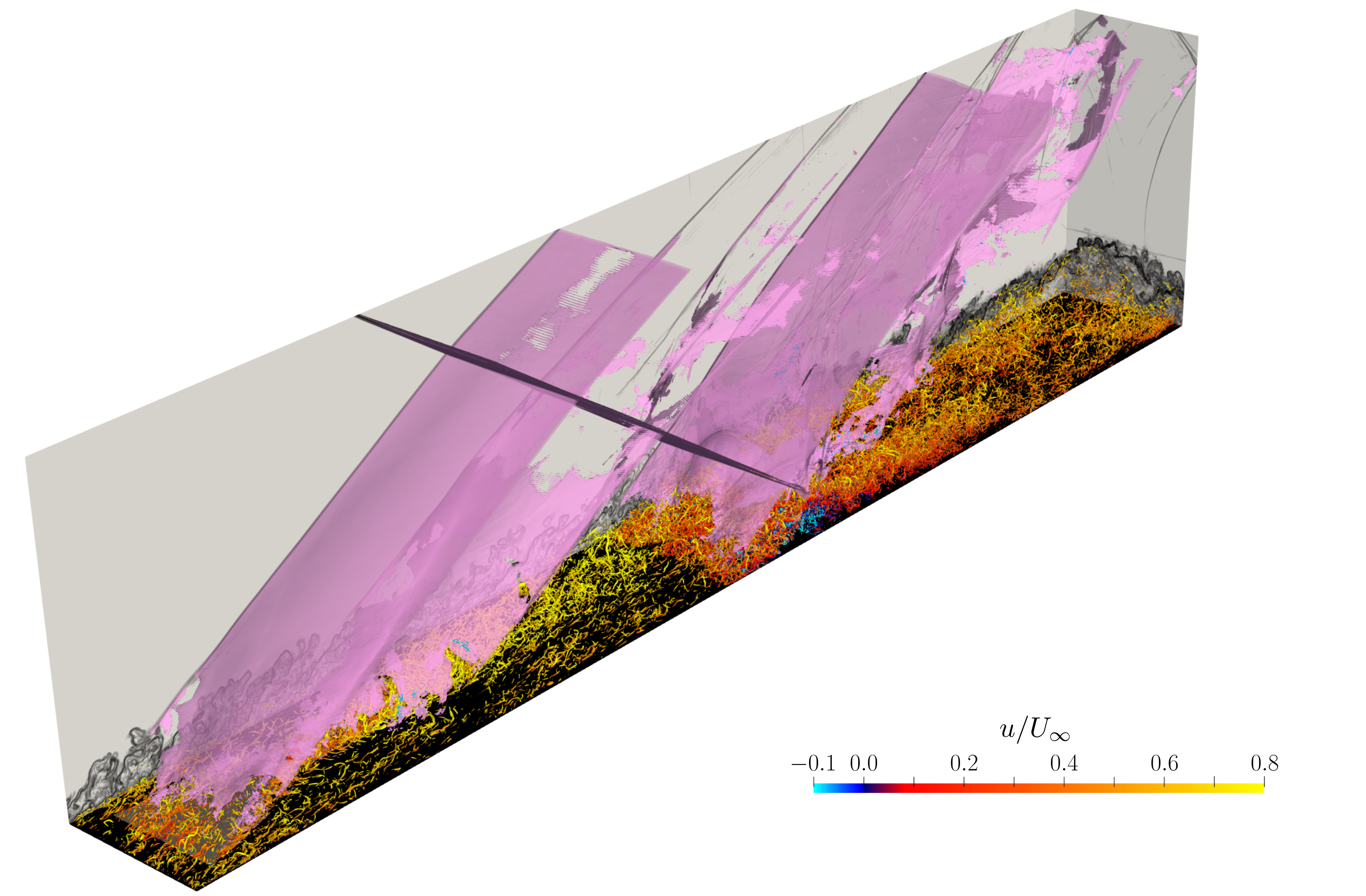} 
     \caption{Instantaneous visualisation of the turbulent and shock structures. 
     Isosurface of the swirling strength coloured by the streamwise velocity component ($\lambda_{ci} L_{sep}^u / U_\infty = 60$), 
     isosurface of the shock sensor in pink ($\theta = 0.9$), numerical schlieren on the xy and yz slices in the background.}
     \label{fig:qcrit}
\end{figure}
\subsection{Qualitative flow organisation}
To first understand the qualitative flow organisation, figure~\ref{fig:qcrit} reports an instantaneous visualisation of the turbulent and shock structures in the flow field
using an isosurface of the swirling strength coloured by the streamwise velocity and an isosurface of the shock sensor. A video of the flow, generated using \textit{in situ} visualisation at runtime \citep{bna2023insitu}, is linked in the supplementary material.
In addition to the shocks associated with the uncontrolled \gls{sbli}, another shock system 
is visible in correspondence with the microramp, which has been extensively characterised in previous works \citep{dellaposta2023direct_reynolds}. In particular, an almost-planar shock wave is generated at the leading edge of the ramp, while a conical shock wave surrounds the wake from the trailing edge. These weak shocks interact with the main impinging shock far from the wall. 
The turbulent structures show, instead, how the incoming flow is first captured at the sides of the microramp, thus generating the primary vortex pair and then the arch-like vortices around the ramp wake. However, the most notable differences with the traditional uncontrolled \gls{sbli} take place in correspondence with the foot of the reflected shock wave, whose shape is significantly altered by the incoming flow. The lower part of the shock is completely disrupted at the centre of the domain, where the ramp wake hits the interaction, and bulges are periodically generated when the \gls{kh} vortices around the wake pass through the shock. The spanwise alteration of the separation region, which is, however, still present as it can be observed from the presence of reversed flow between the shocks, also affects the flow downstream of the interaction region, where the compression wave after the separation is no longer homogeneous in the spanwise direction. 

\begin{figure}[t!]
     \centering
     \subfloat{
     \includegraphics[width=0.36\textwidth]{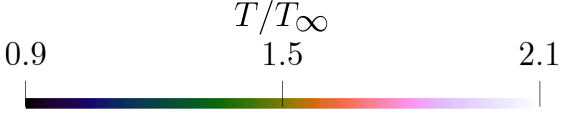}}\\
     \setcounter{subfigure}{0}
     \subfloat[]{
     \includegraphics[width=\textwidth]{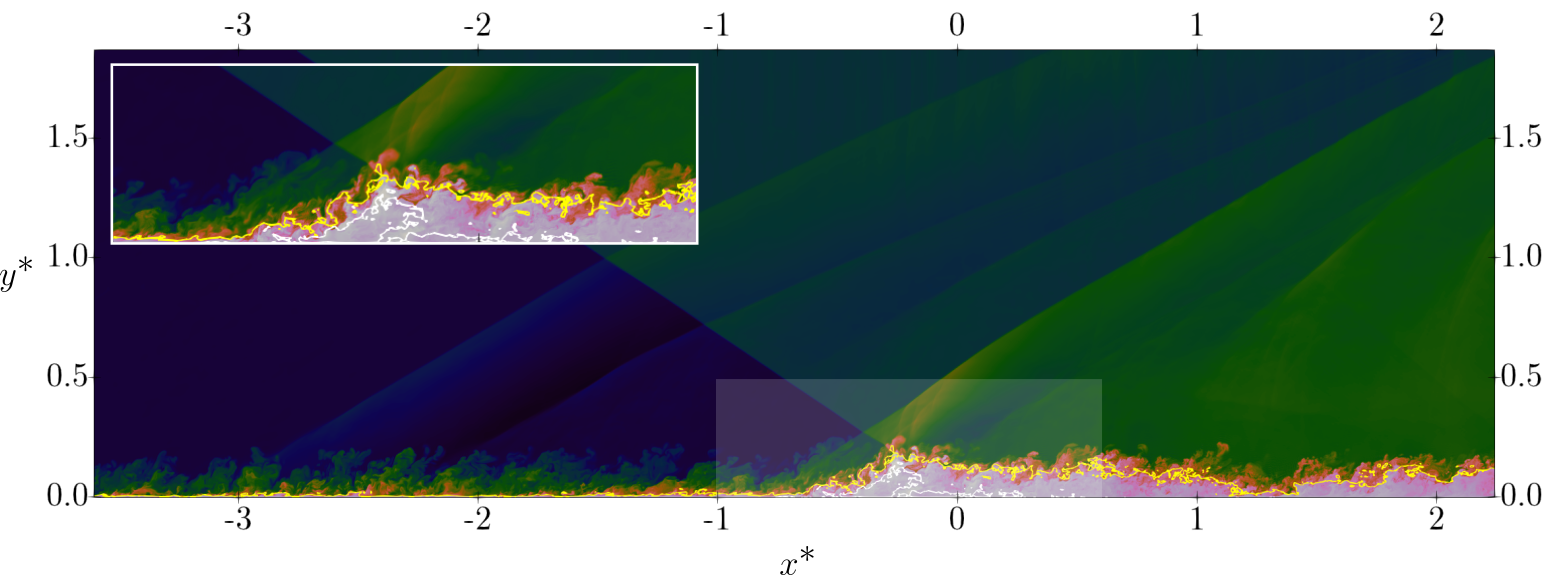}} \\
     \subfloat[]{
     \includegraphics[width=\textwidth]{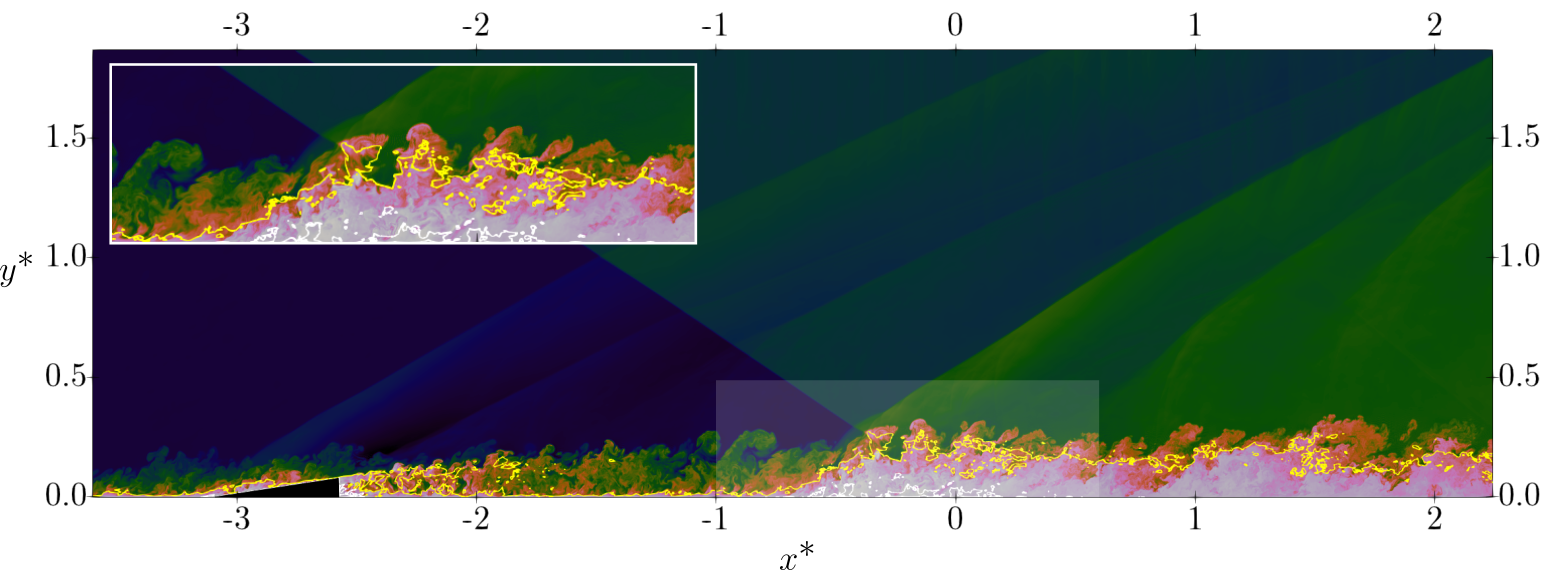}} 
     \caption{Instantaneous temperature on xy planes at (a) $z^* = -0.3$, 
     (b) $z^* = 0.0$. The white lines indicate $u/U_\infty = 0$, while the yellow lines indicate points with $M = 1$. Zoom into the separation region in the boxes.}
     \label{fig:mach_inst}
\end{figure}

The differences along the span of the microramp-controlled \gls{sbli} are also visible in the instantaneous temperature contours at the symmetry and lateral xy planes (figure~\ref{fig:mach_inst}). While the lateral slice reminds of a traditional 2D \gls{sbli}, only with the addition of two extra shocks upstream, the billows of the ramp wake completely alter the incoming flow (as observed in \citet{bo2012experimental}), the interaction region, and the downstream flow at the symmetry plane, with a significantly thicker boundary layer.

\begin{figure}[t!]
     \centering
     \subfloat{
     \includegraphics[width=0.34\textwidth]{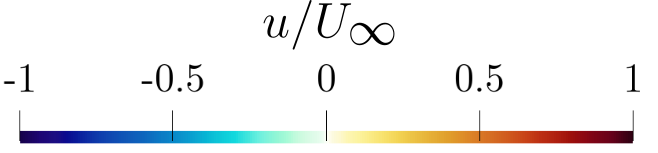}} \\
     \setcounter{subfigure}{0}
     \subfloat[]{
     \includegraphics[width=\textwidth]{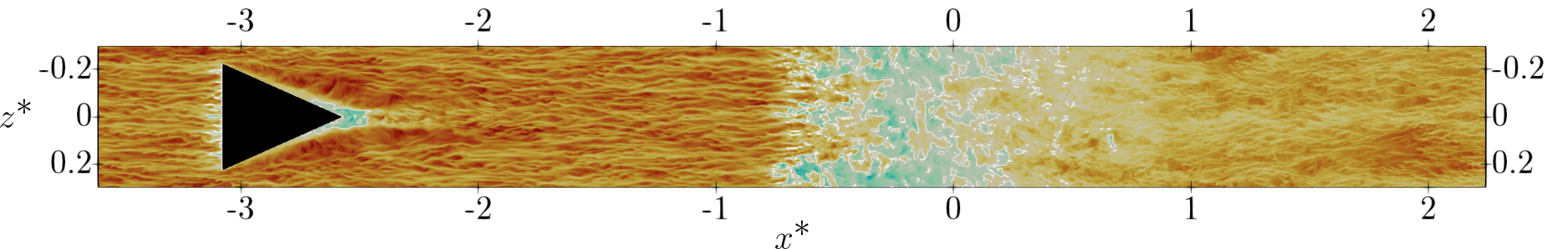}} \\
     \subfloat[]{
     \includegraphics[width=\textwidth]{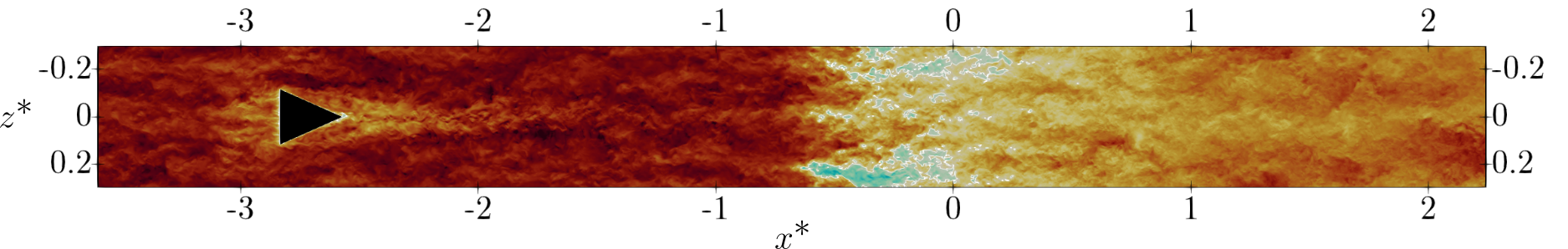}} \\
     \subfloat[]{
     \includegraphics[width=\textwidth]{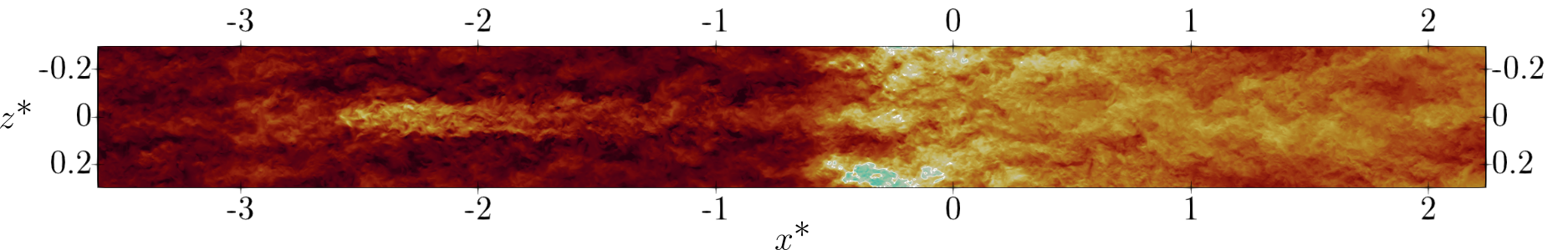}} 
     \caption{Instantaneous streamwise velocity on xz planes at (a) $y^+ = 1$, (b) $y/h = 0.5$, (c) $y/h = 1$. The white lines indicate $u/U_\infty = 0$.}
     \label{fig:streaks}
\end{figure}

As anticipated, the presence of the \gls{mvg} radically affects the incoming boundary layer, which shapes the interaction region in turn. 
Typical near-wall streaks, observable in the velocity contours on xz planes in figure~\ref{fig:streaks}, are now overlapped to the large-scale trace of the ramp wake.
On the one hand, the low-momentum wake decelerates the flow at the centre of the domain, corresponding to an increase in the reversed flow extent. On the other hand, the vortical motion of the primary vortices and the transversal mixing associated with the arch-like vortices promote local accelerations of the flow, corresponding to a decrease in the reversed flow. Before and even after the interaction, the meandering motion of the ramp wake and the spanwise alteration of the separation generate the formation of alternated regions of accelerated and decelerated flows, similar to large-scale streak structures. 

\subsection{A 2D view: a classic 2D SBLI analysis} 

\begin{figure*}[th]
     \centering
     \subfloat[]{
     \includegraphics[width=6cm]{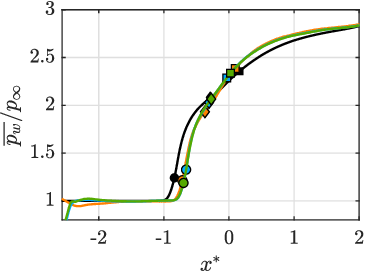}} 
     \qquad
     \subfloat[]{
     \includegraphics[width=6cm]{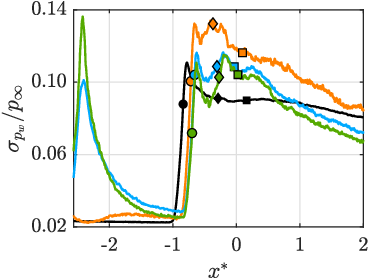}}
     \caption{Streamwise distribution of (a) mean wall pressure and (b) wall-pressure standard deviation. 
     Spanwise average of the uncontrolled case (solid black line), controlled case at \\ $z^* = -0.2986$ (solid orange line), $z^* = -0.0513$ (solid light blue line), $z^* = 0.0$ (solid green line). 
     Symbols indicate the location of the separation point (circles), the reattachment point (squares), and the point with minimum streamwise pressure gradient (diamonds) for the curve of the corresponding colour.}
     \label{fig:p_sigmap}
\end{figure*}

Although the qualitative flow description confirmed the literature findings about the increased geometrical complexity of the 3D flow field for the controlled SBLI, 
figure~\ref{fig:p_sigmap} shows that the streamwise distribution of the wall-pressure rise 
is homogeneous in the span, as also observed in the experimental measurements of \cite{babinsky2009microramp}. The figure reports the wall pressure at three notable sections in the span, which are: the symmetry plane $z^* = 0$, the lateral plane $z^* \approx - 0.30$, and the spanwise section corresponding to the minimum streamwise extent of the separation, $z^* = -0.0513$ (see section~\ref{subsec:CSBLI_separation}). The curves show that, compared with the USBLI case, the presence of the microramp induces a forward shift in the separation onset (indicated with circles) and a slight backward shift in the reattachment (indicated with squares), which thus entails an overall shorter separation along the span. 
The wall-pressure standard deviation $\sigma_{p_w}$ at the same sections shows instead that wall-pressure fluctuations are stronger for CSBLI. As in the case of non-adiabatic wall conditions \citep{bernardini2016heat} or transitional \glspl{sbli} \citep{quadros2018numerical}, a reduction (increase) of the separation length corresponds to an increase (decrease) in the intensity of the wall-pressure fluctuations. 
The CSBLI curves also present another peak in the rear part of the separation and, most importantly, vary considerably with the span (see section~\ref{subsec:CSBLI_separation}). In particular, the standard deviation distribution is minimum at the symmetry plane and maximum at the lateral plane. After the separation, the decay rate of $\sigma_{p_w}/p_\infty$ is larger in CSBLI than in USBLI, and rather homogeneous in the span. 

\begin{figure}[t!]
     \centering
     \includegraphics[width=6cm]{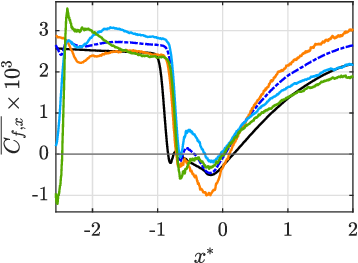}
     \caption{Streamwise distribution of the streamwise skin friction coefficient.\\
     Spanwise average of the uncontrolled case (solid black line), spanwise average of the controlled case (dashed-dotted blue line), controlled case at $z^* = -0.2986$ (solid orange line), $z^* = -0.0513$ (solid light blue line), $z^* = 0.0$ (solid green line).}
     \label{fig:cfx}
\end{figure}

Figure~\ref{fig:cfx} reports instead the distribution of the time-averaged streamwise skin friction component $\overline{C_{f,x}}$ at different sections in the span. 
As shown by the spanwise-averaged curve of the CSBLI case, the extent of the region with negative $\overline{C_{f,x}}$ is significantly shortened compared to the USBLI case. However, we also notice in this case that conditions vary considerably with the span when the microramp is present, suggesting that the streamwise skin friction component alone may provide misrepresentative information regarding the extent and the nature of the separation, because of the non-negligible contribution of the spanwise skin friction component. 

\begin{figure}[p]
     \centering
     \subfloat{
     \includegraphics[width=0.32\textwidth]{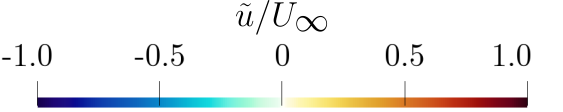}}\\
     \setcounter{subfigure}{0}
     \vspace{-5pt}
     \subfloat[]{
     \includegraphics[width=0.8\textwidth]{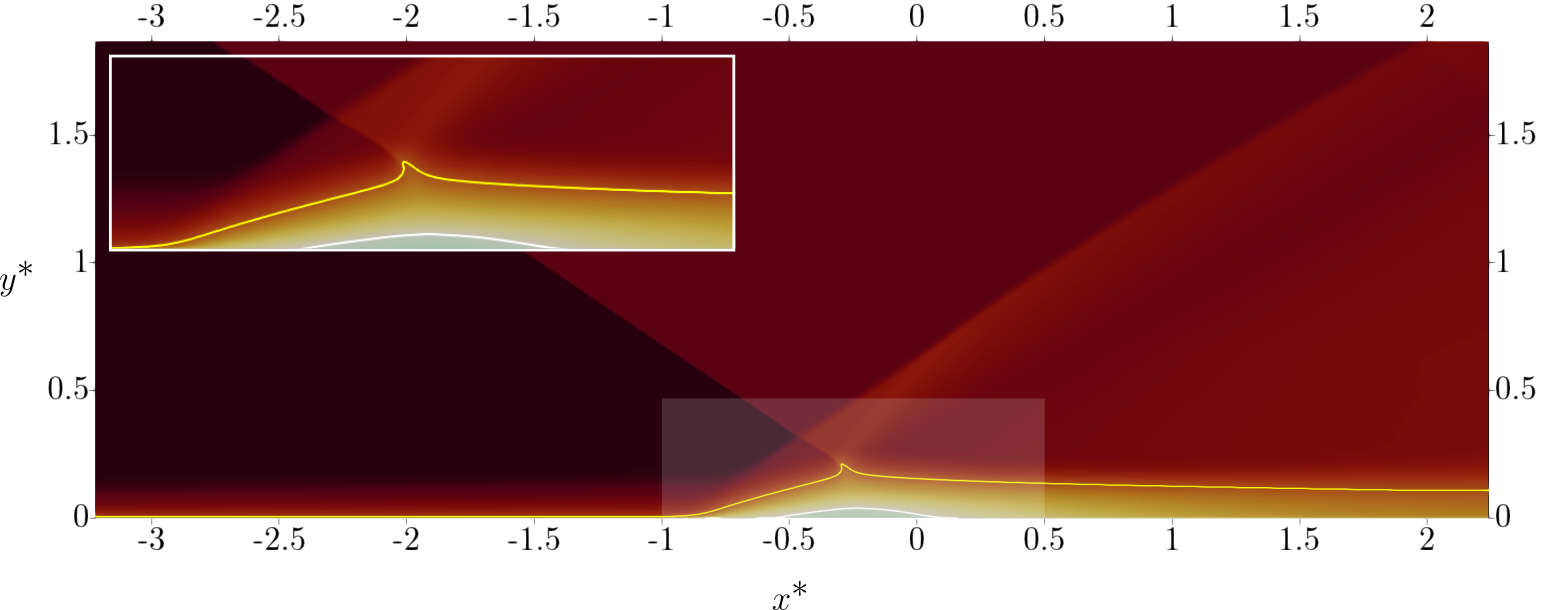}}\\ 
     \vspace{-5pt}
     \subfloat[]{
     \includegraphics[width=0.8\textwidth]{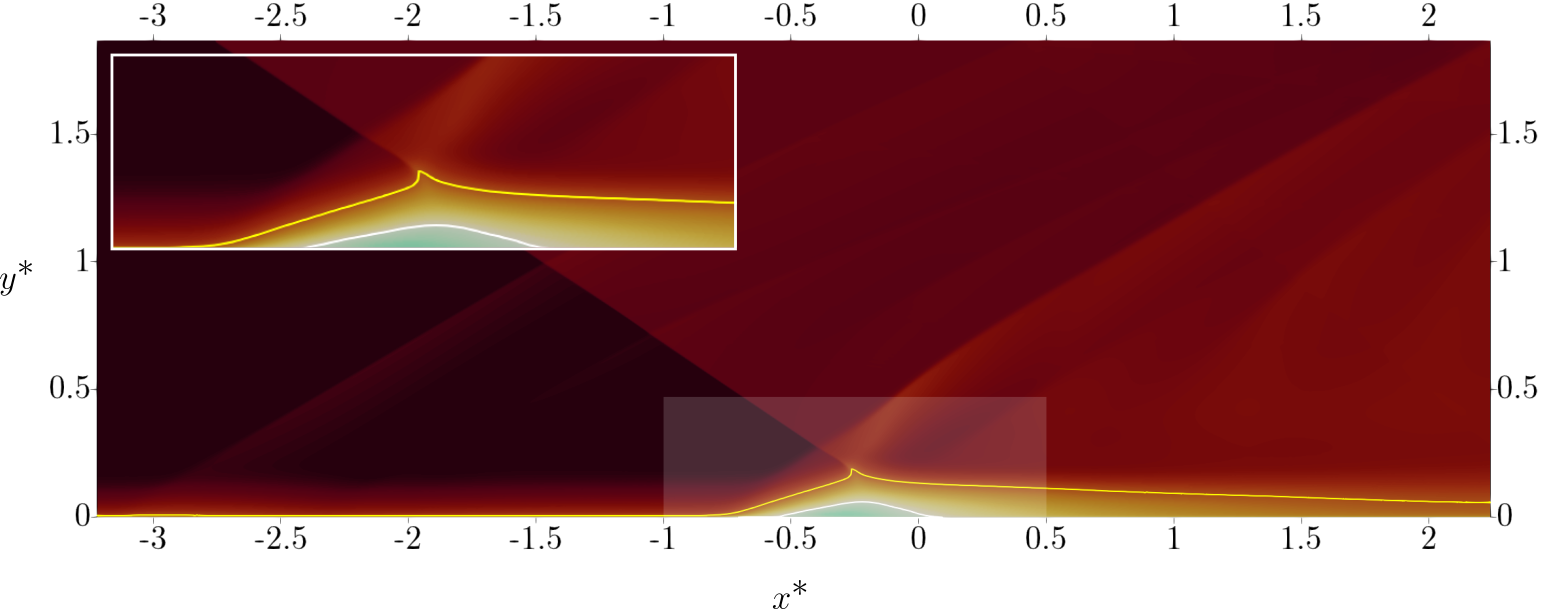}} \\
     \vspace{-5pt}
     \subfloat[]{
     \includegraphics[width=0.8\textwidth]{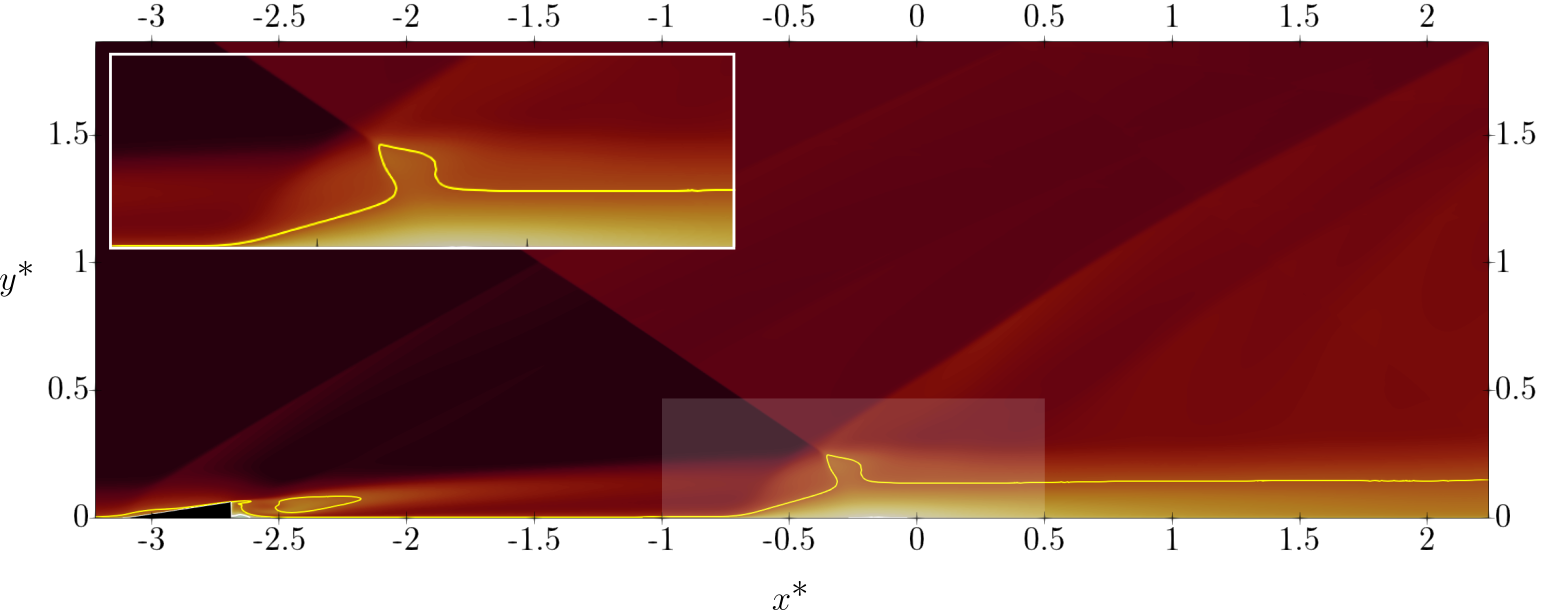}} \\
     \vspace{-5pt}
     \subfloat[]{
     \includegraphics[width=0.8\textwidth]{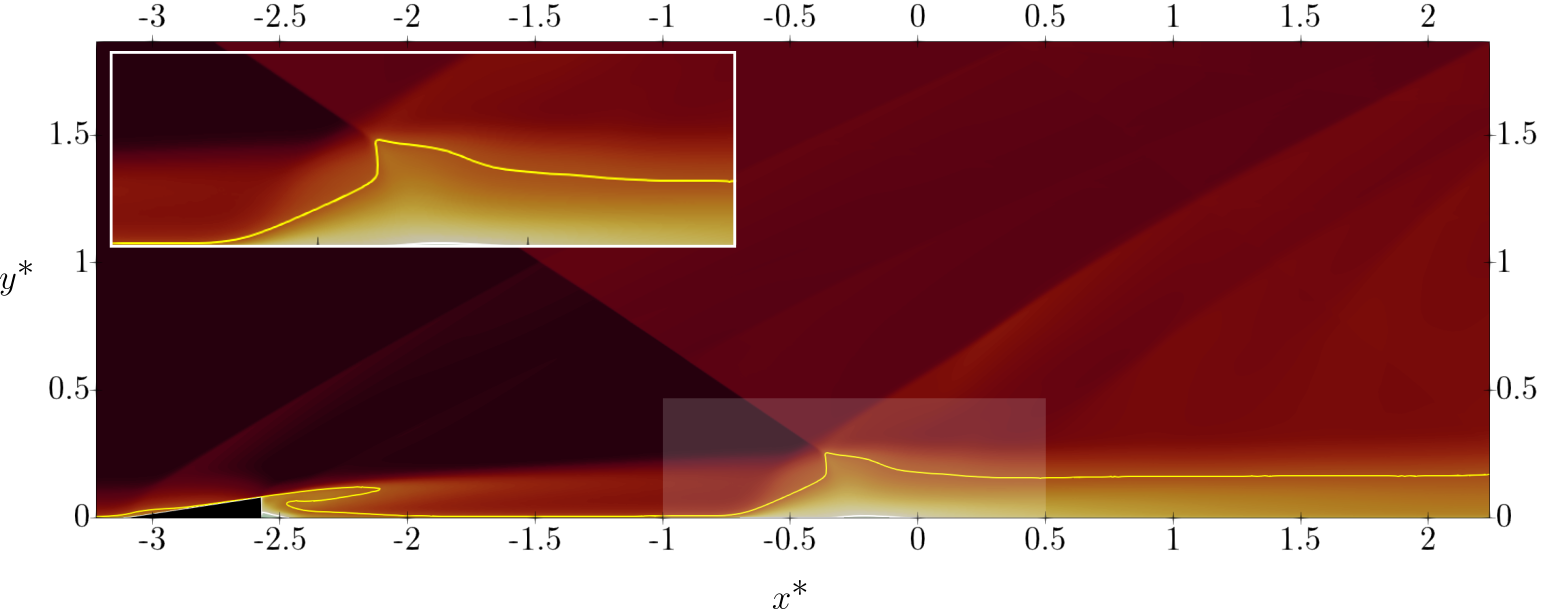}} 
     \caption{Favre-averaged streamwise velocity on xy planes: 
     (a) USBLI, spanwise-averaged, and CSBLI at (b) $z^* \approx -0.3$, 
     (c) $z^* = -0.05$, (c) $z^* = 0$.  Yellow lines indicate points with $\widetilde{M_\infty} = 1$, white lines indicate points with $\tilde{u}/U_\infty = 0$.}
     \label{fig:mean_u_slices}
\end{figure}

The significant spanwise modulation of the flow is also observed in the xy-slices of 
figure~\ref{fig:mean_u_slices}, which compares the Favre-averaged streamwise velocity component for the USBLI case (spanwise-averaged) and for the three above-defined spanwise sections of the CSBLI case. 
To consider compressibility, Favre averages are used, which are defined as $\widetilde{\phi} = \overline{\rho\,\phi}/\overline{\rho}$ for a generic variable $\phi$.
The reversed flow region is highlighted and hints that for the two inner sections, the separation is reduced relevantly. 
However, it is important to notice that especially for strongly three-dimensional flow, as in the case under consideration, the reversed flow is not indicative of the actual separation taking place. 

In conclusion, the wall-pressure standard deviation, the streamwise skin friction component, and the xy slices of the mean velocity proved that for many features -- but not all -- the flow organisation is fully three-dimensional, especially close to the wall and in the interaction region. The analysis of single streamwise slices of the streamwise components of vectorial quantities may be thus insufficient and even misleading to understand the overall interaction topology and dynamics. 

%
%


\subsection{More than 2D: the controlled SBLI separation \label{subsec:CSBLI_separation}}
Given the three-dimensionality of the separation, we must resort to more sophisticated 
tools for its analysis. Following the work of Legendre and Werl{\'e} \citep{delery2001robert, delery2013three}, the topology of the separation can be examined by studying the organisation of the skin friction lines at the wall. Skin friction lines can be viewed as the limit towards the wall of the streamlines and, if a mean flow is well-defined, they can provide insightful information about the actual location of the separation and reattachment. 

\begin{figure}[th]
     \centering
     \subfloat{
     \includegraphics[width=0.34\textwidth]{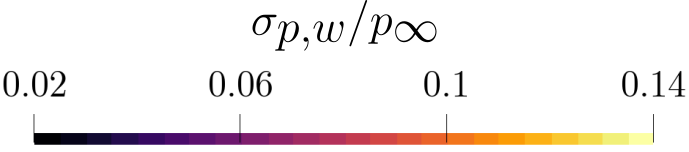}}\\
     \setcounter{subfigure}{0}
     \subfloat[]{
     \includegraphics[width=\textwidth]{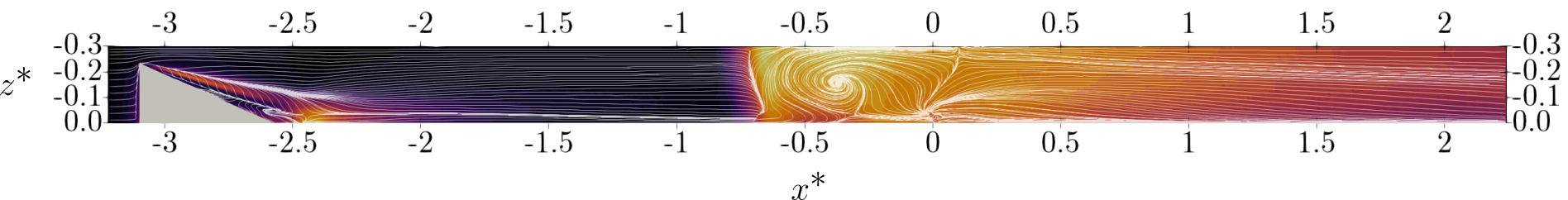}} \\
     \subfloat[]{
     \includegraphics[width=0.9\textwidth]{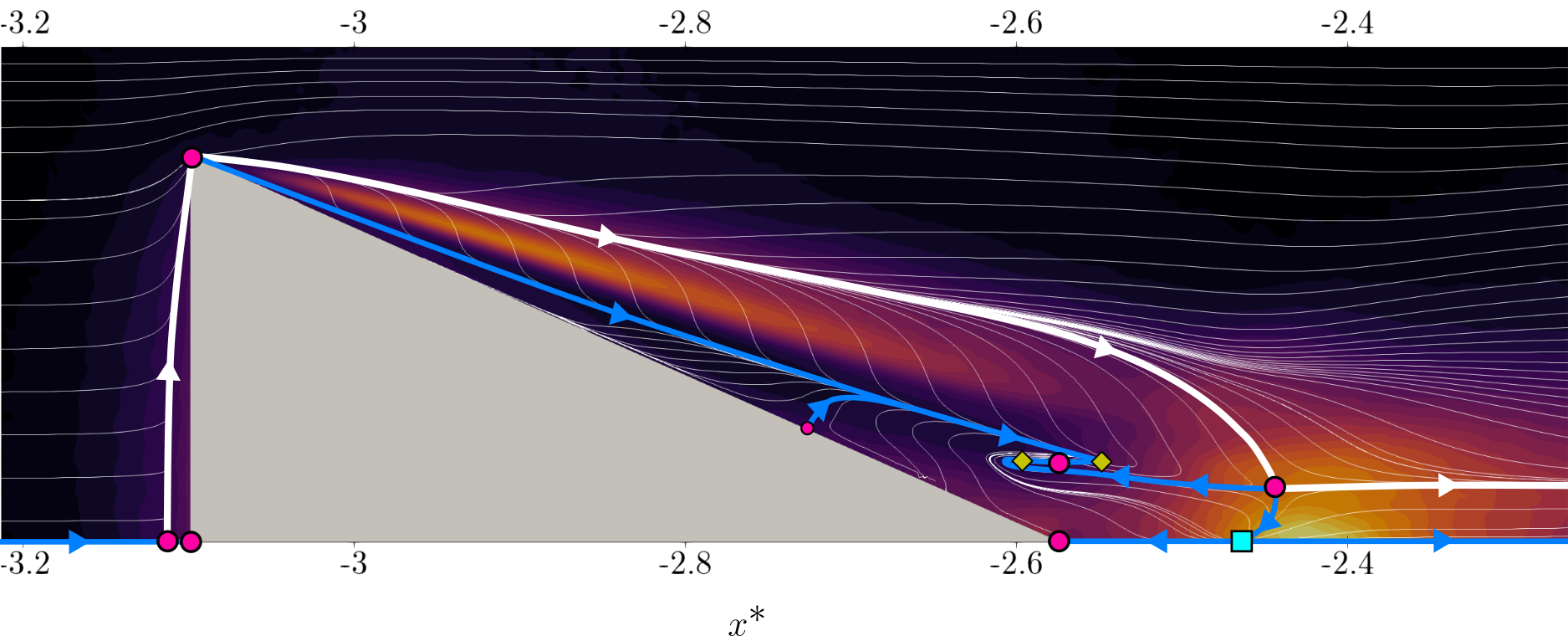}}\\
     \subfloat[\label{fig:sep_peak_locus}]{
     \includegraphics[width=0.9\textwidth]{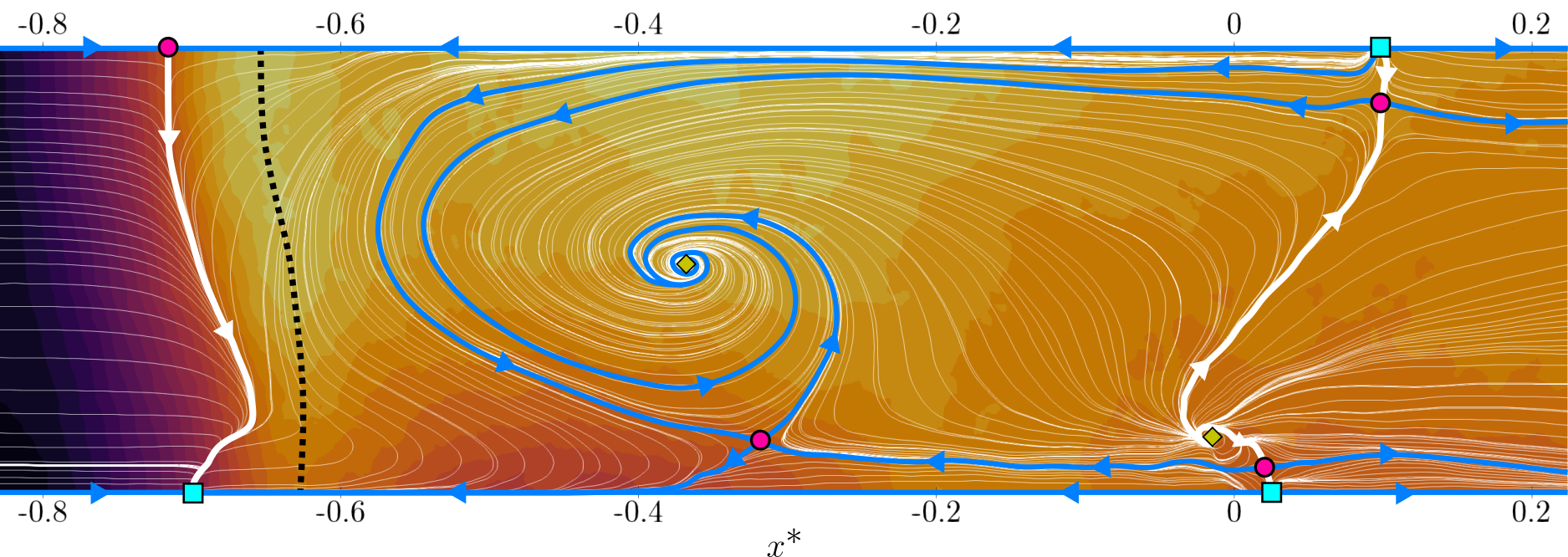}}\\
     \caption{Skin friction lines overlapped to a contour of the wall-pressure standard deviation on the xz plane for the CSBLI case: (a) overall domain, (b) zoom into ramp region, (c) zoom into separation region. Half of the domain is shown for symmetry. 
     Saddle points are indicated with circles in magenta, nodes with squares in light blue, and foci with diamonds in orange. The main separation and reattachment lines are highlighted in white, other relevant critical lines in blue, while the foremost peak of $\sigma_{p_w}$ is indicated with the dashed black line.}
     \label{fig:skin_friction_lines}
\end{figure}

Figure~\ref{fig:skin_friction_lines} reports the skin friction lines for the CSBLI case 
on half of the domain, taking advantage of the flow case symmetry, and highlights
the formation of curved separation and reattachment lines along the span, 
where skin friction lines converge to and depart from respectively. 
Critical points ($C_{f,x} = C_{f,z} = 0$) are also highlighted according to their nature:
saddle points indicated as light green circles, nodes as light blue squares, and foci as orange diamonds. 
For the sake of completeness, we remind the reader that only two skin friction lines run through a saddle point, while all the others avoid it adopting the shape of a hyperbolic curve, that all the skin friction lines have a common tangent at a node, except for one of them, and that skin friction lines end at a focus after spiralling around it.
Considering also the symmetric half not shown in the figure, the periodic boundary conditions, and the two nodes at infinite upstream and downstream, our configuration for the interaction region (figure~\ref{fig:sep_peak_locus}) has a total of five nodes ($n=5$), four foci ($f = 4$), and seven saddle points ($s=7$), which satisfies the necessary topological condition $n+f-s = 2$ \citep{delery2013three}. Our arrangement has one node, two foci, and three saddle points more than the one in \citet{grebert2023microramp}, as the separation line is split into more than one single segment in our case.

\begin{figure}[th]
     \centering
     \includegraphics[width=\textwidth]{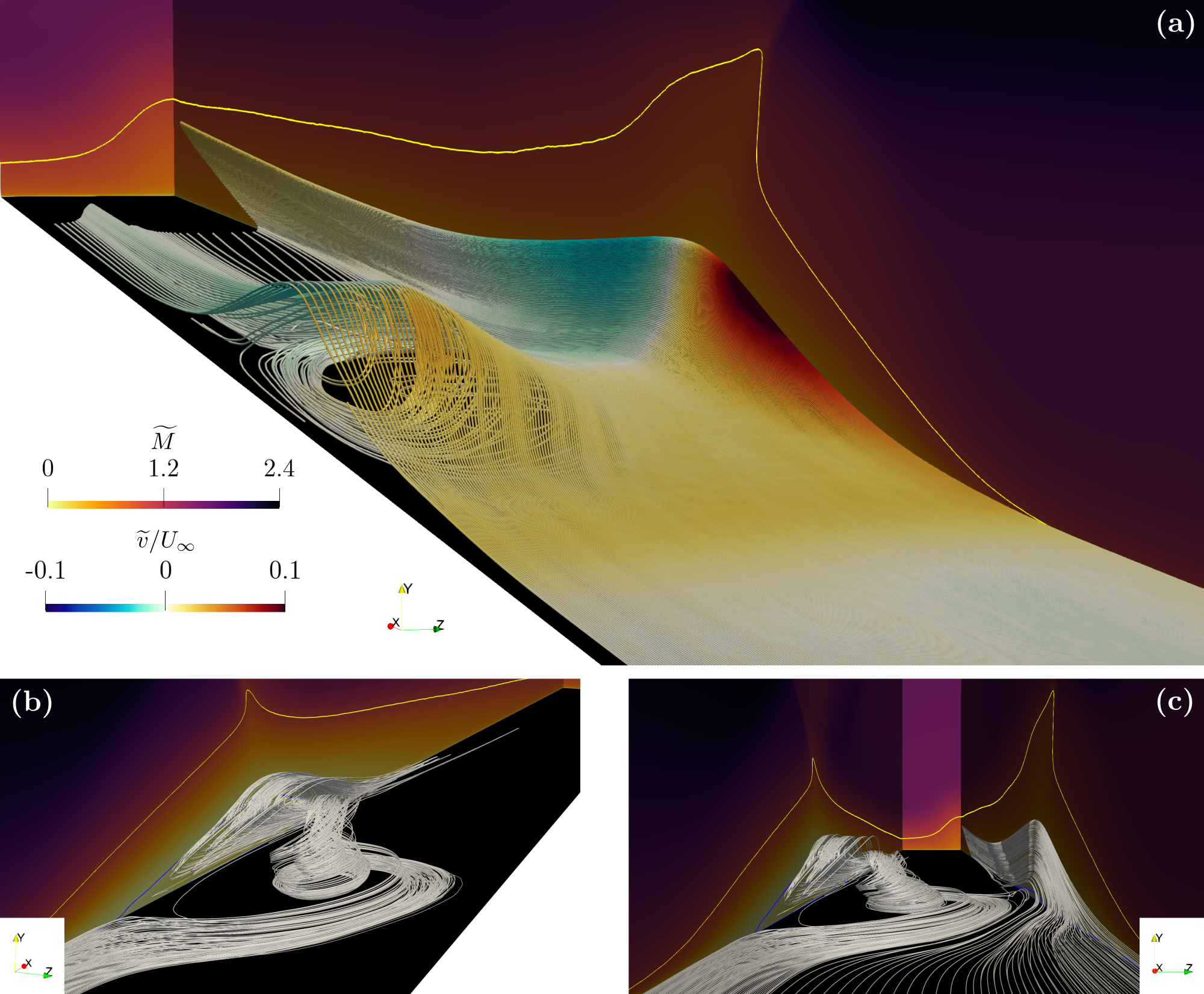}
     \caption{Streamlines in the separation region of the CSBLI case: qualitative edge of the recirculation bubble (streamlines coloured by mean vertical velocity) (a),  
     the tornado-like structure (b), and the internal structure of the bubble (c).
     Vertical slices report the Favre-averaged Mach number $\widetilde{M}$, yellow lines indicate $\widetilde{M} = 1$, and blue lines indicate $\tilde{u}/U_\infty = 0$ on the vertical slices. Half of the domain is shown for symmetry. \label{fig:streamlines_3d}}    
\end{figure}

The focus, already observed also in oil-flow visualisations \citep{babinsky2009microramp}, indicates the presence of a tornado-like structure which captures the flow close to the wall and pushes it upwards.
A visualisation of the streamlines associated with the tornado in figure~\ref{fig:streamlines_3d}b shows that this structure tilts outwards to join 
the wall region closer to the symmetry plane and the external sides of the separation bubble.
Indeed, its axis is first normal to the wall in correspondence with the focus and ends up being aligned with the spanwise direction towards the side. 

\begin{figure}[t!]
     \centering
     \includegraphics[width=6cm]{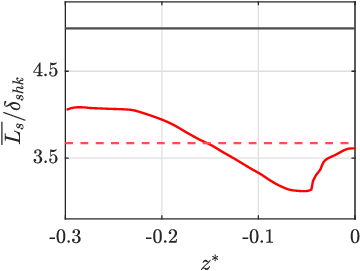}
     \caption{Spanwise distribution of the time-averaged separation length 
     with respect to the boundary layer thickness at the shock impingement location.
     Spanwise average of the uncontrolled case (solid black line), 
     local controlled case (solid red line), and 
     spanwise average of the controlled case (dashed red line).}
     \label{fig:Lsep_z}
\end{figure}

Compared to the classical 2D recirculation bubble, the region of separated flow is now more complicated and not identified by the simple condition of negative $\overline{C_{f,x}}$. 
Moreover, reversed flow is now even less indicative of the separation, which makes it difficult to identify the edge of the recirculating bubble. Figure~\ref{fig:streamlines_3d}a uses streamlines to show the qualitative shape of the bubble, which presents a characteristic saddle shape with a higher separated region at the symmetry and the lateral planes and a lower separation region in correspondence with the minimal streamwise separation extent (see figure~\ref{fig:Lsep_z}).  

Significant 3D effects are also visible in the distribution of the wall-pressure standard deviation $\sigma_{p,w}/p_\infty$ in figure~\ref{fig:skin_friction_lines}. 
Close to the microramp, we can notice regions with relevant pressure fluctuations associated with i) the impingement of the flow of the primary vortices at the sides of the ramp towards the wall, ii) the reattachment of the flow after the trailing edge and the following formation of the parallel, primary vortex pair. In the interaction region instead, we observe a first, stronger peak after the separation onset, whose streamwise position follows the separation front and whose magnitude is greater close to the lateral boundaries. The distribution is compatible with the observation that the shock is disrupted at the centre of the domain and, hence, weaker on average. As a result of the reduced shock intensity, the amplification of the pressure fluctuations induced by the shock wave is weaker in the central region of the domain. A second peak, with a spanwise-modulated amplitude as well, takes place at approximately half of the separation, in correspondence with the peak height of the bubble. After the second peak, the standard deviation decays, maintaining a non-uniform spanwise distribution even far from the interaction.

Given the separation and reattachment line in figure~\ref{fig:skin_friction_lines}, it is possible to estimate the extent of the time-averaged streamwise separation $\overline{L_s}$ along the span reported in figure~\ref{fig:Lsep_z}. The spanwise average $\langle \overline{L_s} \rangle/\delta_{shk}$ shows that the mean separation in the presence of the microramp shrinks by 26.47\% the value observed in the USBLI case. Moreover, the distribution shows that the separation is largely modulated in the span, $(\overline{L_s}_{max} - \overline{L_s}_{min})/ \langle \overline{L_s} \rangle = 26.31\%$ along the span, with a minimal separation at $z^* \approx -0.05$ ($\overline{L_s}_{min}/\delta_{shk} = 3.12$), 

\begin{figure}[t!]
     \centering
     \subfloat{
     \includegraphics[width=0.34\textwidth]{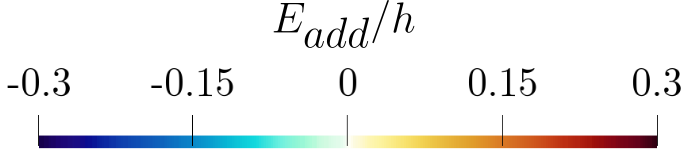}}\\
     \setcounter{subfigure}{0}
     \subfloat[\label{fig:added_momentum_long}]{
     \includegraphics[width=\textwidth]{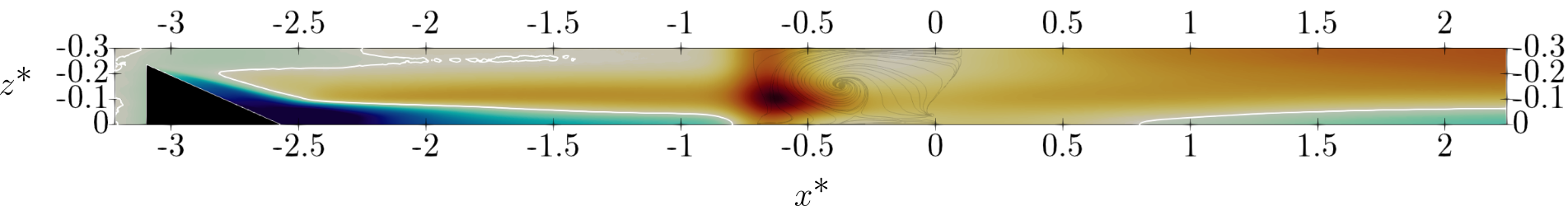}} 
     \\
     \subfloat[]{
     \includegraphics[width=\textwidth]{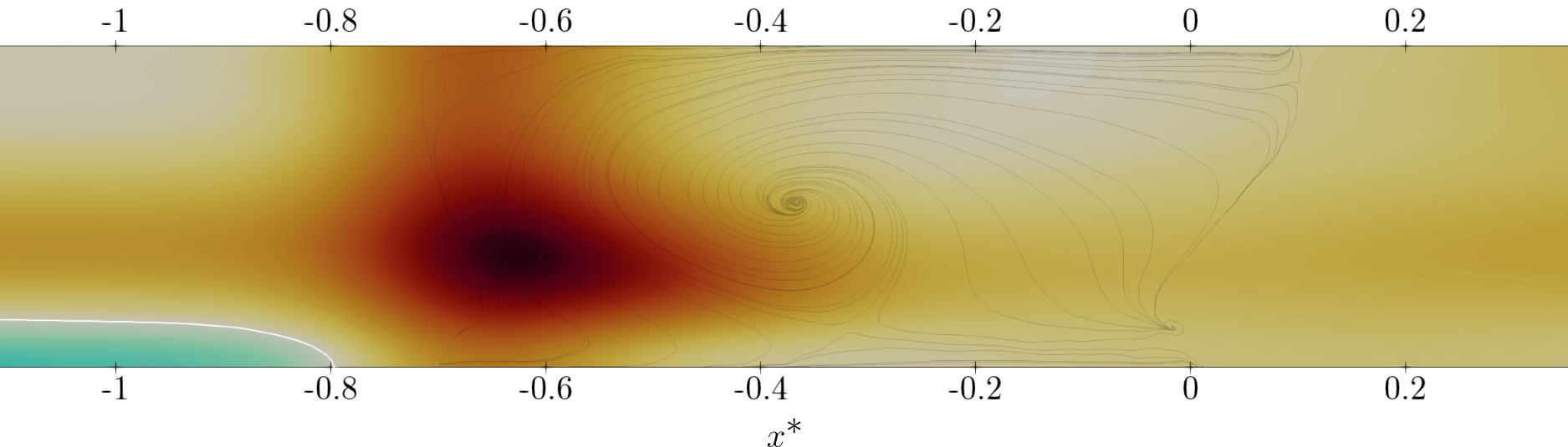}}
     \caption{Normalised compressible added momentum $E_{add}/h$ along the xz plane (a) and zoom into the separation region (b). Skin friction lines in the separation region are indicated in grey. Half of the domain is shown for symmetry.}
     \label{fig:added_momentum}
\end{figure}

To understand the relationship between the spanwise modulation of the separation and the addition of momentum by the primary vortex pair, figure~\ref{fig:added_momentum} reports the distribution of the compressible added momentum in the xz plane. 
First defined in its incompressible version by \citet{giepman2014flow}, the compressible added momentum is defined as
\begin{equation}
    E_{add} = \int_{0}^{\overline{y^*}} \frac{\overline{\rho u^2}_{CSBLI} - \overline{\rho u^2}_{USBLI}}{\rho_\infty U_\infty^2} \, \mathrm{d} y
\end{equation}
and tracks the addition of streamwise momentum towards the wall. Following \citet{giepman2014flow}, the upper bound of integration $\overline{y^*}$ is taken equal to $0.43\,\delta$.
The distribution of $E_{add}/h$ along the entire wall-parallel plane allows us to appreciate 
that the primary vortex pair brings fresh momentum towards the wall in a large spanwise range, even quite far from the symmetry plane. 
The peak added momentum just before the interaction is at $z^*\approx -0.1$, just at the side of the minimum separation, which thus suggests that the onset of the separation -- and its local extent -- follows strictly the addition of momentum associated with the primary vortex pair helical motion. 

\begin{figure}[t!]
     \centering
     \subfloat[\label{fig:tornado_streamlines}]{
     \includegraphics[width=0.45\textwidth]{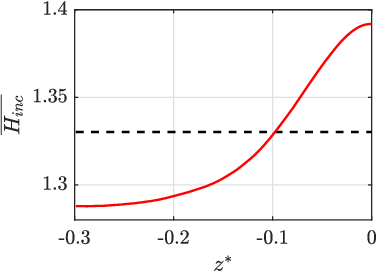}} 
     \qquad
     \subfloat[\label{fig:streamlines_bubble}]{
     \includegraphics[width=0.445\textwidth]{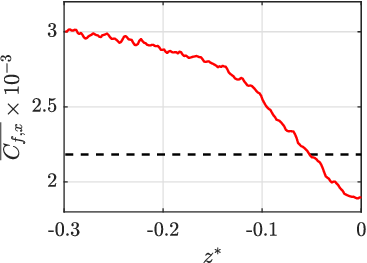}}
     \caption{Spanwise distribution at $x^* = 2$ of 
     (a) incompressible shape factor and (b) streamwise skin friction coefficient. USBLI (dashed black line) and CSBLI (solid red line).
     \label{fig:spanwise_Hi_Cfx}}    
\end{figure}

After a transitory region following the interaction, we can observe that the trace of the primary streamwise vortices becomes once again more visible, indicating that these are able to withstand the shock waves and that their signature lasts for a long streamwise distance (figure~\ref{fig:added_momentum_long}). Even far downstream of the separation, we thus have non-uniform conditions as highlighted by figure~\ref{fig:spanwise_Hi_Cfx} reporting the spanwise distribution at $x^* = 2$ of the incompressible shape factors and the streamwise skin friction coefficient for the USBLI and CSBLI cases. The shape factor denotes a fuller and healthier boundary layer even after the interaction at the sides of the symmetry plane, which however corresponds to increased skin friction. We point out that the associated increased drag and the potential consequences of a long-lasting, non-uniform flow for the engine components following the inlet may limit the benefits of a reduced separation and thus that these aspects should be addressed carefully when considering \glspl{mvg} for real applications. 

\subsection{Vortical structures: a mean view} 

An interesting issue regarding the effects of microramps on \gls{sbli} is the mutual interaction between the 
arch-like vortical structures induced by \glspl{mvg} and the interaction region.

According to the literature and the results in the previous sections, 
the main effect on the \gls{sbli} of the arch-like vortices is 
that they periodically disrupt the impinging and reflected shock waves, leading to regions of large curvature in the shock surfaces (the ``bump'' in figure~\ref{fig:qcrit}). 
The effect of the shock waves on the arch-like vortices is instead less clear. 
Some works \citep{yan2013study} suggest that these vortices travel rather undisturbed across the shock, however, quantitative information supporting this statement, regarding for example the effect of \gls{sbli} on the trajectory and intensity of the arch-like vortices, is lacking. 
To assess the streamwise development of the vortical structures related to the microramp and the \gls{sbli}, we consider the behaviour of the Favre-averaged vorticity, which we define in this work as the curl of the Favre-averaged velocity ($\widetilde{\boldsymbol{\zeta}} = \nabla \times \widetilde{\boldsymbol{u}}$). In particular, we can assume that, at the symmetry plane, the negative peaks of the spanwise component $\widetilde{\zeta_z}$ correspond to the trace of the \gls{kh} vortices associated with the arch-like vortices and with the shear layer delimiting the separation. 

\begin{figure}[th]
     \centering
     \subfloat[]{
     \includegraphics[width=\textwidth]{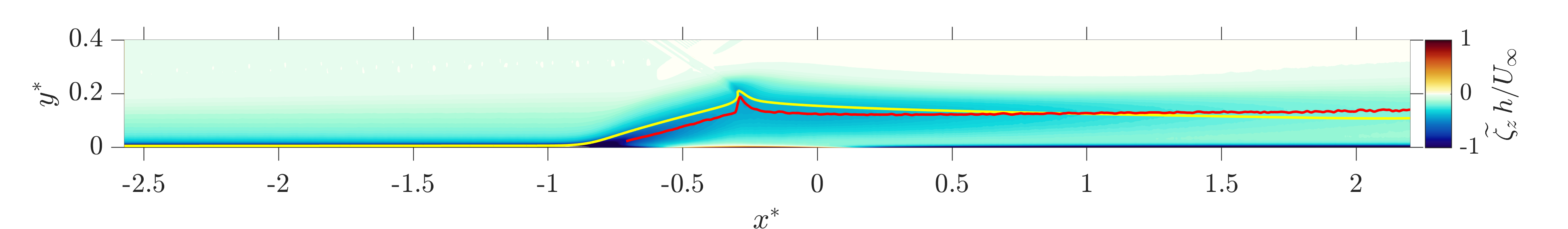}} 
     \\
     \subfloat[]{
     \includegraphics[width=\textwidth]{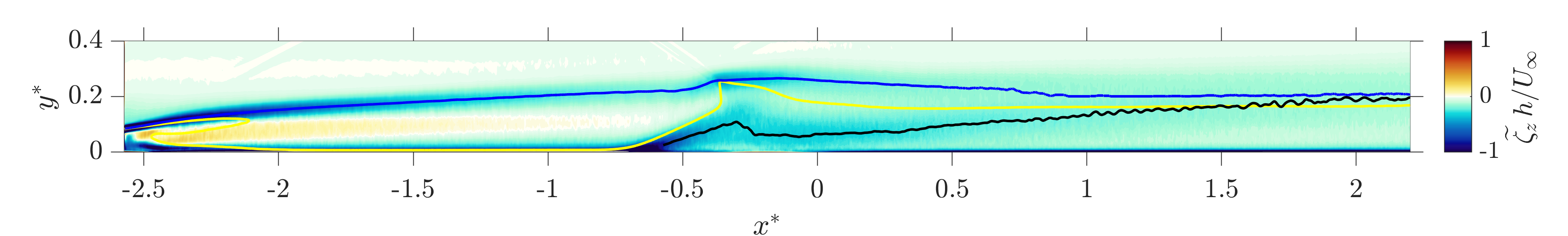}} 
     \\
     \subfloat[]{
     \includegraphics[width=\textwidth]{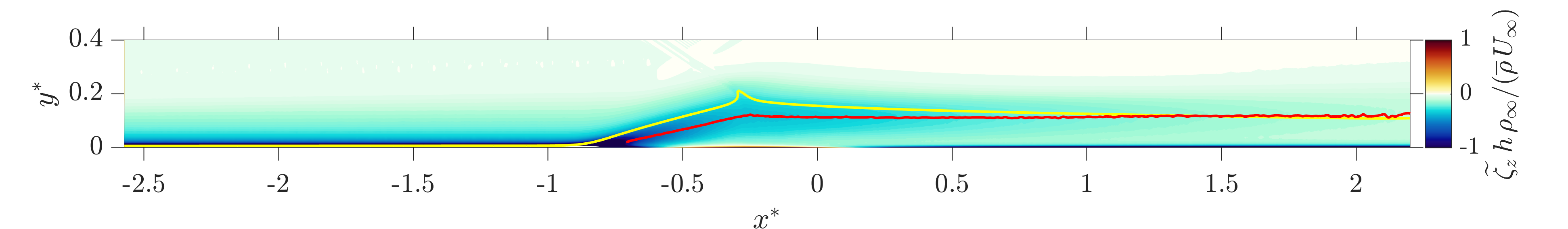}} 
     \\
     \subfloat[]{
     \includegraphics[width=\textwidth]{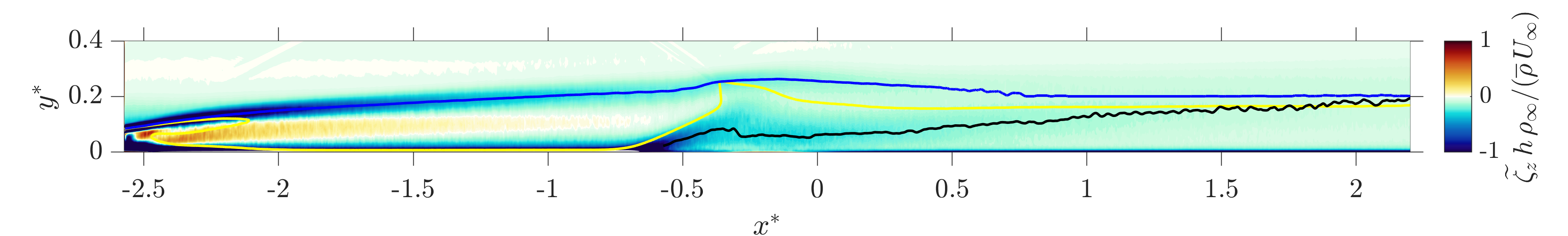}} 
     \caption{Contours of the spanwise component of the Favre-averaged vorticity (a-b) and the density-weighted Favre-averaged vorticity (c-d) for the USBLI (a-c) and the CSBLI (b-d) cases. 
     Yellow lines indicate points at $\widetilde{M} = 1$, blue and black lines indicate respectively the position of the top and bottom shear layers of the CSBLI case, while red lines indicate the shear layer of the USBLI case.}
     \label{fig:vortz_contours}
\end{figure}

\begin{figure}[th]
     \centering
     \subfloat[]{
     \includegraphics[width=0.45\textwidth]{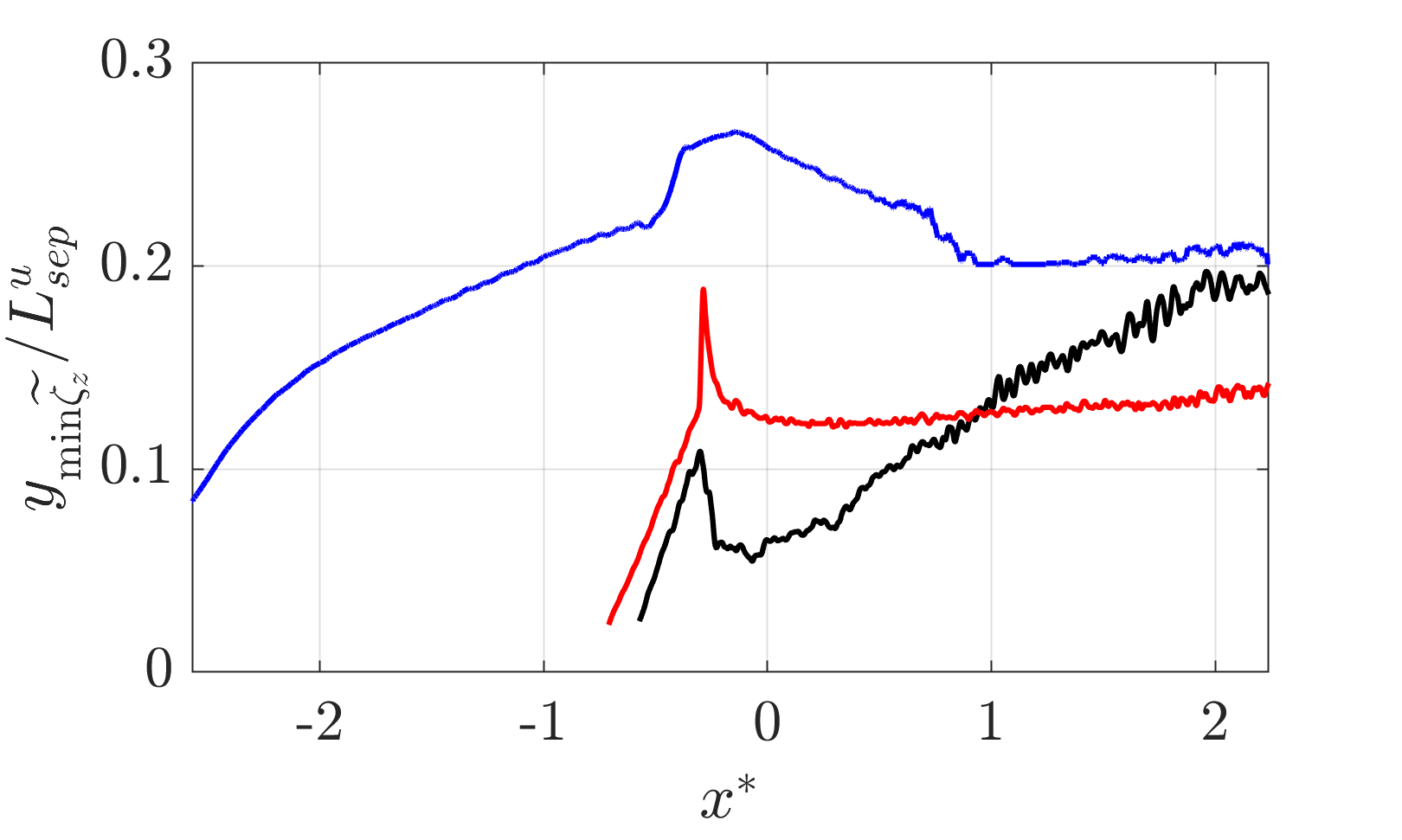}} 
     \,
     \subfloat[]{
     \includegraphics[width=0.45\textwidth]{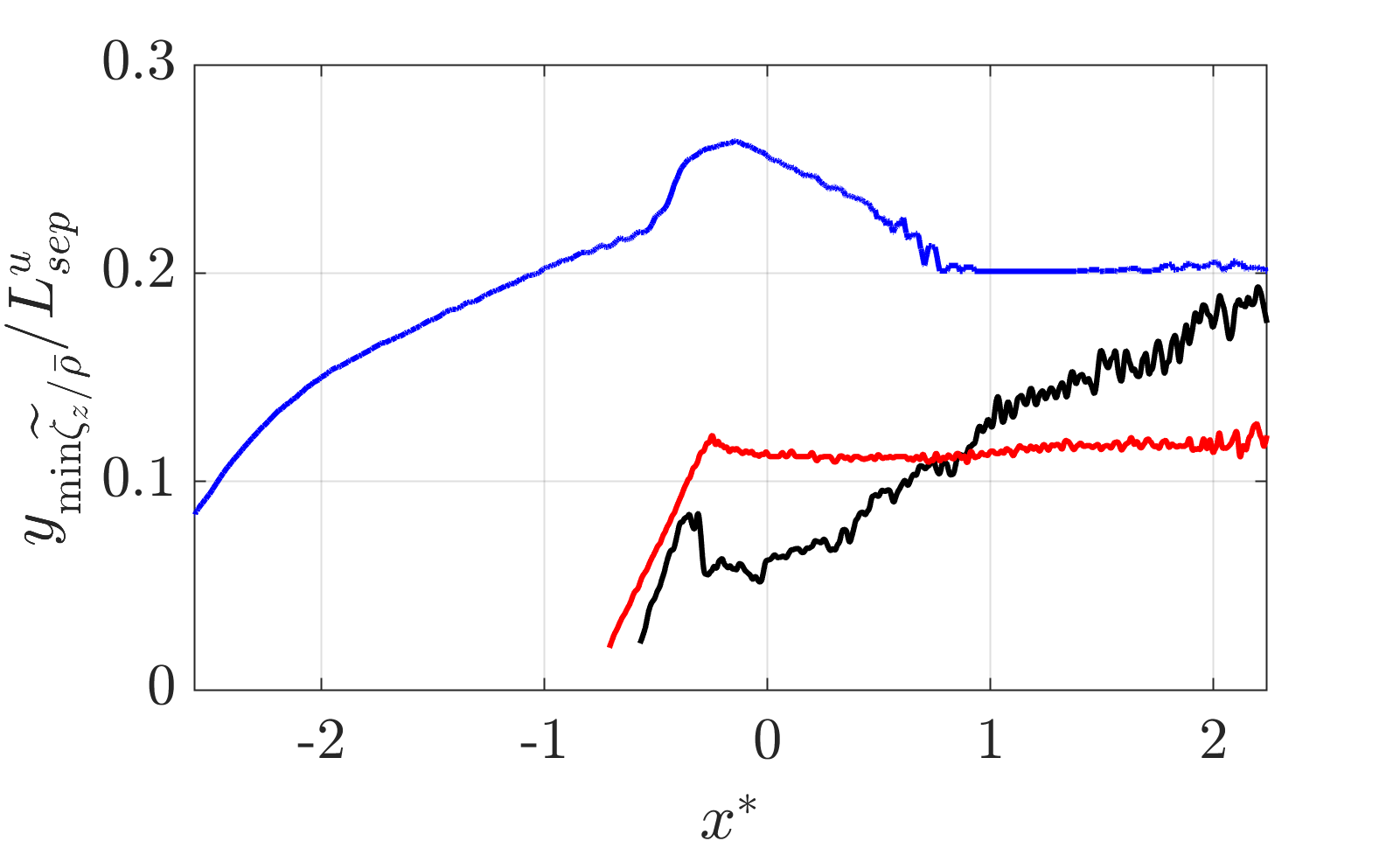}} 
     \caption{Wall-normal position of shear layers at the symmetry plane based on the spanwise Favre-averaged vorticity without (a) and with (b) density weighting: USBLI shear layer (red), CSBLI bottom shear layer (black), CSBLI top shear layer (blue). }
     \label{fig:y_vortz}
\end{figure}

\begin{figure}[th]
     \centering
     \subfloat[]{
     \includegraphics[width=0.45\textwidth]{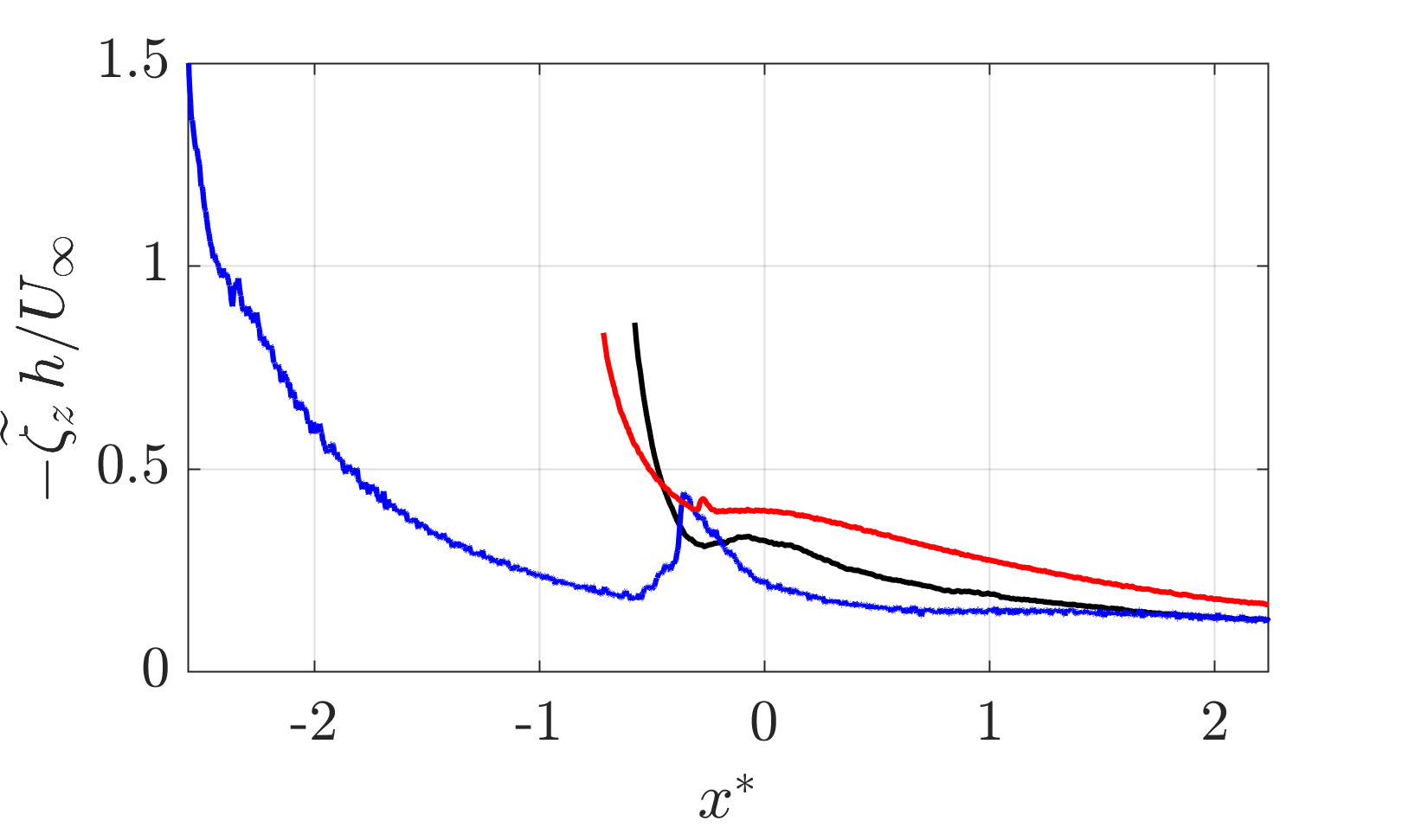}} 
     \,
     \subfloat[]{
     \includegraphics[width=0.45\textwidth]{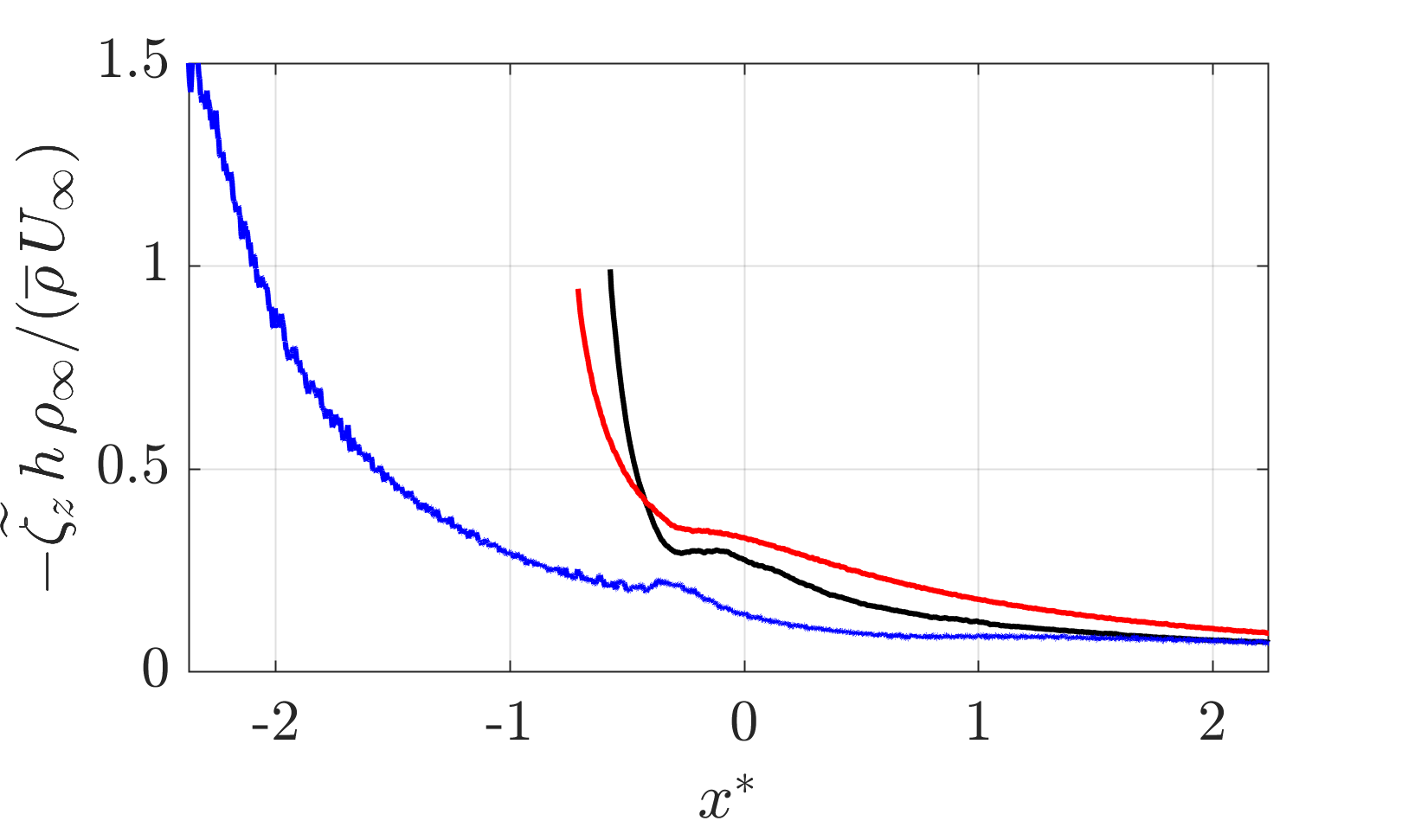}} 
     \caption{Spanwise component of the Favre-averaged vorticity without (a) and with (b) density weighting along the respective shear layers at the symmetry plane: USBLI shear layer (red), CSBLI bottom shear layer (black), CSBLI top shear layer (blue).}
     \label{fig:zeta_vortz}
\end{figure}

Locating these minima and recording the magnitude of the vorticity along these loci allows us to track the trajectory of the main vortical structures and investigate the streamwise evolution of their intensity. Figure~\ref{fig:y_vortz} reports the wall-normal coordinate of the points with minimal Favre-averaged spanwise vorticity for a given streamwise section, whereas figure~\ref{fig:zeta_vortz} reports the corresponding value of the mean spanwise vorticity.\\
Observing the last figure, the peaks in correspondence with the shock position suggest an increase in the intensity of the external shear layers that directly intersect the impinging shock in both the USBLI and CSBLI cases (red and blue curves). Whether this increase is associated with the generation of vorticity induced by baroclinic, diffusive, turbulent, or other effects is not clear at this stage. However, if we observe the equation for the evolution of the Favre-averaged vorticity (see Appendix~\ref{sec:appendix}), we obtain that
\begin{equation}
\frac{\mathrm{D}}{\mathrm{D}t}\,\left( \frac{\widetilde{\boldsymbol{\zeta}}}{\overline{\rho}} \right) = \left(\frac{\widetilde{\boldsymbol{\zeta}}}{\overline{\rho}}\right) \cdot \nabla \widetilde{\boldsymbol{u}} + 
    \frac{\nabla \, \overline{\rho} \times \nabla (\overline{p} + 2/3 \, \overline{\rho}\,\widetilde{k})}{\overline{\rho}^3} 
    + \frac{1}{\overline{\rho}} \nabla \times \left( \frac{\nabla \cdot \tau^{t,d}}{\overline{\rho}} \right)    
\end{equation}
where the first term at the right-hand side is the compressible vortex stretching and tilting, the second term is the baroclinic term, and the third term is the diffusion of vorticity by the action of viscous and turbulent stresses. 
We indicated with $\mathrm{D}/\mathrm{D}t$ the material derivative, and with $\tau_{ij}^t = \overline{\tau_{ij}} + \widetilde{\tau_{ij}}^R$ the total stress tensor including both the Reynolds-averaged molecular stress tensor $\overline{\tau_{ij}}$ and the Favre-averaged Reynolds stress tensor $\widetilde{\tau_{ij}}^R = - \overline{\rho u_i'' u_j''}$, while the Reynolds-averaged pressure is indicated as $\overline{p}$, the Favre-averaged turbulent kinetic energy as $\widetilde{k}$ and the deviatoric part of the total stress tensor as $\tau_{ij}^{t,d}$. 
Compressibility effects in the conservation of mass for a generic fluid element are thus correctly accounted for in the Favre-averaged vorticity equation only if we introduce density-weighting. Indeed, by dividing the Favre-averaged vorticity by the Reynolds-averaged density, the classical equations describing vorticity dynamics are recovered.
Therefore, if we observe the contours of the specific vorticity $\widetilde{\zeta_z}/\overline{\rho}$ (figure~\ref{fig:vortz_contours}), we can see that the sudden jump observed in correspondence with the incident shock is mainly an effect of the conservation of mass associated with the sudden density rise, which is also observable in the USBLI case. 
Figure~\ref{fig:zeta_vortz} further confirms this conclusion, suggesting that only a limited generation of mean vorticity in the top shear layer of the CSBLI case can be related to baroclinic or turbulent effects across the shock, as the vortex stretching and tilting term is null for the spanwise vorticity component because of the flow symmetry at $z^* = 0$. Looking at the specific vorticity, the decay of the shear layer intensity is now almost undisturbed by the presence of the shock, which confirms the qualitative belief that the arch-like vortices are robust enough not to be affected by \gls{sbli}. 
Figure~\ref{fig:y_vortz}, however, shows that there is no relevant difference between the trajectories of the wall-normal coordinates of the $\widetilde{\zeta_z}$ and $\widetilde{\zeta_z}/\overline{\rho}$ peaks. 
Considering the CSBLI case, the bottom shear layer surrounding the separation region follows the behaviour of the one in the USBLI case up to reattachment ($x^* \approx 0$). 
Conversely, the upper shear layer defined by the arch-like vortices and delimiting the edge of the boundary layer, first rises slowly until the interaction region, because of the known lift-up at the symmetry plane in the microramp wake, and then follows the triangular shape of the separation region with a slow recover after reattachment. Indeed, the wall-normal position is affected by the separation up to $x^* \approx 1$. After the reattachment point, the bottom shear layer converges to the upper one, which thus delimits the new boundary layer downstream of the interaction, significantly thicker than in the USBLI case. 

\subsection{Flow unsteadiness}

Except for very few recent studies \citep{grebert2018simulations, dong2018spectrum, sun2020wake, grebert2023microramp}, the features of the flow unsteadiness associated with \gls{mvg}-controlled \gls{sbli} have been scarcely considered in the literature, despite the shock low-frequency unsteadiness being one of the most critical aspects for \gls{sbli}. Indeed, long integration periods to capture this spectral property are computationally demanding for high-fidelity methods, and \gls{rans} methods are notoriously unable to properly describe flow unsteadiness. To shed light on the spectral characteristics of the flow under consideration, in the following, we analyse the time evolution of the pressure at the wall. We first present a Fourier analysis of the pressure along the xz and xy planes and then a wavelet analysis of $p_w(t)/p_\infty$ for selected probes. 



\subsubsection{Spectra}

\begin{description}
\item[Streamwise spectra]\,
First, we consider the overall picture given by the spanwise-averaged premultiplied spectra of the wall pressure along the streamwise coordinate. Spectra have been evaluated with the Welch method, using 16 (USBLI) and 8 (CSBLI) segments with 50\% overlap and a rectangular window, and have then been smoothed using a Konno-Ohmachi filter \citep{konno1998ground}. A summary of the main results is reported in table~\ref{tab:spectra_recap}, where $x^*_{low}$ is the location of the low-frequency peak, $x^*_{sep}$ is the separation point, $x^*_{peak}$ is the foremost wall-pressure standard deviation peak, $x^*_{reat}$ is the reattachment point, $L_{sep}/\delta_{shk}$ is the separation length with respect to the boundary layer thickness at inviscid shock impingement, $f\, L_{sep}^u/U_\infty$ is the non-dimensional value of the peak low frequency with respect to the free-stream velocity and the USBLI separation length, $f\, L_{sep}/U_\infty$ is the non-dimensional value of the peak low frequency with respect to the free-stream velocity and the local separation length.

\begin{table}[t]
 \begin{center}
\def~{\hphantom{0}}
  \begin{tabular}{l c c c c c c c}
    \hline\noalign{\smallskip}
 	Case & $x^*_{sep}$ & $x^*_{low}$ & $x^*_{peak}$ & $L_{sep}/\delta_{shk}$ & $f\,L_{sep}^u/U_\infty$ & $f\,L_{sep}/U_\infty$\\
    \noalign{\smallskip}\hline\noalign{\smallskip}
    USBLI -- span-av.           & -0.838 & -0.792 & -0.779 & 4.994 & 0.052 & 0.052 \\
    CSBLI -- span-av.           & -0.694 & -0.666 & -0.626 & 3.672 & 0.079 & 0.058 \\
    CSBLI ($z^* \approx -0.30)$ & -0.717 & -0.645 & -0.661 & 4.061 & 0.082 & 0.067 \\
    CSBLI ($z^* \approx -0.05)$ & -0.659 & -0.647 & -0.625 & 3.120 & 0.082 & 0.051 \\
    CSBLI ($z^* \approx  0.00)$ & -0.699 & -0.642 & -0.628 & 3.619 & 0.079 & 0.057 \\
    \noalign{\smallskip}\hline
  \end{tabular}
  \caption{Main streamwise locations of interest, separation length, and value of the peak low frequency.
  \label{tab:spectra_recap} }
 \end{center}
\end{table}

The contours in figure~\ref{fig:streamwise_spectra_spanwise_av} show the typical behaviour of the wall-pressure spectra. A broadband low-frequency peak in correspondence with the onset of the separation is associated with the oscillations of the separation shock foot. The low-frequency peak is followed by an intermediate region with mid-to-high frequencies associated with the development of the separation bubble. Here, the dominant frequencies drop in the first half of the bubble and then rise in correspondence with the reattachment. 
After this point, high-frequency pressure fluctuations indicate the increased turbulent activity in the boundary layer following the interaction. 
In the CSBLI case, we can also notice a high-frequency content before the interaction onset due to the microramp wake flow. The energetic region, sharp in space at $x^* \approx -2.4$ and broadband in frequency, corresponds to the peak of the wall-pressure standard deviation observed just after the ramp trailing edge (see figure~\ref{fig:p_sigmap}) and is associated with the trace on the wall of the conical shock around the microramp wake \citep{dellaposta2023direct_reynolds}.
Comparing the USBLI and CSBLI cases, we can notice that the entire frequency content is slightly shifted towards higher frequencies. As a result, the low-frequency broadband peak, whose absolute maximum is located at $f\, L_{sep}^u/U_\infty \approx 0.052$ for the USBLI case, increases slightly up to $f\, L_{sep}^u/U_\infty \approx 0.079$ for the CSBLI case. In addition, another low-frequency energetic content is visible at $x^* \in [-0.5,-0.25]$, where the second peak of wall-pressure standard deviation is present in figure~\ref{fig:p_sigmap}. This interval corresponds to the first half of the interaction, where the height of the separation bubble is increasing and the impinging shock interacts with the shear layer surrounding the recirculating flow. 
Finally, although an increased energy content is observable at $f\, L_{sep}^u/U_\infty \in [3,4]$, corresponding to the shedding frequency of the arch-like vortices \citep{bo2012experimental, dellaposta2023direct_mach}, we can not conclude that this is a trace on the wall of these structures, as an analogous increase at the same frequencies is also observable in the uncontrolled case.

\begin{figure}[t!]
     \centering
     \subfloat[]{
     \includegraphics[width=0.49\textwidth]{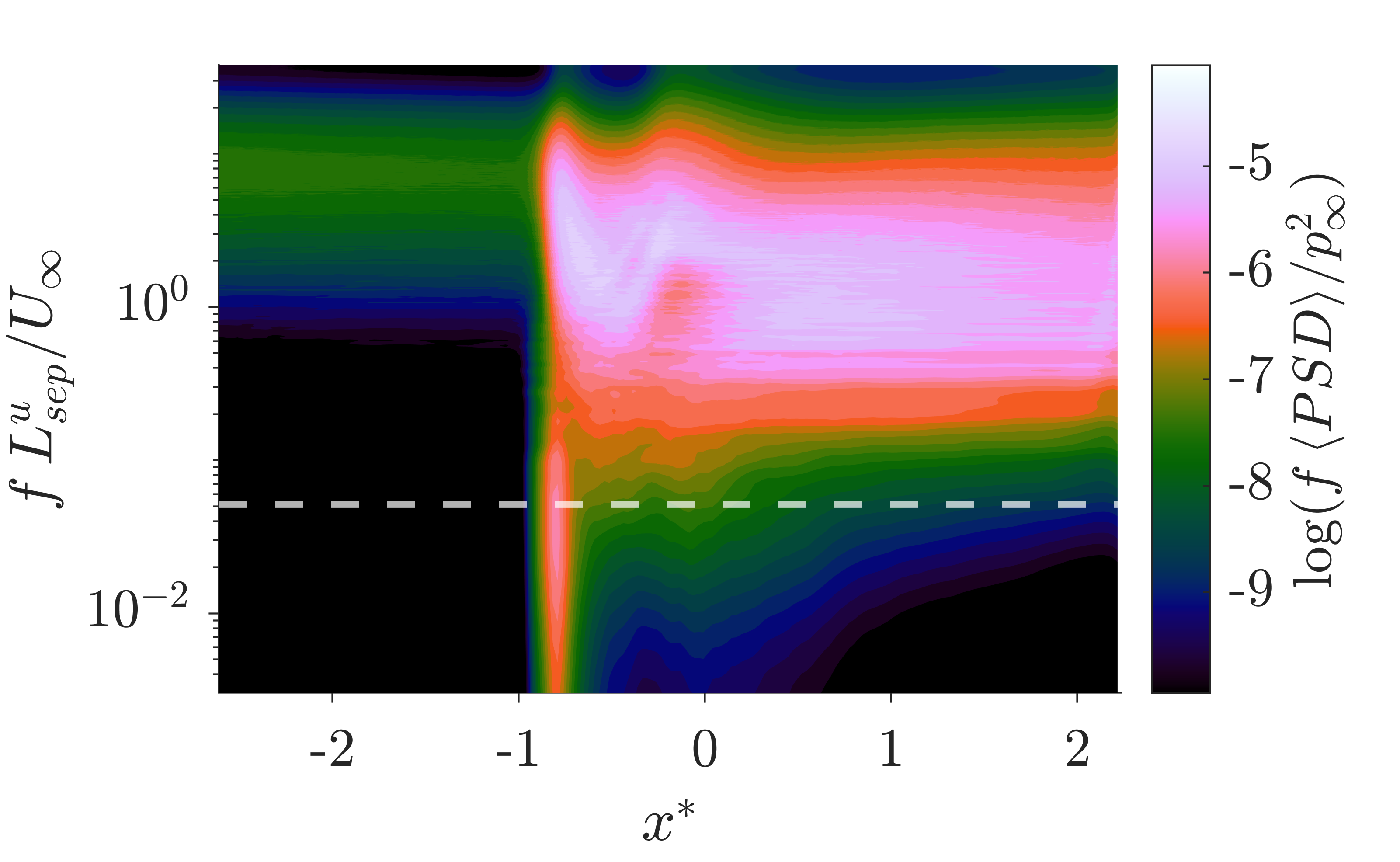}} 
     \subfloat[]{
     \includegraphics[width=0.49\textwidth]{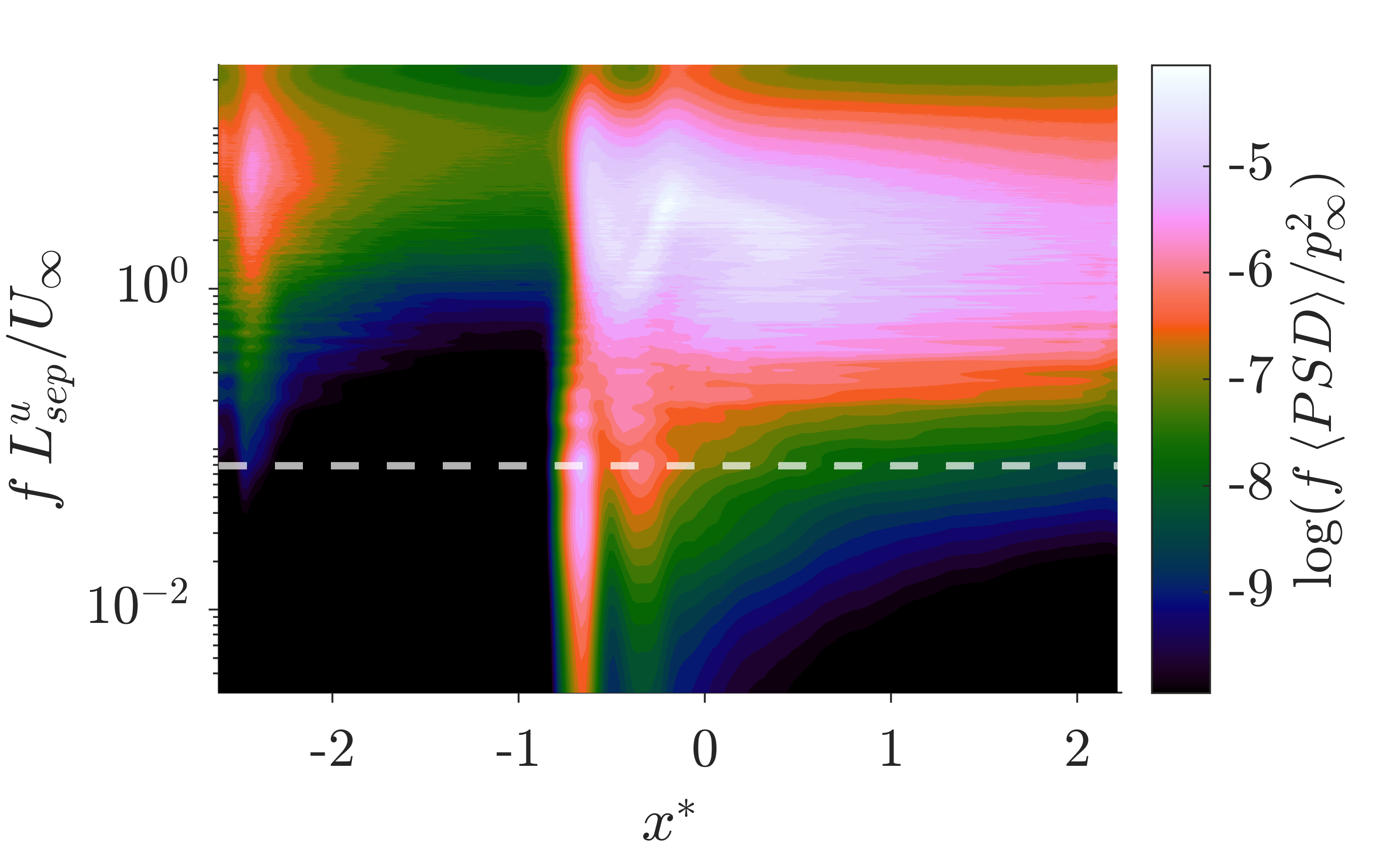}} 
     \caption{Streamwise distribution of the spanwise-averaged premultiplied wall-pressure spectra: (a) USBLI, (b) CSBLI. The dashed white line indicates the frequency of the low-frequency peak in the shock region.}
     \label{fig:streamwise_spectra_spanwise_av}
\end{figure}

Given the relevant spanwise modulation of the flow induced by the microramp, 
it is worth considering the premultiplied-spectra along the streamwise direction at single spanwise sections. As in the previous sections, we examine the notable stations at $z^* \approx -0.3$, $-0.05$, and $0$. 
Figure~\ref{fig:streamwise_spectra_local_span} shows that there is no qualitative distinction between the spectra in the span and also the low-frequency peak takes place at approximately the same Strouhal number. Slight quantitative differences however are present in the reattachment region and beyond ($x^* \approx [-0.25, 0.5]$) for high frequencies ($f\, L_{sep}^u/U_\infty \in [2,5]$). At $z^* \approx -0.05$, the intensity of the spectra is higher, suggesting an increased activity of the eddies in the boundary layer. In this section, the local separation length is minimum and the height of the recirculation bubble is smaller (see figure~\ref{fig:streamlines_3d}). Given this observation, we could speculate that the action of the arch-like vortices, whose characteristic shedding frequency is precisely in this range, is here felt stronger close to the wall and thus contributes to the surge in the spectra. Moreover, looking at the distribution of the skin friction lines in figure~\ref{fig:skin_friction_lines}, the spanwise section with minimum separation length corresponds approximately to the location of a focus on the reattachment line, where the convergence of the flow could further contribute to the observed increase in high-frequency wall-pressure fluctuations. Unsteady data in this spanwise section have not been considered in this work and will be the object of future research. 
An increased intensity is also visible close to the ramp trailing edge at $z^* \approx -0.05$. Here, at the sides of the primary vortex pair, the conical shock wave around the microramp wake can penetrate the flow close to the wall, thus leaving a stronger imprint on the spectra.

Finally, it is worth noticing that, despite its increase in absolute terms, 
if the local peak frequency is scaled by the local separation length (see table~\ref{tab:spectra_recap}), the resulting non-dimensional frequency goes back approximately to the value of the 2D USBLI. The slight differences that are still present may be due to three-dimensional issues that make the 2D scaling not completely effective. 
%
\begin{figure}[t!]
     \centering
     \subfloat[]{
     \includegraphics[width=0.49\textwidth]{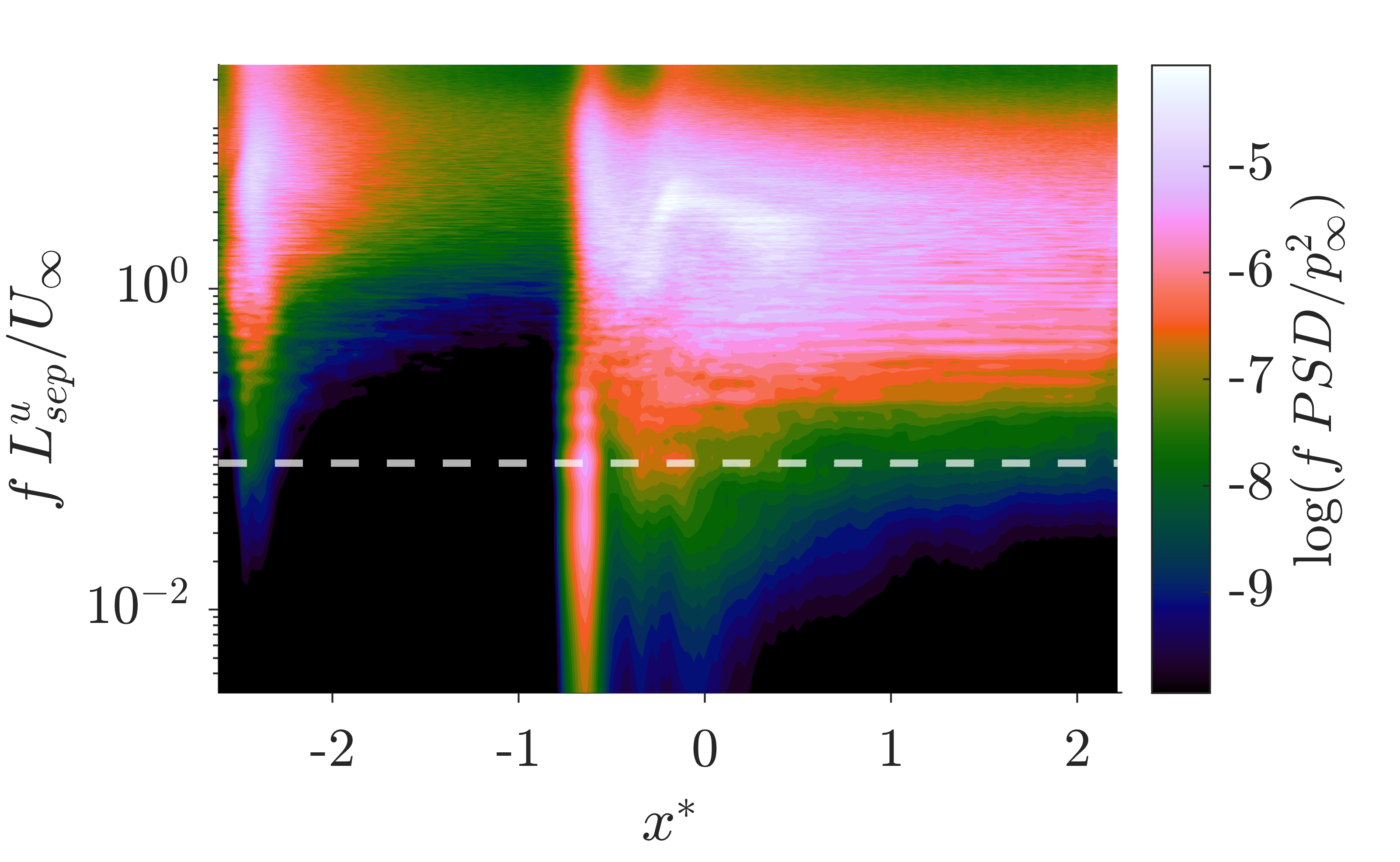}}
     \subfloat[]{
     \includegraphics[width=0.49\textwidth]{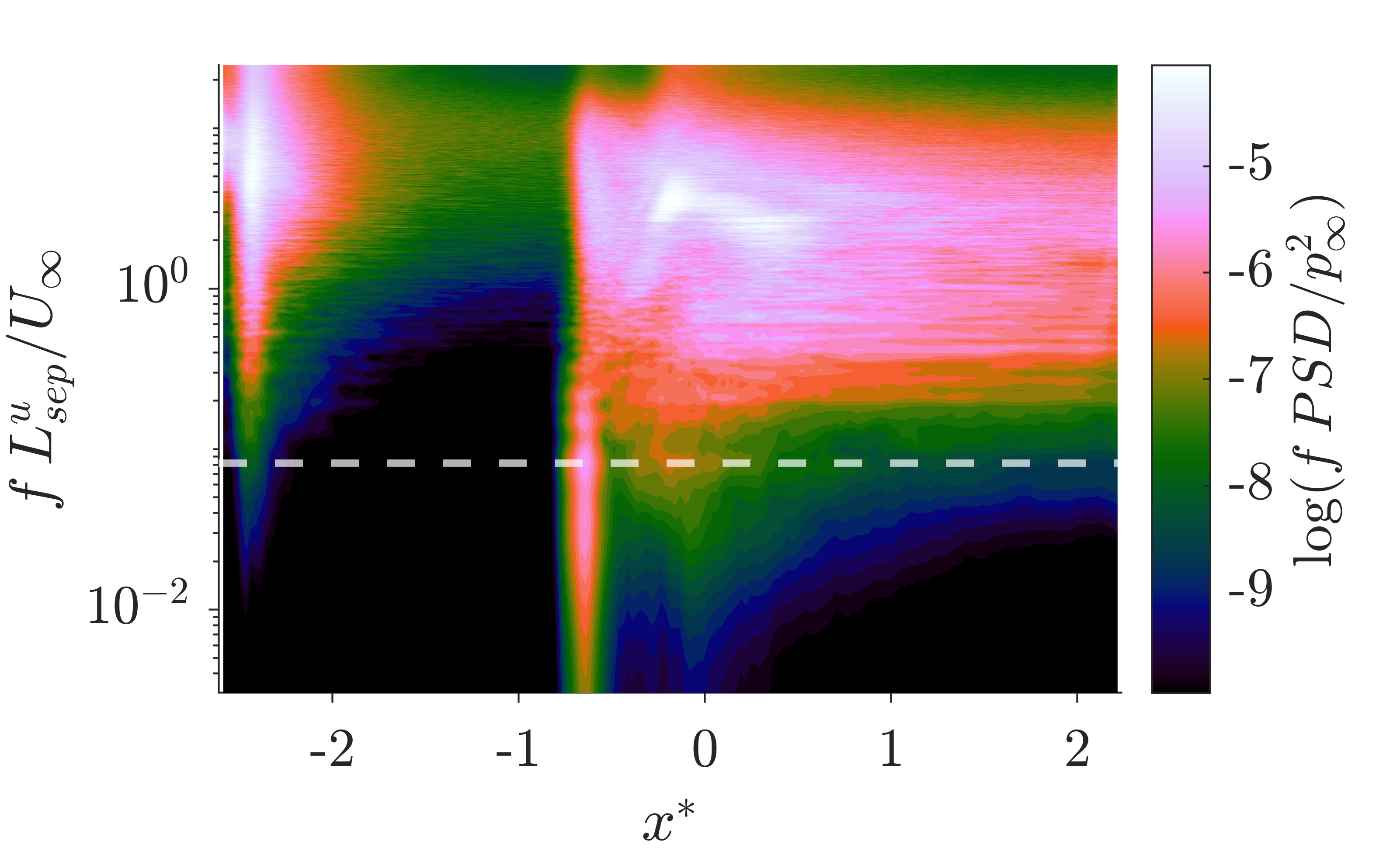}}\\ 
     \subfloat[]{
     \includegraphics[width=0.49\textwidth]{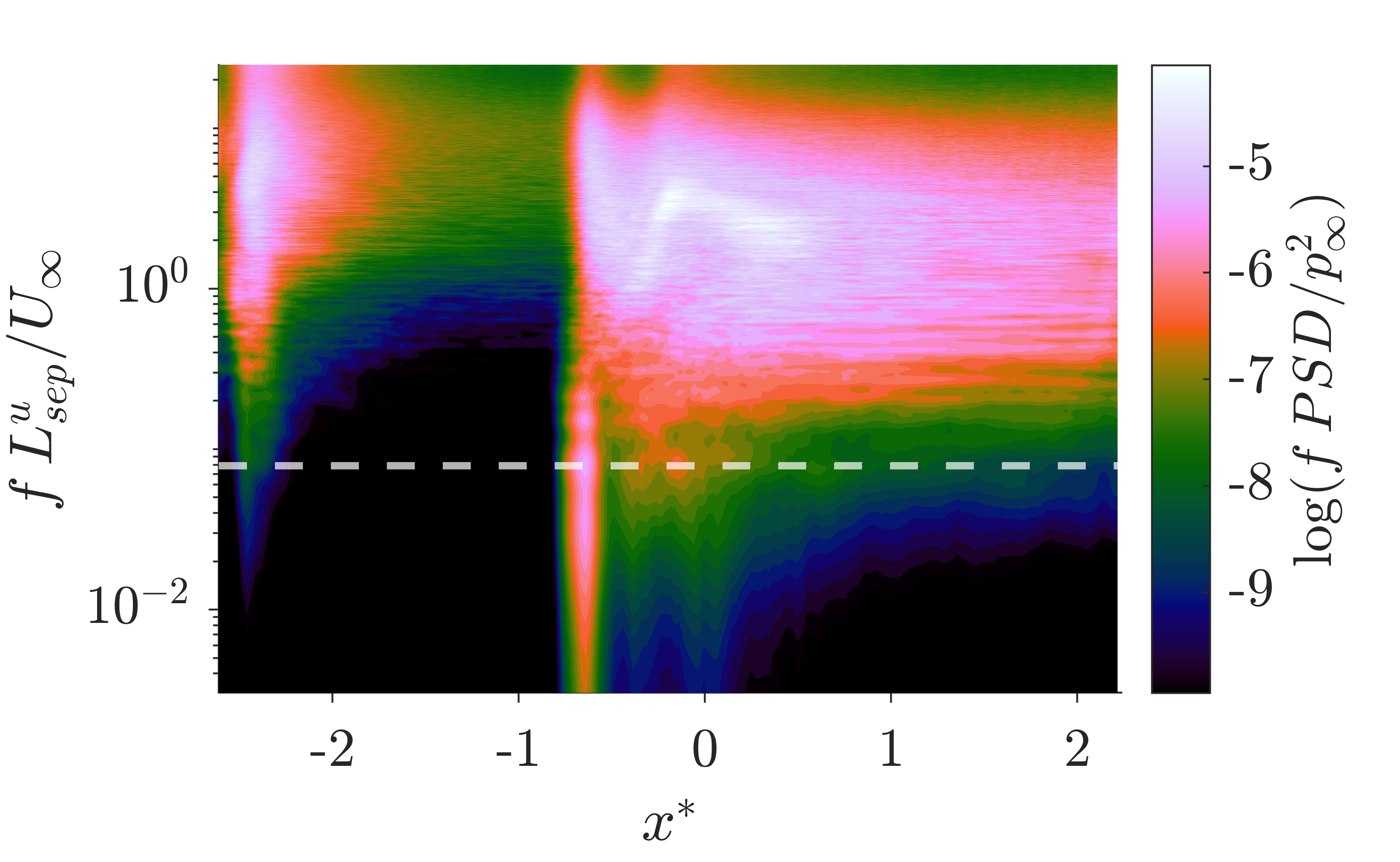}}    
     \caption{Streamwise distribution of the premultiplied wall-pressure spectra of the CSBLI case at: (a) $z^* \approx -0.3$, (b) $z^* \approx -0.05$, and (c) $z^* \approx 0$. The dashed white line indicates the frequency of the local low-frequency peak in the shock region.}
     \label{fig:streamwise_spectra_local_span}
\end{figure}

\item[Spanwise spectra]\,
To better investigate the relative distribution in frequency in the $z^*$ direction, 
figure~\ref{fig:spanwise_spec_along_sep_peak} reports the premultiplied spectra along notable curves in the span. 
The previous analysis confirmed that the microramp wake modulates the flow in the spanwise direction, with a mild variation also in the spectral features of the \gls{sbli}.
For this reason, to compare analogous conditions along $z^*$, we sampled the wall-pressure signals 
following the separation line, the foremost wall-pressure standard deviation peak location, and the reattachment line
(see figure~\ref{fig:sep_peak_locus}). 
However, the integral in frequency of the spectra is not constant along these curvilinear coordinates, as it corresponds to the variance of the local wall-pressure signals (Parseval's theorem), which changes along the xz plane in the CSBLI case.
Since we want to compare only the relative distribution in frequency at different sections, we normalised the spectra with the local wall-pressure variance. The results allow us to examine how the relative magnitude of the spectral content associated with the shock oscillation, turbulent fluctuations, and other features varies along the span.

From the contours in the first row, as expected, the USBLI results show that the relative distribution is approximately constant along the span. Along the -- straight -- separation line, the energy of the wall-pressure signal is distributed evenly between the low- and high-frequency ranges. Moving downstream, towards the standard deviation peak and then the reattachment -- straight -- lines, the dominant spectral content shifts to higher frequencies, as the influence of the reflected shock unsteadiness vanishes progressively. At the reattachment, the peak frequency related to the eddies in the reattaching boundary layer is $f\, L_{sep}^u /U_\infty \approx 2$. 

The contours of the second rows report, instead, the results of the CSBLI case. 
Along the separation, compared to the USBLI case, the scenario at the side and central regions is different. Close to the lateral boundaries, the energy of the signal at the low-frequency peak is larger than the energy at high-frequencies. The opposite takes place at the centre of the domain, behind the microramp, where the contribution from the low frequencies is dampened and the high-frequency contributions are stronger. According to the previous results, the spectra in the span confirm that although microramps do not cancel the global unsteadiness of the shock, they provide a local attenuation of its relative impact in the region of their wake. Moreover, as the frequency associated with the shock unsteadiness stays almost constant, we can deduce that, despite its geometry being altered by the impingement of the ramp wake, the front of the reflected shock remains coherent in the span during its low-frequency oscillation. The distribution along the peak standard deviation is similar to that along the separation, although we can notice that the contribution from the low-frequency range in the CSBLI case is still relevant compared to the USBLI case. 
Interesting differences are present instead along the reattachment line. Similarly to the USBLI case, the energy is here concentrated in the high-frequency range, and especially at the lateral boundaries the contour is almost the same as without microramp. However, close to the symmetry plane, the dominant contribution increases slightly in magnitude and frequency, reaching exactly the shedding frequency of the arch-like vortices in a peaked fashion. 

A trace at the wall of the \gls{kh} vortices around the wake is noticeable only at the reattachment, while the typical tonal signature at $f\, L_{sep}^u/U_\infty \in [3,4]$ is absent in the first part of the interaction. Considering also their reduced spanwise vorticity compared to the one of the bottom shear layer (see figure~\ref{fig:zeta_vortz}), it is possible to believe that the action of the arch-like vortices is first shielded by the 3D shear layer developing around the separation region, and thus that they may play a limited role in the delay of the separation. Their role may instead be more relevant in the mechanism to close the recirculation bubble.

\begin{figure}[th]
     \centering
     \subfloat[]{
     \includegraphics[width=0.325\textwidth]{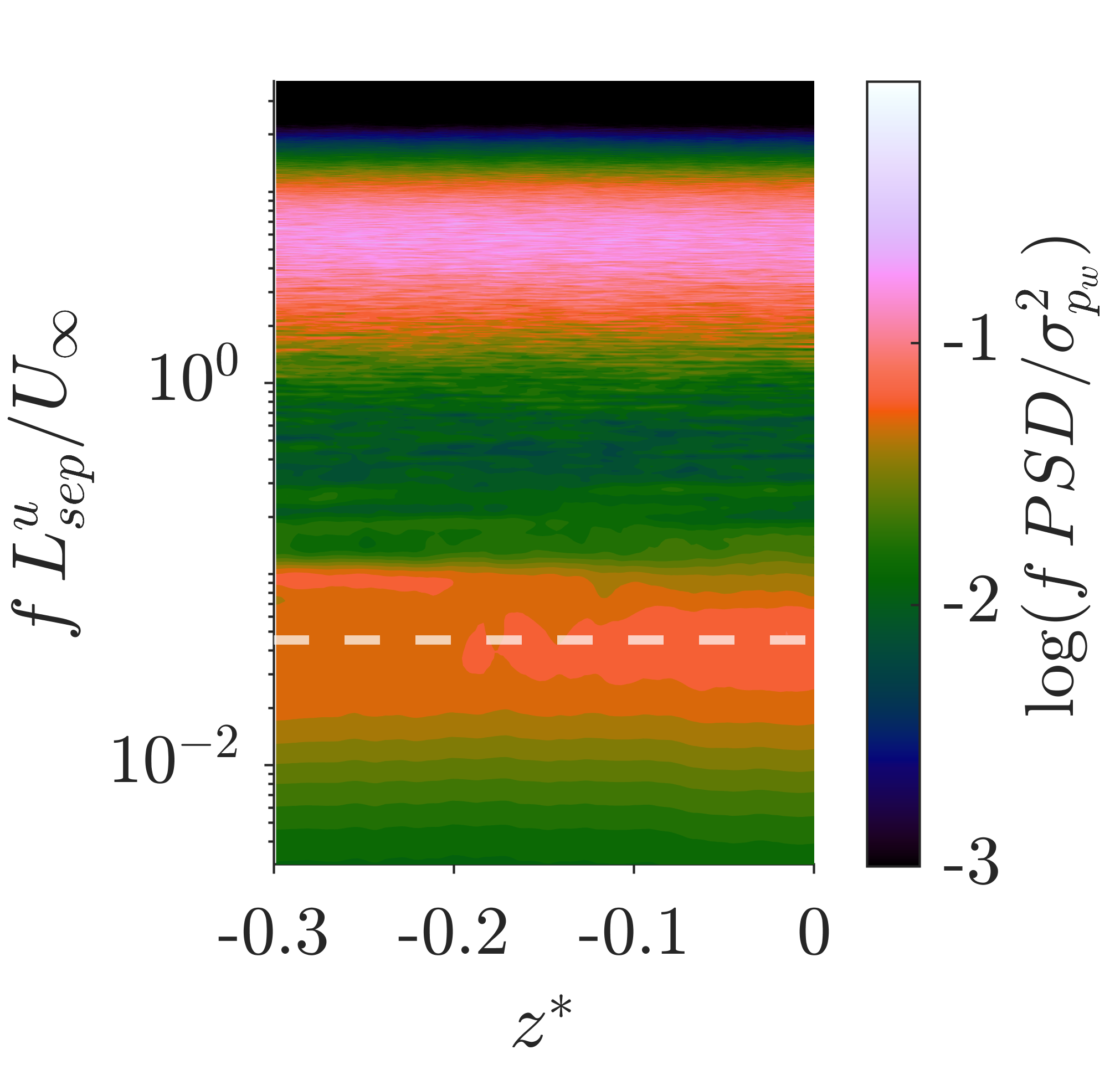}}
     \subfloat[]{
     \includegraphics[width=0.325\textwidth]{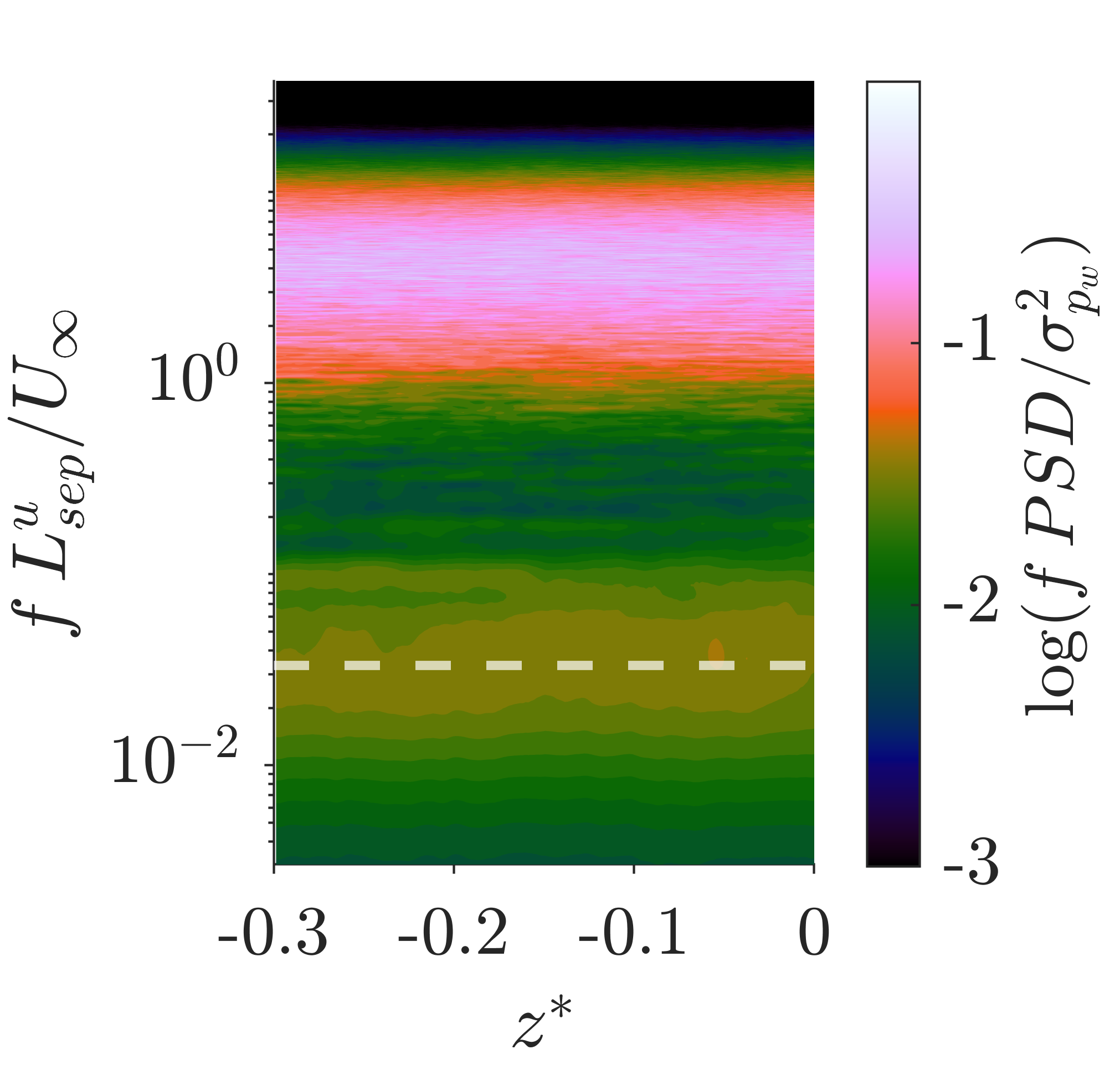}}
     \subfloat[]{
     \includegraphics[width=0.325\textwidth]{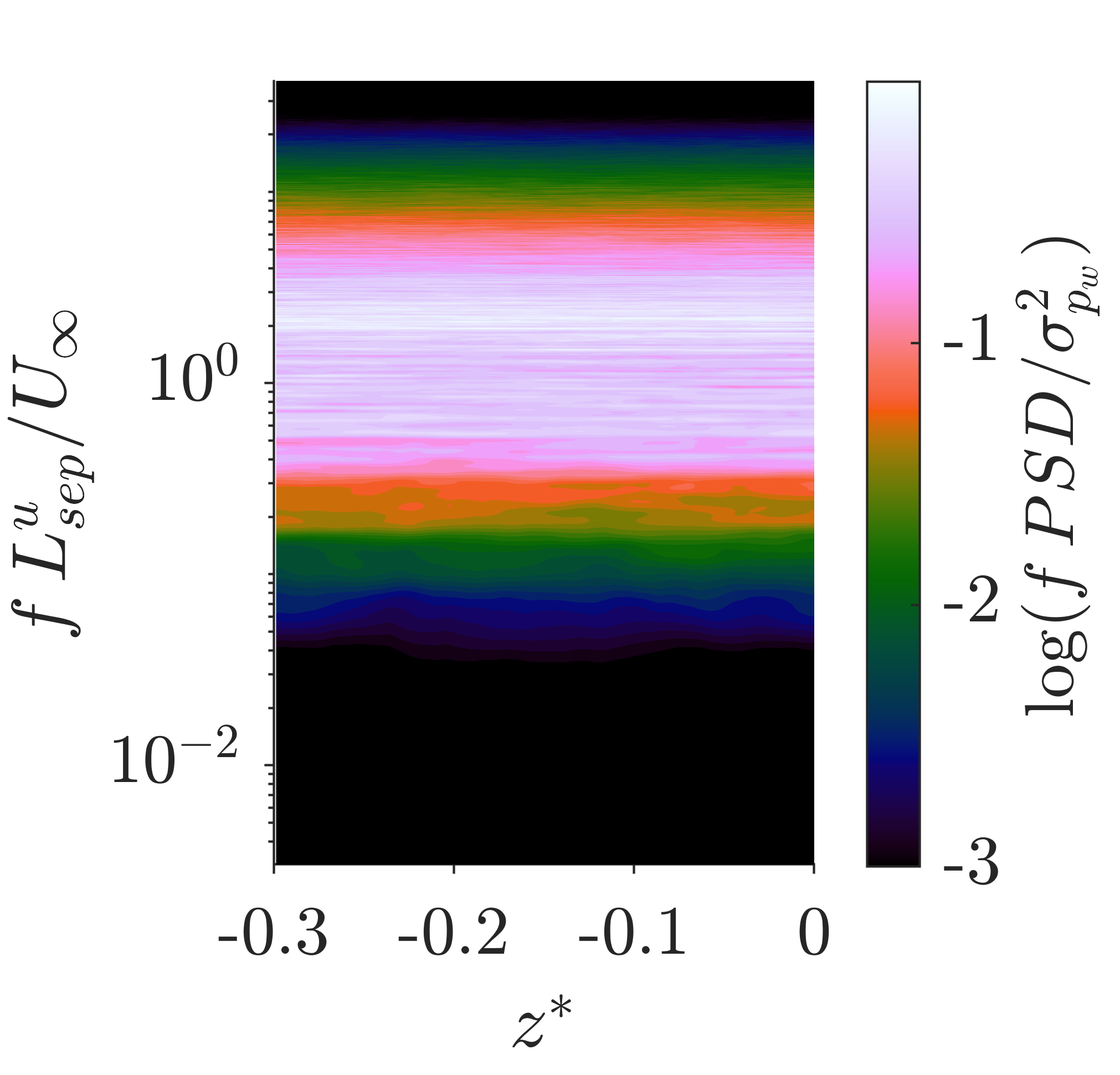}}
     \\
     \subfloat[]{
     \includegraphics[width=0.325\textwidth]{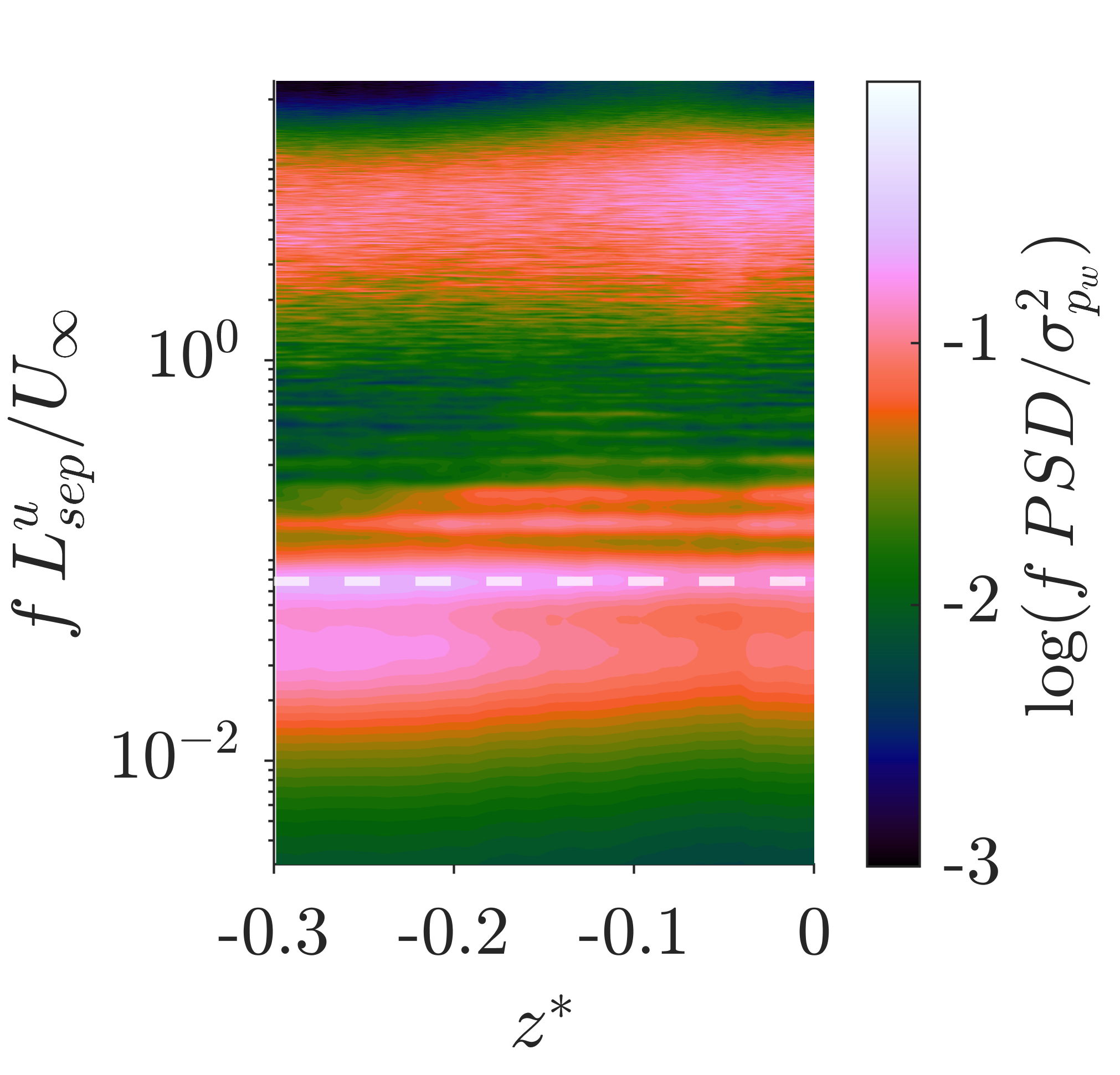}}
     \subfloat[]{
     \includegraphics[width=0.325\textwidth]{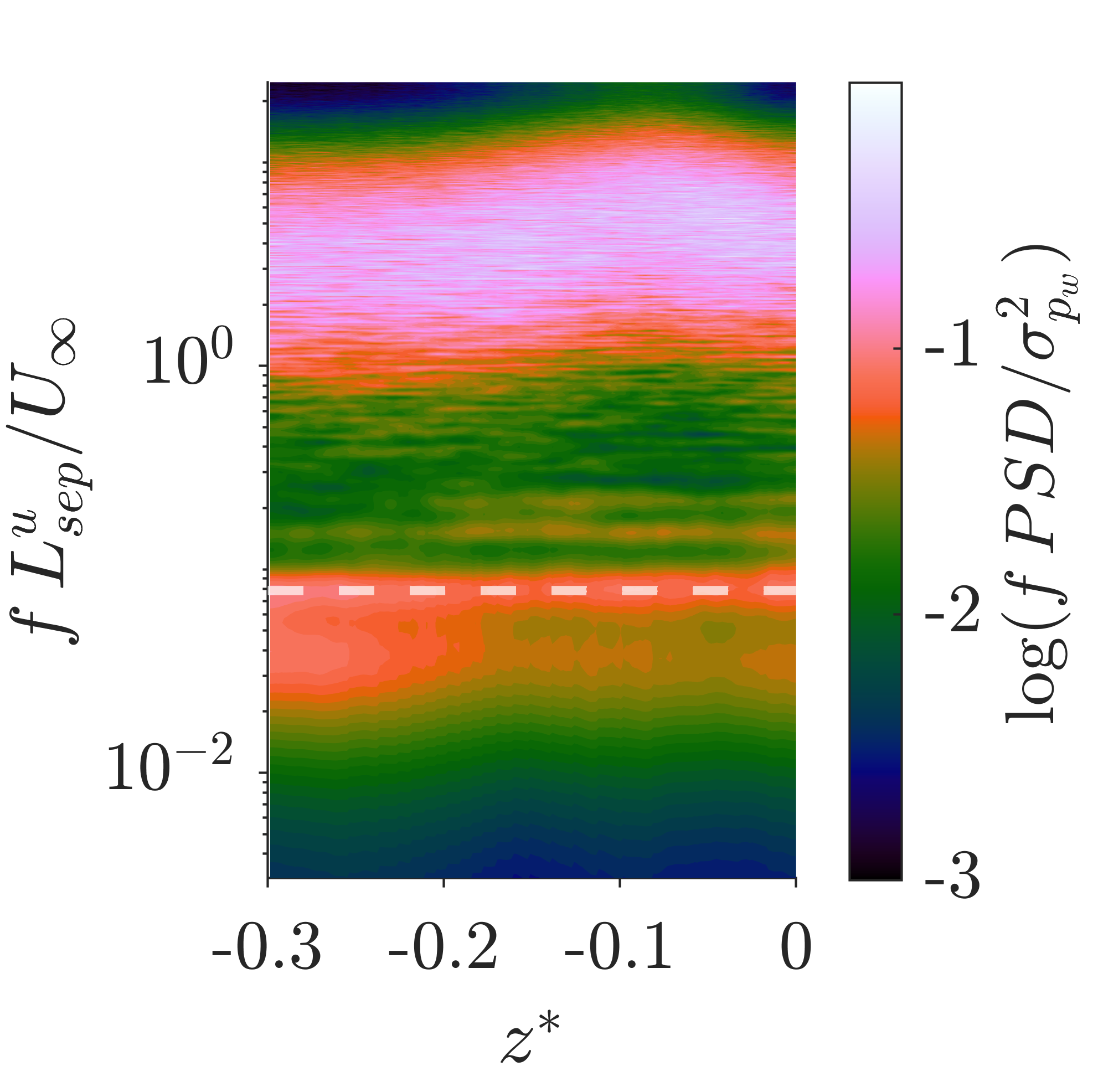}}
     \subfloat[\label{fig:spanwise_spec_along_reat}]{
     \includegraphics[width=0.325\textwidth]{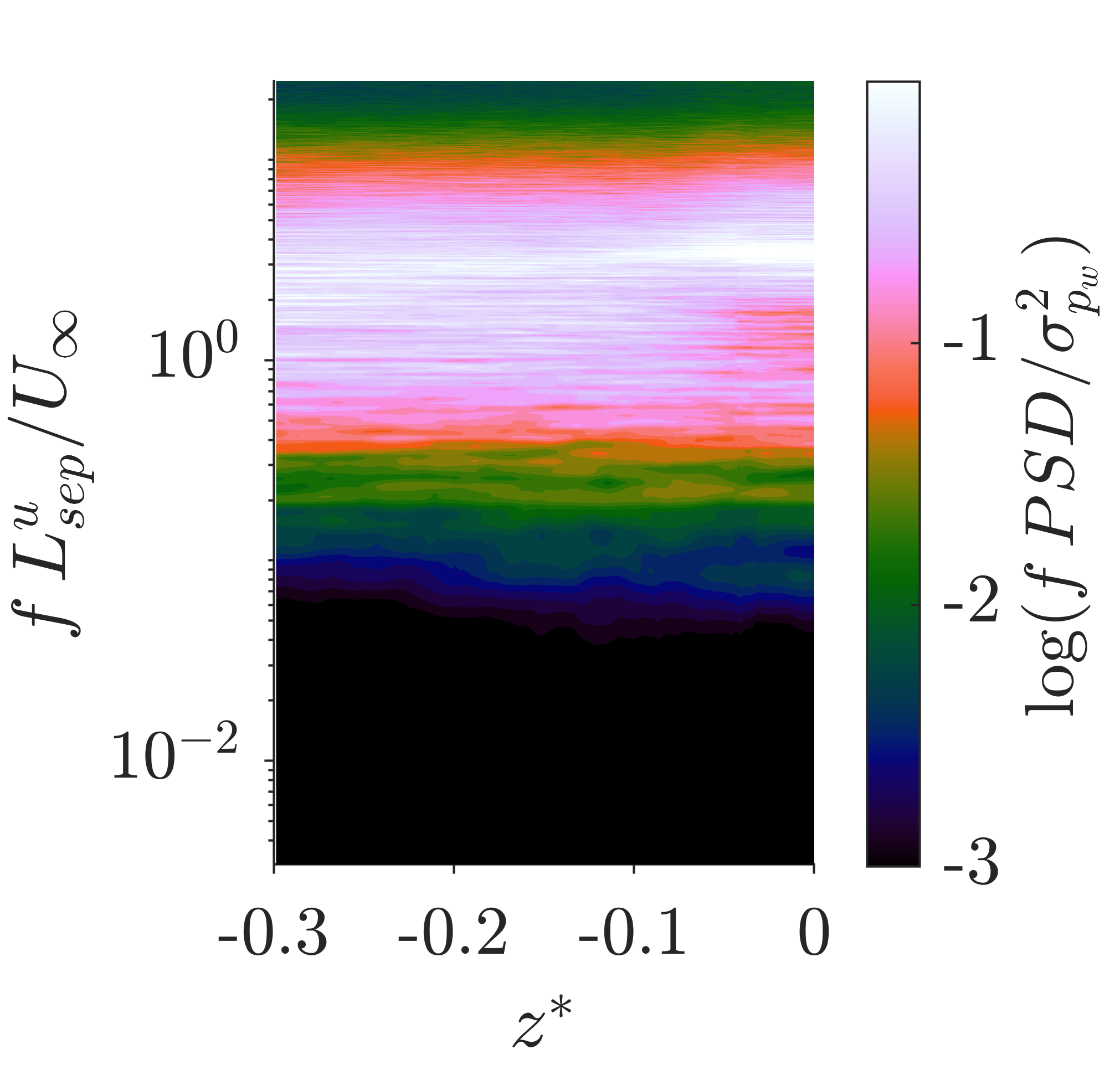}}
     \caption{Spanwise distribution of locally-normalised premultiplied wall-pressure spectra. USBLI (first row) and CSBLI (second row), along the separation (first column), along the wall-pressure standard deviation peak (second column), and along the reattachment (third column).}
     \label{fig:spanwise_spec_along_sep_peak}
\end{figure}

\item[Wall-normal spectra]\,
We complete the picture of the Fourier analysis with figure~\ref{fig:wallnormal_spectra_local_span}, which shows the premultiplied spectra of the pressure along the wall-normal coordinate, at the symmetry plane. Time signals are sampled in correspondence with a streamwise location close to separation (a-b) 
and reattachment (c-d) for the USBLI (a-c) and CSBLI (b-d) cases.
These spectra allow us to locate the characteristic frequencies of the flow in the wall-normal direction, 
and thus relate the behaviour of the fluctuations at the wall with those far from the wall.

The USBLI results at the separation document the energetic fluctuations associated with the low-frequency motion of the reflected shock wave, which initiates the separation penetrating the boundary layer up to a short wall-normal distance. 
The high-frequency content close to the wall is instead the trace of the vortical structures living at the edge of the recirculation bubble. Indeed, their intensity is small at separation, as the separation shear layer is here still at its onset, while being diffused across the vertical direction at reattachment, where pressure fluctuations are associated with the eddies of the new boundary layer downstream of the interaction. 
Finally, the contour at reattachment shows also an interesting feature related to the reattachment shock taking place downstream of the separation, observable at  $y^* \approx 0.38$. This shock is characterised by a broadband shape, as for the separation shock, despite its peak non-dimensional frequency being considerably higher (at least a decade more than the traditional low-frequency unsteadiness). 

The CSBLI contours show analogous results for what concerns the trace of the low-frequency peak of the separation shock, 
although the trace now covers a larger wall-normal range, suggesting an increased shock smearing.


The most interesting difference however, is the strong peak at $y^* \in [0.2,0.3]$ and $f\, L_{sep}^u / U_\infty \in [2,5]$, which marks the spectral content of the arch-like vortices. 
This contribution remains generally separate from the high-frequency one located close to the wall -- especially in the separation region -- and its intensity is comparable with that of the most energetic features in the spectra. At reattachment, the rise in intensity of the fluctuations in the high-frequency range induced by the shocks all along the wall-normal direction causes a larger peak in correspondence with the arch-like vortices, whose influence is able to reach the wall, as we observed in the spanwise spectra reported in figure~\ref{fig:spanwise_spec_along_reat}. 
The spectral analysis thus confirms that the arch-like vortices' shedding frequency is in the same range as the turbulent fluctuations taking place in traditional \gls{sbli} and that their influence directly reaches the wall only at reattachment. 
However, despite providing interesting information about the spatial and spectral features of the flow, spectra do not provide information about the actual reattachment mechanism in microramp-controlled \glspl{sbli} and about the role arch-like vortices play in it, which thus remains an open question for future work. 

%
\begin{figure}[t!]
     \centering
     \subfloat[]{
     \includegraphics[width=0.49\textwidth]{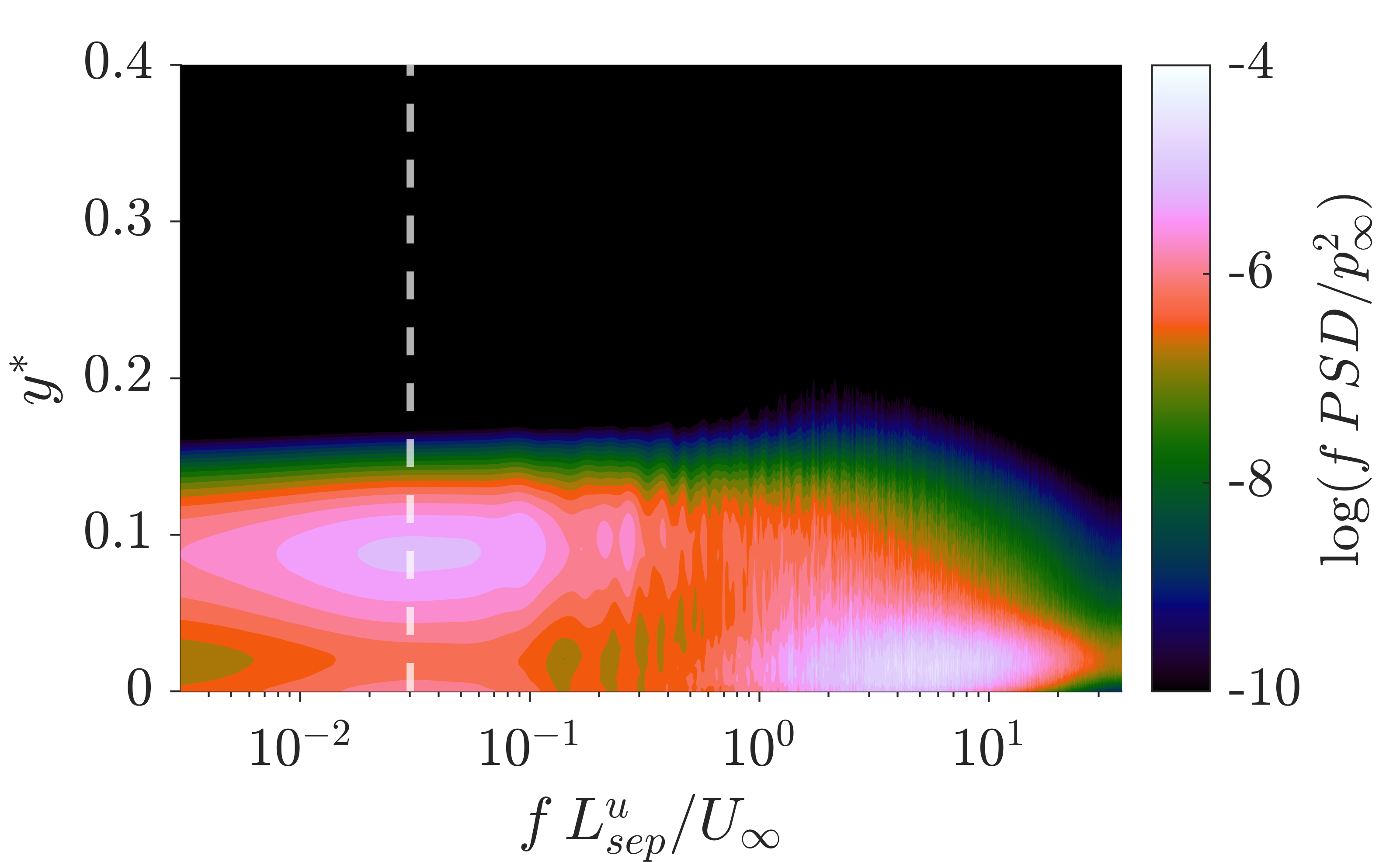}}
     \subfloat[]{
     \includegraphics[width=0.49\textwidth]{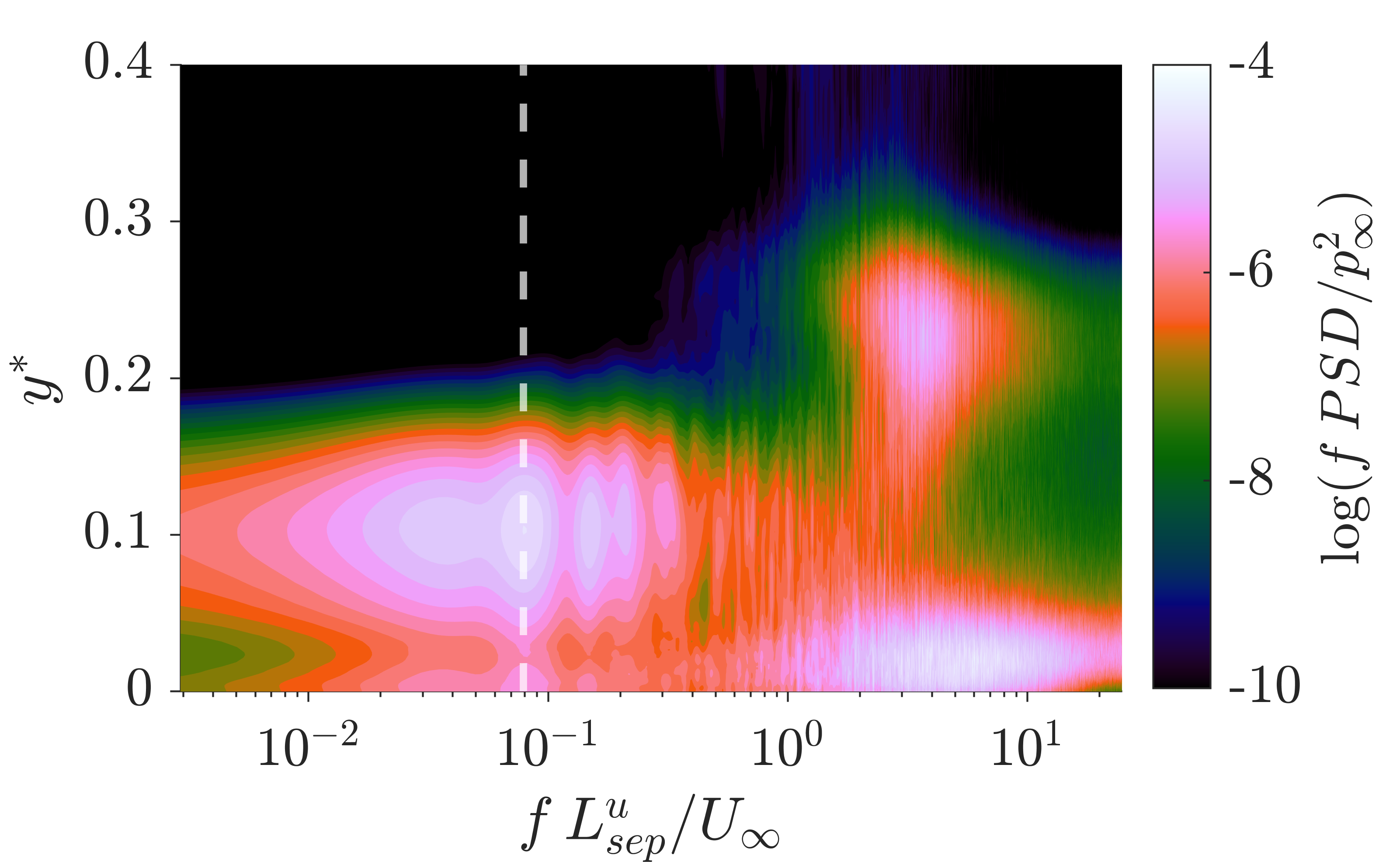}}\\
     \subfloat[]{
     \includegraphics[width=0.49\textwidth]{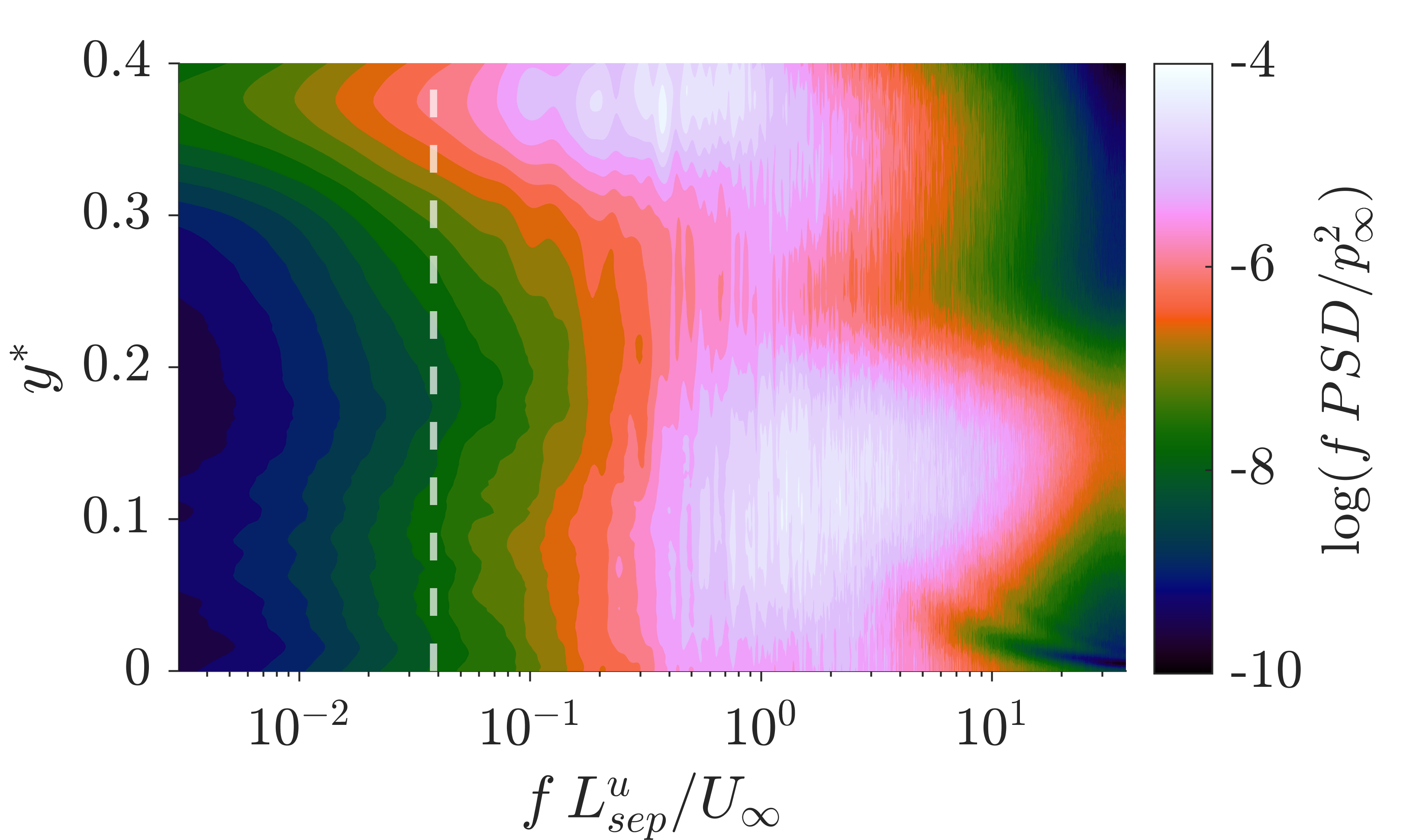}}
     \subfloat[]{
     \includegraphics[width=0.49\textwidth]{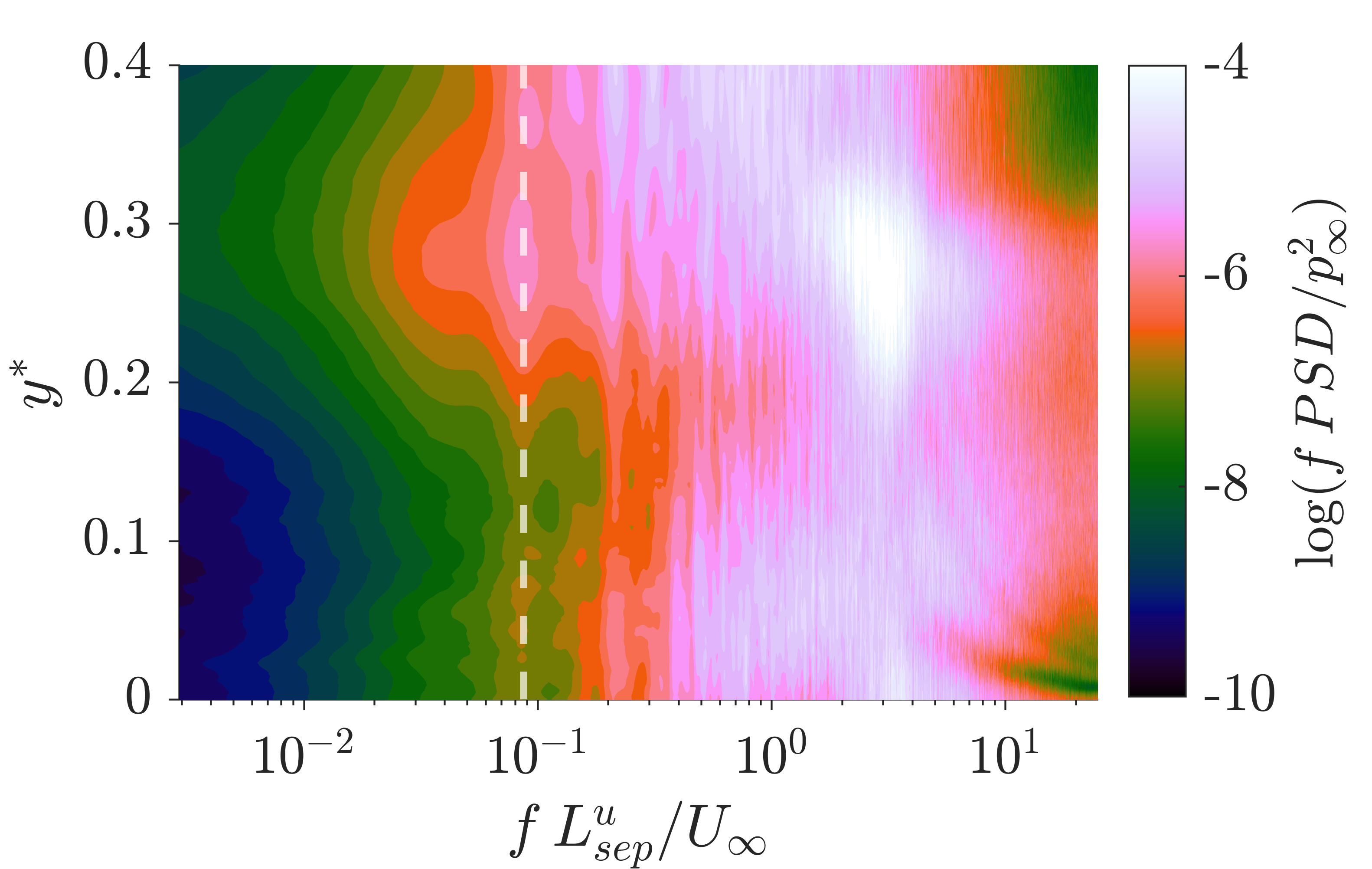}}
     \caption{Wall-normal distribution of the premultiplied wall-pressure spectra at the symmetry plane in the (a-b) separation 
     and (c-d) reattachment regions. USBLI on the left column, CSBLI on the right column. The dashed white line indicates the local low-frequency peak associated with the oscillation of the separation shock.}
     \label{fig:wallnormal_spectra_local_span}
\end{figure}
\end{description}

\subsubsection{Wavelet analysis}


In this section, we want to characterise the degree of intermittency of the wall-pressure signals in the interaction region, in particular at the onset of the separation and at reattachment. Similarly to the approach adopted in \citet{bernardini2023unsteadiness}, wavelet analysis is applied to extract energetic intermittent events. Indeed, this approach provides a more direct measure of the degree of intermittency of the wall-pressure field and allows for the extraction of local -- in time -- features that may be partially lost using Fourier analysis. 

The wavelet transform is computed by the convolution of the wall-pressure signal $p_{w}(t)$ 
with the dilated (by the factor $k$) and translated (by the factor $t$) 
complex conjugate counterpart of a so-called mother wavelet, according to the following formalism:
\begin{equation}
G_{\Psi}(k,t) = \frac{1}{\sqrt{k}} \int_{-\infty}^{+\infty} p_w(\tau)\Psi^{*}\left(\frac{\tau-t}{k}\right)\mathrm{d}\tau \,,
\label{eq:def}
\end{equation}
where $\Psi$ is the wavelet mother function, $k$ is a dilatation parameter indicating the time scale of the event under consideration, $t$ 
is the time-translation parameter and $*$ denotes the complex conjugate. 
A detailed theoretical framework can be found in \citet{farge1992wavelet} and \citet{mallat1999wavelet}. In this study, the Morlet wavelet has been chosen, as a higher resolution in frequency can be achieved compared to other mother functions. 
Owing to the wavelet admissibility condition~\citep{torrence1998practical}, the frequency associated with the dilatation parameter for the Morler wavelet is defined as $f = 0.97/k$.

In statistics, intermittency denotes the rare occurrence of exceptionally spiky events which are patchy and bursty.  These instances cause higher-order moments to converge with greater difficulty, suggesting a significant departure from Gaussian statistics and hence non-homogeneous distribution of energy in time.

Once computed the wavelet transform coefficients $G_{\Psi}(k,t)$, it is possible to obtain the scale-time distribution of the energy density $|G_{\Psi}(k,t)|^2$ of the wall-pressure signal. Thanks to this property, \citet{meneveau1991analysis} and \citet{camussi2021intermittent} suggested that an 
effective indicator of the intermittency is the squared local intermittency measure, denoted as $LIM2$:
\begin{equation}
LIM2(k,\tau)=\frac{|G_{\Psi}(k,t)|^4}{\langle|G_{\Psi}(k,t)|^2\rangle_{t}^2}\,. 
\label{eq:LIM2}
\end{equation}
where $\langle \bullet \rangle_t$ indicates the time average of the considered quantity.
$LIM2$ can be interpreted as a time-scale dependent measure of the flatness factor or kurtosis of wall-pressure signals. Therefore, the $LIM2$ parameter will be equal to 3 when the probability distribution is Gaussian, while the condition $LIM2>3$ identifies only those rare bursts of energy contributing to the deviation of the wavelet coefficients from a normal, Gaussian distribution. 

\begin{figure}[t!]
     \centering
     \subfloat[\label{fig:lim2_sep_USBLI}]{
     \includegraphics[width=0.49\textwidth]{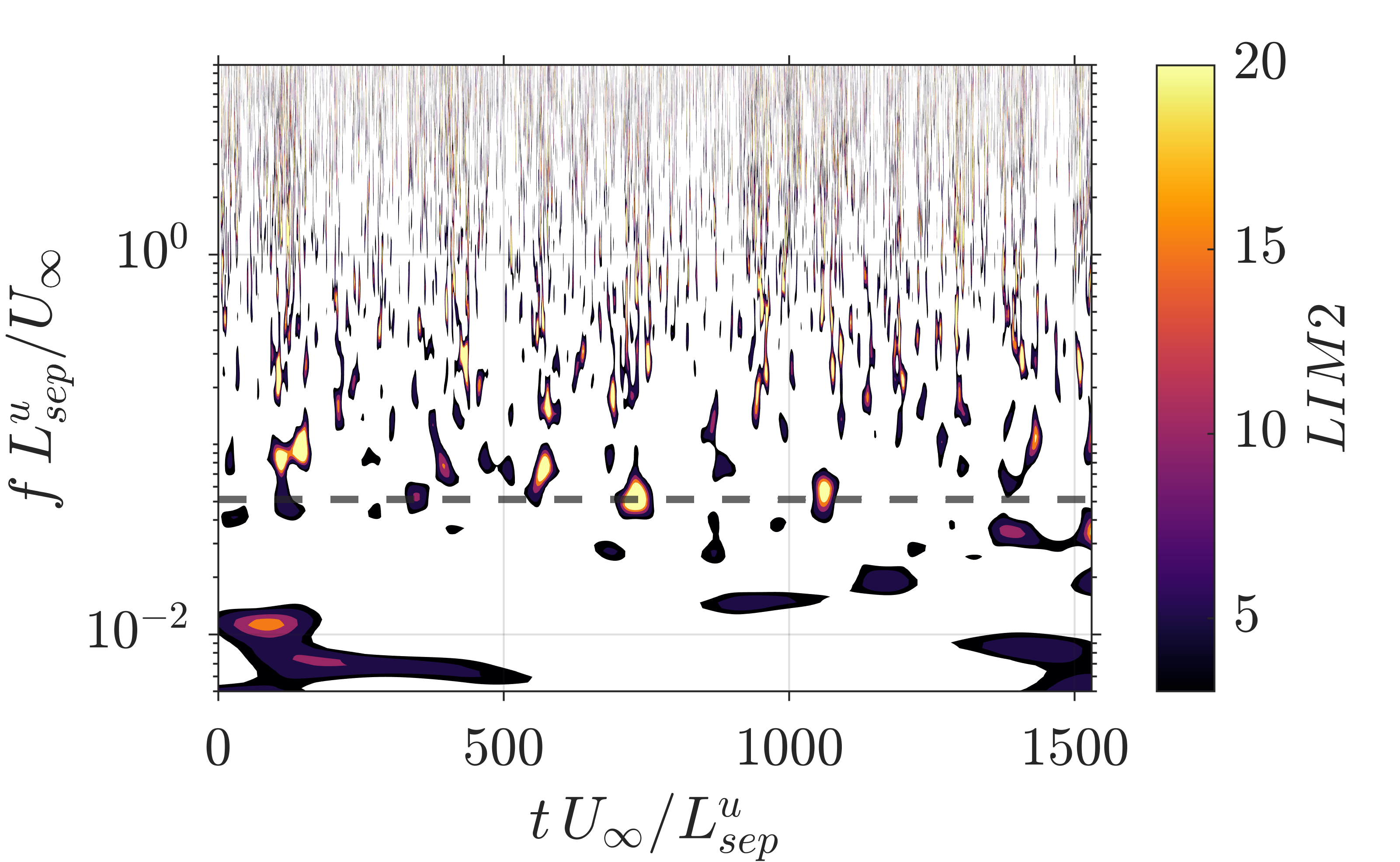}}
     \subfloat[\label{fig:lim2_sep_CSBLI}]{
     \includegraphics[width=0.49\textwidth]{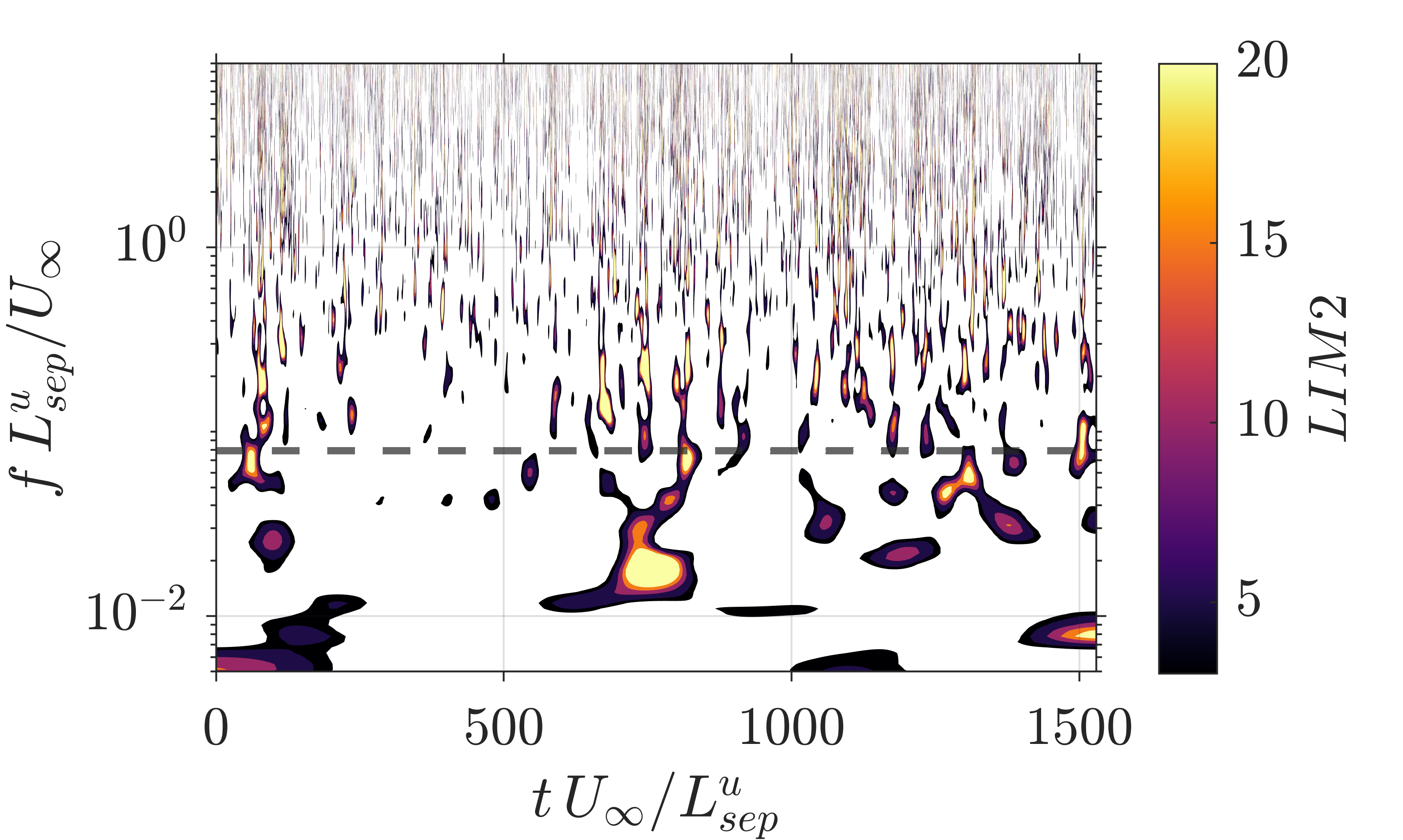}}
     \\
      \subfloat[\label{fig:lim2_re_USBLI}]{
     \includegraphics[width=0.49\textwidth]{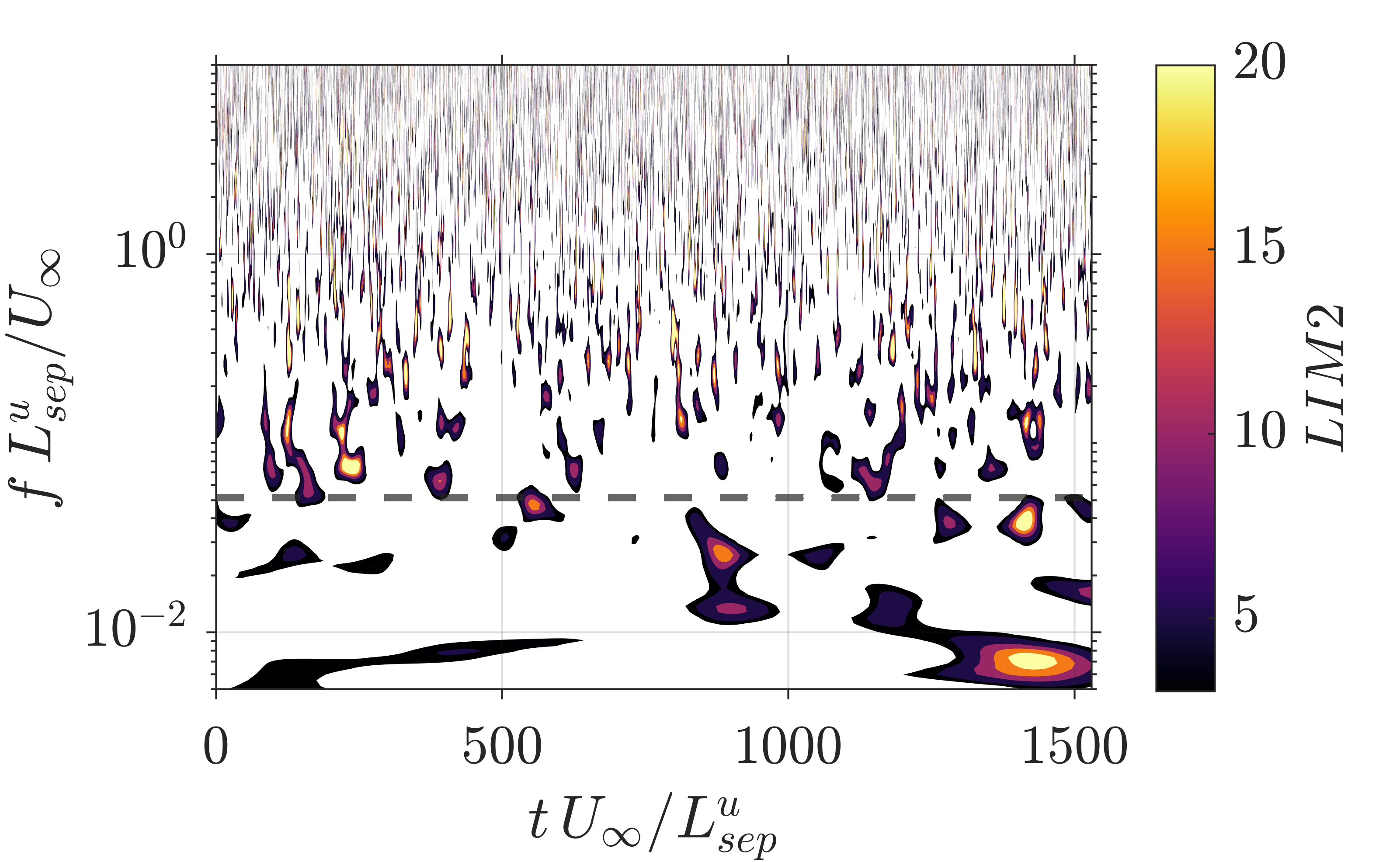}}
     \subfloat[\label{fig:lim2_re_CSBLI}]{
     \includegraphics[width=0.49\textwidth]{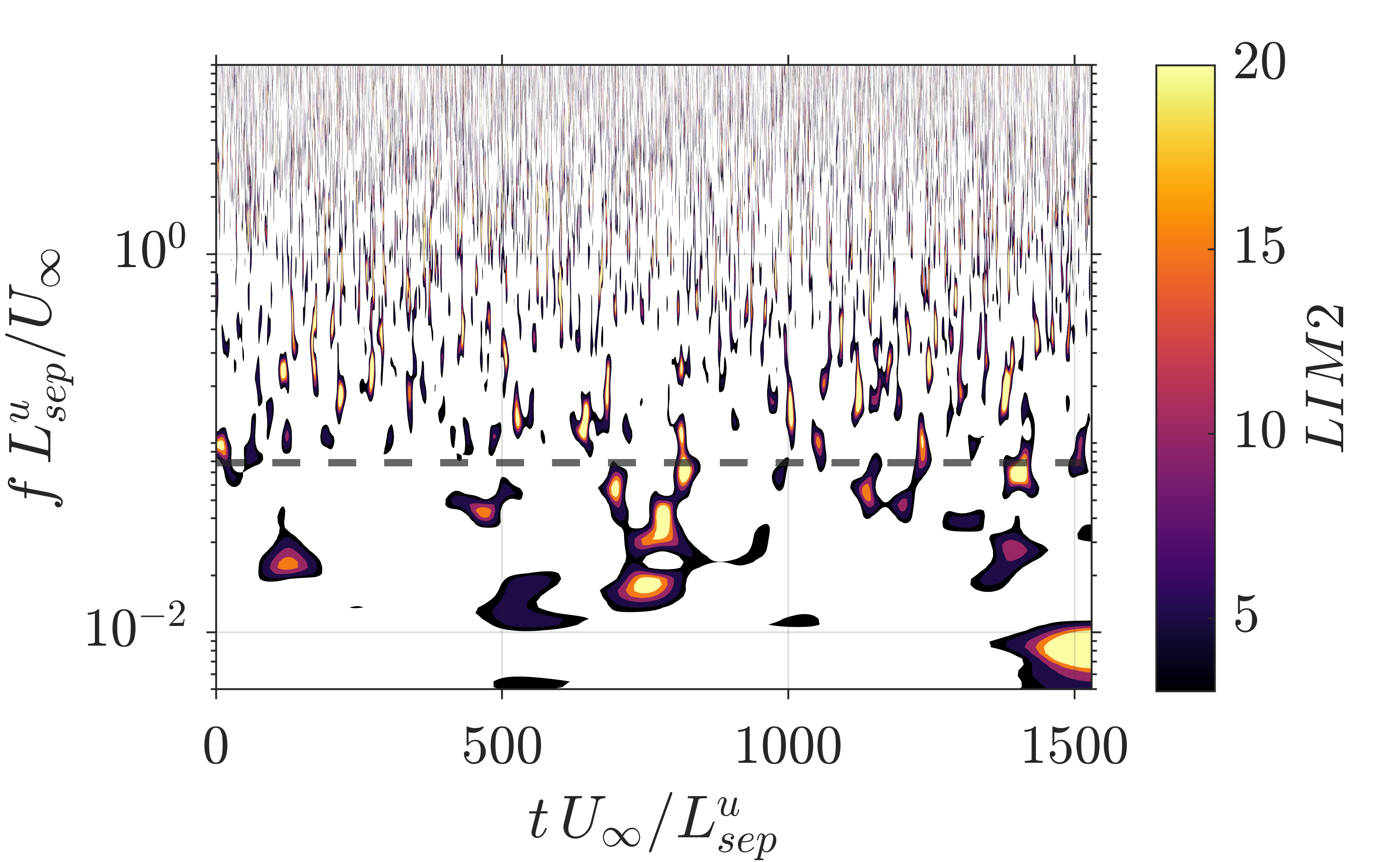}}
     \caption{Contours of $LIM2$ of $p_w/p_{\infty}$ at the separation (a-b) and reattachment points (c-d) at the symmetry plane. USBLI (left column) and CSBLI (right column). Only $LIM2$ values above 3 are shown. Dashed black lines indicate the dominant low-frequency peak of each case.}
     \label{fig:lim2}
\end{figure}

In the following, we apply the wavelet analysis to the wall-pressure signal  
at two relevant stations for both the USBLI and CSBLI cases: the separation point and the reattachment point at the symmetry plane. 
Figure~\ref{fig:lim2_sep_USBLI} shows the LIM2  maps (only the levels greater than 3) at the separation point for USBLI. The occurrence of intermittent events in the frequency range characterising the shock unsteadiness -- whose Fourier frequency peak is indicated with a dashed horizontal line -- is immediately apparent. The most relevant characteristic feature of these bursts of energy is that they are rather scattered in time and clustered around the Fourier peak frequency, but with some very energetic events at lower frequencies. The long simulation time allows our \gls{dns} to capture these rare events. This picture again confirms that the low-frequency unsteadiness of \glspl{sbli} is composed of a series of rare end energetic events and it is this rarity that causes the difficulty of convergence of higher-order statistics. At higher frequencies, it is possible to see the intermittency given by turbulence which is quite separate in the frequency space from the events caused by the shock/boundary layer interaction.
If we now look at the CSBLI case~\ref{fig:lim2_sep_CSBLI}, it is evident that the intermittent events are grouped around a higher frequency -- as indicated by the Fourier analysis --
with the additional occurrence of a single extremely energetic low-frequency event at $ t L^{u}_{sep}/U_{\infty} \approx 750$.
The analysis of the LIM2 maps at the reattachment points for both cases, figures~\ref{fig:lim2_re_USBLI} and \ref{fig:lim2_re_CSBLI}, shows a very similar behaviour. We can just notice that the energetic bursts at lower frequencies survive for all the interaction length.
It may be concluded that, at least for these test case conditions, microramps do not alter the highly intermittent nature of \gls{sbli}, even if a period longer than the one considered may be necessary to detect appreciable differences in higher-order statistics.


\section{Conclusions}
\label{sec:conclusions}


Given the increased interest in manoeuvrable aerospace systems flying at supersonic and hypersonic speeds, studying possible control strategies to remedy the adverse effects of often occurring \glspl{sbli} is crucial.
A promising strategy to reduce the boundary layer separation induced by an impinging shock is the use of \glspl{mvg}, sub-boundary layer passive vortex generators able to retard the separation. Although several studies described many aspects of the mean and instantaneous features induced by \glspl{mvg}, also in the presence of \glspl{sbli}, many questions remain open.
Moreover, only a few studies have used experimental or numerical methodologies that can provide reliable, high-fidelity data with full accessibility of the results like \glspl{dns} do,
which is especially worthy for such a three-dimensional and unsteady wall flow like the one under consideration. 

Given this scenario, this study uses direct numerical simulations to assess the effects of a ramp-type \gls{mvg} on a \gls{sbli} generated by an oblique shock wave impinging on a turbulent boundary layer over a flat plate. 
In particular, we compare a baseline simulation considering an uncontrolled \gls{sbli}, named USBLI, with another considering a microramp-controlled \gls{sbli}, named CSBLI, whose geometrical setup is based on the optimisation study of \citet{anderson2006optimal}. 
A remarkably long integration period is considered for both cases, allowing us to examine the effects of microramps on the characteristic low-frequency unsteadiness of \glspl{sbli}. 

A qualitative analysis of the results shows that, besides two additional shocks, the microramp induces a significant spanwise modulation of the flow. Alternated regions of accelerated and decelerated flows retard the separation depending on the span location, and large-scale streak structures associated with the ramp wake overlap the boundary layer streaks typically observed in wall-bounded turbulent flows, even after the interaction. Bumps in the separation shock surface are also generated by the periodic impingement of the arch-like vortices around the ramp wake, which are clearly visible in the billows at the symmetry plane not only upstream of the interaction but also far downstream. 

The mean wall pressure shows that the onset of separation is effectively shifted downstream, with a corresponding increase in the overall pressure gradient across the interaction. Despite what is typically observed for 2D flows, where the onset of separation is relatively insensitive to changes in shape factor (the H-paradox, \citet{babinsky2011shock}), the fuller -- 3D -- boundary layer induced by the microramp wake upstream of the interaction successfully reduces the separation length in the case under consideration. Indeed, although the mean wall pressure is approximately constant along the span, the streamwise distributions of the wall-pressure standard deviation and of the time-averaged streamwise skin friction coefficient confirm that the flow cannot be studied using traditional tools used for 2D \glspl{sbli} because of the non-negligible three-dimensionality of the flow, even far from the wall. Finally, corresponding to an overall decrease in the separation length and an increase in the pressure gradient, the intensity of the pressure fluctuations increases, as already observed in transitional \glspl{sbli} and \glspl{sbli} on non-adiabatic walls.

The analysis of the mean skin friction lines provides then valuable information to fully comprehend the changes in the separation bubble, whose topology is largely three-dimensional. For example, the convergence of the lines into critical loci on the wall 
indicates the formation of tornado-like structures redistributing the flow in both the wall-normal and transversal directions. The confluence and departure of the lines at the forefront and rear parts of the interaction allowed us to identify the separation and reattachment line respectively, and thus to estimate the precise variation along the span of the streamwise separation length. The CSBLI case shows a mean separation shorter than 30\% the value of the USBLI case but presents variations of up to 30\% along the span compared to the mean value. Moreover, thanks to the analysis of the added momentum \citep{giepman2014flow}, we show that the region of minimum separation is strictly associated with that of maximum momentum transport towards the wall by the helical motion of the ramp wake. We notice, however, that the effects of the primary vortices and the spanwise non-uniformity of the flow last even far downstream of the interaction, which, for example, may be inconvenient for the components following the separation in supersonic inlets. 

Another debated aspect we deal with is the characterisation of the mutual interaction between the arch-like vortices and the interaction region. The distribution of spanwise vorticity at the symmetry plane allowed us to examine the effects of the interaction on the vortices and shows that the trajectory of the arch-like vortices follows the edge of the separation. Their intensity, instead, is almost unaffected by the shocks if one accounts correctly for compressibility effects by considering a density scaling of the vorticity defined with the Favre-averaged velocity, following our analytical derivation. On the other hand, we show that the arch-like vortices directly affect the interaction by periodically altering the shape of the separation shock. In addition, the spectral analysis in the streamwise, spanwise, and wall-normal directions, shows that a trace in the wall pressure of the \gls{kh} shedding frequency defined by \citet{bo2012experimental} and \citet{dellaposta2023direct_mach} is only visible at reattachment. Results thus suggest that the arch-like vortices may be less relevant for the separation delay, although their role in the unsteady closure of the bubble is still unrecognised. 

We then present a thorough analysis of the Fourier spectra in the streamwise direction, following the separation and the reattachment along the span, and in the wall-normal direction at the separation and reattachment at the symmetry plane. Results show that microramps induce an increase in the magnitude and the non-dimensional frequency, scaled by the separation length of the USBLI case, of the typical low-frequency \gls{sbli} unsteadiness. The rise is approximately constant along the span, although the relative contribution of the low-frequency fluctuations is attenuated behind the ramp compared to what happens at the lateral boundaries. Results thus indicate that, although the arch-like vortices periodically modify its shape, the separation shock front oscillates coherently at the increased frequency. The physical explanation for this increase in absolute terms may be related to existing models of \gls{sbli} low-frequency unsteadiness and will be the topic of future work. However, we also highlight that, despite the three-dimensionality of the interaction, using the local separation length to make the frequency non-dimensional is still effective in recovering the typical peak low frequency. 

Finally, we also used wavelet analysis to characterise how microramps affect the intermittency of the wall-pressure signal at separation and reattachment using the $LIM2$, which is a time-frequency decomposition of the flatness factor. Results show that there are no particular differences between the controlled and the uncontrolled cases, even if an even longer time integration period may be necessary to detect the differences in the higher-order statistics because of the significant intermittency of the flow also at low frequencies.

%
\backsection[Supplementary data]{\label{SupMat}A supplementary movie in high-resolution with a comparison of the controlled and uncontrolled \glspl{sbli}, generated at runtime by \textit{in situ} visualisation, is available at \url{https://www.youtube.com/@streamscfd6365}.}

\backsection[Acknowledgements]{We acknowledge CINECA Casalecchio di Reno (Italy) for providing us with the computational resources 
required by this work in the framework of the Early Access Phase of LEONARDO (LEAP\_082).}

\backsection[Funding]{
This research received financial support from ICSC – Centro Nazionale di Ricerca in "High Performance Computing, Big Data and Quantum Computing", funded by European Union – NextGenerationEU.}

\backsection[Declaration of interests]{The authors report no conflict of interest.}

\backsection[Data availability statement]{
The post-processing codes developed to read and analyse the data set can be found on the GitHub page \url{https://github.com/GiacDP/POSTPRO_MVG.git}. The full data set of the DNS simulations is on the order of hundreds of terabytes. By contacting the authors, a small subset of the data can be made available.}

\backsection[Author ORCID]{
G. Della Posta, \url{https://orcid.org/0000-0001-5516-9338}; 
E. Martelli, \url{https://orcid.org/0000-0002-7681-3513};
F. Salvadore, \url{https://orcid.org/0000-0002-1829-3388}; 
M. Bernardini, \url{https://orcid.org/0000-0001-5975-3734}
}

%
 \appendix

 \section{Time evolution of the density-weighted Favre-averaged vorticity}\label{sec:appendix}
If we define the Favre average of a generic variable $\phi$ as $\widetilde{\phi} = \overline{\rho\,\phi}/\overline{\rho}$, and we average the governing equations, the conservation of mass and momentum without external forces can be written as 
\begin{subequations}
\begin{align}
    \frac{\partial \overline{\rho}}{\partial t} + \nabla \cdot (\overline{\rho}\, \widetilde{\boldsymbol{u}}) 
    = 0 & \quad \Rightarrow \quad 
    \frac{\mathrm{D} \overline{\rho}}{\mathrm{D} t} = 0 \\
    \frac{\partial \overline{\rho} \, \widetilde{\boldsymbol{u}}}{\partial t} + \nabla \cdot (\overline{\rho} \, \widetilde{\boldsymbol{u}} \, \widetilde{\boldsymbol{u}}) 
    = \nabla \cdot (\overline{\tau} + \widetilde{\tau}^R) & \quad\Rightarrow \quad 
    \frac{\mathrm{D} \overline{\rho} \, \widetilde{\boldsymbol{u}}}{\mathrm{D} t} = \nabla \cdot \tau^t
\end{align}    
\end{subequations}
where $\mathrm{D}/\mathrm{D}t$ indicates the material derivative, $\tau_{ij}^t = \overline{\tau_{ij}} + \widetilde{\tau_{ij}}^R$ is the total stress tensor including both the Reynolds-averaged molecular stress tensor $\overline{\tau_{ij}}$ and the Favre-averaged Reynolds stress tensor $\widetilde{\tau_{ij}}^R = - \overline{\rho u_i'' u_j''}$.
By exploiting the conservation of mass, the momentum equation can be written in non-conservative form as
\begin{equation}
 \overline{\rho} \frac{\mathrm{D} \widetilde{\boldsymbol{u}}}{\mathrm{D} t} = \nabla \cdot \tau^t
\end{equation}
By taking the curl of the derived equation divided by $\overline{\rho}$, considering that $\nabla \cdot (\nabla \times \phi) = 0$ and the relation derivable using the Levi-Civita tensor for two generic vectors $\boldsymbol{a}$ and $\boldsymbol{b}$
\begin{equation}
    \nabla \times (\boldsymbol{a} \times \boldsymbol{b}) = 
    \boldsymbol{b} \cdot \nabla \boldsymbol{a} + 
    \boldsymbol{a} \nabla \cdot \boldsymbol{b} -
    \boldsymbol{b} \nabla \cdot \boldsymbol{a} -    
    \boldsymbol{a} \cdot \nabla \boldsymbol{b}
\end{equation}
we obtain that 
\begin{equation}
    \frac{\mathrm{D} \widetilde{\boldsymbol{\zeta}}}{\mathrm{D}t} =  \widetilde{\boldsymbol{\zeta}} \cdot \nabla \widetilde{\boldsymbol{u}} 
    - \widetilde{\boldsymbol{\zeta}} \, \nabla \cdot \widetilde{\boldsymbol{u}} + 
    \frac{\nabla \,\overline{\rho} \times \nabla (\overline{p} + 2/3 \, \overline{\rho}\,\widetilde{k})}{\overline{\rho}^2} 
    + \nabla \times \left( \frac{\nabla \cdot \tau^{t,d}}{\overline{\rho}} \right)
    \label{eq:vort}
\end{equation}
where we indicate with $\overline{p}$ the Reynolds-averaged pressure, with $\widetilde{k}$ the Favre-averaged turbulent kinetic energy equal to half the trace of the Favre-averaged Reynolds stress tensor, and with $\tau^{t,d}$ the deviatoric part of the total stress tensor given by $\tau_{ij}^{t,d} = \tau_{ij}^t - (\overline{p} + 2/3 \, \overline{\rho}\,\widetilde{k}) \delta_{ij}$ neglecting bulk viscosity. 

Using the expression of the mean divergence from the conservation of mass in non-conservative form 
\begin{equation}
    \nabla \cdot \widetilde{\boldsymbol{u}} = -\frac{1}{\overline{\rho}} \frac{\mathrm{D} \overline{\rho}}{\mathrm{D}t}
\end{equation}
we can write
\begin{equation}
    \frac{\mathrm{D} \widetilde{\boldsymbol{\zeta}}}{\mathrm{D}t} + \widetilde{\boldsymbol{\zeta}} \, \nabla \cdot \widetilde{\boldsymbol{u}} = \overline{\rho} \frac{\mathrm{D}}{\mathrm{D}t}\,\left( \frac{\widetilde{\boldsymbol{\zeta}}}{\overline{\rho}} \right)
\end{equation}
which substituted into equation~\ref{eq:vort} gives the equation of conservation for the density-weighted Favre-averaged vorticity
\begin{equation}
    \frac{\mathrm{D}}{\mathrm{D}t}\,\left( \frac{\widetilde{\boldsymbol{\zeta}}}{\overline{\rho}} \right) = \left(\frac{\widetilde{\boldsymbol{\zeta}}}{\overline{\rho}}\right) \cdot \nabla \widetilde{\boldsymbol{u}} + 
    \frac{\nabla \, \overline{\rho} \times \nabla (\overline{p} + 2/3 \, \overline{\rho}\,\widetilde{k})}{\overline{\rho}^3} 
    + \frac{1}{\overline{\rho}} \nabla \times \left( \frac{\nabla \cdot \tau^{t,d}}{\overline{\rho}} \right)    
\end{equation}
where the first term is the compressible vortex stretching and tilting, the second term is the baroclinic term due to modified pressure gradients, and the third term is the diffusion of vorticity by the action of both viscous and deviatoric, anisotropic turbulent stresses. 



\bibliographystyle{jfm}
\bibliography{main}

\end{document}